\documentclass[prd,aps,showpacs,epsf,floats]{revtex4}%
\usepackage{amssymb}
\usepackage{amsfonts}
\usepackage{amsmath}
\usepackage{graphicx}%
\providecommand{\U}[1]{\protect\rule{.1in}{.1in}}
\begin{document}
\title{\textbf{Approximate quantum error correction for generalized amplitude damping
errors}}
\author{Carlo Cafaro$^{1,2}$ and Peter van Loock$^{2}$}
\affiliation{$^{1}$Max-Planck Institute for the Science of Light, 91058 Erlangen, Germany }
\affiliation{$^{2}$Institute of Physics, Johannes-Gutenberg University Mainz, 55128 Mainz, Germany}

\begin{abstract}
We present analytic estimates of the performances of various approximate
quantum error correction schemes for the generalized amplitude damping (GAD)
qubit channel. Specifically, we consider both stabilizer and nonadditive
quantum codes. The performance of such error-correcting schemes is quantified
by means of the entanglement fidelity as a function of the damping probability
and the non-zero environmental temperature. The recovery scheme employed
throughout our work applies, in principle, to arbitrary quantum codes and is
the analogue of the perfect Knill-Laflamme recovery scheme adapted to the
approximate quantum error correction framework for the GAD error model. We
also analytically recover and/or clarify some previously known numerical
results in the limiting case of vanishing temperature of the environment, the
well-known traditional amplitude damping channel. In addition, our study
suggests that degenerate stabilizer codes and self-complementary nonadditive
codes are especially suitable for the error correction of the GAD noise model.
Finally, comparing the properly normalized entanglement fidelities of the best
performant stabilizer and nonadditive codes characterized by the same length,
we show that nonadditive codes outperform stabilizer codes not only in terms
of encoded dimension but also in terms of entanglement fidelity.

\end{abstract}

\maketitle

\section{Introduction}

Quantum computers are especially sensitive to noise. Therefore, any
technological implementation of such a machine requires the use of suitable
error mitigation techniques. Quantum error correction (QEC) represents one of
the most efficient available techniques capable of giving us the realistic
hope of building practical quantum computers \cite{shor1995}. For a detailed
overview of the basic working principles of QEC, we refer to \cite{gotty2010}.

In general, it is a highly nontrivial task to design quantum codes for any
given noise model. In the majority of cases, researchers have focused on the
error correction of Pauli-type errors. This type of error model certainly
constitutes the worst possible scenario to be considered and a quantum code
that can correct all Pauli errors $\left\{  I\text{, }X\equiv\sigma_{x}\text{,
}Y\equiv\sigma_{y}\text{, }Z\equiv\sigma_{z}\right\}  $,%
\begin{equation}
I\overset{\text{def}}{=}\left(
\begin{array}
[c]{cc}%
1 & 0\\
0 & 1
\end{array}
\right)  \text{, }\sigma_{x}\overset{\text{def}}{=}\left(
\begin{array}
[c]{cc}%
0 & 1\\
1 & 0
\end{array}
\right)  \text{, }\sigma_{y}\overset{\text{def}}{=}\left(
\begin{array}
[c]{cc}%
0 & -i\\
i & 0
\end{array}
\right)  \text{, }\sigma_{z}\overset{\text{def}}{=}\left(
\begin{array}
[c]{cc}%
1 & 0\\
0 & -1
\end{array}
\right)  \text{,} \label{paolino}%
\end{equation}
can also provide protection against arbitrary qubit noise, since the Pauli
operators form a basis of the $2\times2$ matrices. However, the worst possible
scenario is not necessarily the most realistic one in actual experimental
laboratories. Furthermore, aiming at designing quantum codes capable of
error-correcting general noise errors may not be the most fruitful way to
combat decoherence and noise in quantum computers.

A very common type of noise that appears in realistic settings is the
so-called amplitude damping (AD) noise model \cite{nielsen-book}. For
instance, the AD noise model is employed to describe the photon loss in an
optical fiber. The AD channel is the simplest channel whose Kraus operators
cannot be described by unitary Pauli operations. The two non-unitary Kraus
operators for the qubit AD channel are given by \cite{nielsen-book},%
\begin{equation}
A_{0}\overset{\text{def}}{=}\frac{1}{2}\left[  \left(  1+\sqrt{1-\gamma
}\right)  I+\left(  1-\sqrt{1-\gamma}\right)  \sigma_{z}\right]  \text{ and,
}A_{1}\overset{\text{def}}{=}\frac{\sqrt{\gamma}}{2}\left(  \sigma_{x}%
+i\sigma_{y}\right)  =\sqrt{\gamma}\left\vert 0\right\rangle \left\langle
1\right\vert \text{,} \label{a0}%
\end{equation}
where $\gamma$ denotes the\textbf{ }amplitude damping probability parameter.
For the sake of completeness, we point out that AD channels acting on states
characterized by higher photon numbers in combination with a finite number of
modes can also be considered. For instance, qubits living in a two-dimensional
Hilbert space can be replaced by bosonic states of higher photon numbers in a
finite number of optical modes \cite{isaac}. In general, the operator-sum
decomposition of such higher-dimensional noise models is characterized by
error operators\textbf{ }$A_{k}$\textbf{ }with\textbf{ }$k=1$\textbf{,...,
}$N$\textbf{ }and, in principle,\textbf{ }$N$\textbf{ }can approach infinity
\cite{nielsen-book}. We observe that there is no simple way of reducing
$A_{1}$ in Eq. (\ref{a0}) to one Pauli error operator, since $\left\vert
0\right\rangle \left\langle 1\right\vert $ is not normal. The Pauli operators
are remarkable in that they are unitary and Hermitian at the same time and, in
addition, both unitary and Hermitian operators are normal. Although
the\textbf{ }five-qubit code \cite{laflamme1, bennett1} can be successfully
used for the error correction of AD\ errors, since it is a\textbf{ }universal
$1$-error correcting quantum code, Leung et \textit{al}. were capable of
designing a four-qubit quantum code especially suitable for the error
correction of arbitrary single-AD errors with a higher encoding rate equal to
$1/4$ (greater\textbf{ }than $1/5)$ \cite{debbie}. A quantum code with higher
encoding rate is very welcome, since it would require less resources for its
implementation. The two main points advocated in \cite{debbie} were the
following: first, when dealing with specific error models, better codes may be
uncovered; second, the fulfillment of the approximate (relaxed) QEC conditions
enlarges the realm of possible useful quantum error correcting codes.
Therefore, it can simplify the code construction process.

Both\textbf{ }the five-qubit and four-qubit codes are nondegenerate stabilizer
(additive) quantum codes \cite{nielsen-book}. However, the scientific
literature accommodates a fairly wide variety of additional AD error
correction schemes where neither additive nor nondegenerate quantum codes are
employed. For instance, one of the very first QEC scheme used to combat
AD\ errors was quite unconventional, since it used bosonic states of higher
photon numbers in a finite number of optical modes (the so-called bosonic
quantum codes, \cite{isaac}). More conventional and recent works are inspired
by the seminal work presented in \cite{debbie}. In \cite{andy1}, using the
stabilizer formalism, generalizations of the Leung's et \textit{al}.
four-qubit code for higher rates were constructed. Specifically, a class of
$\left[  \left[  2\left(  m+1\right)  \text{, }m\right]  \right]  $
channel-adapted quantum codes for integers $m\geq1$ with encoding rate
arbitrarily close to $1/2$ are generated. In \cite{andy2}, the performance of
QEC schemes for the AD model has been investigated via semidefinite programs
(that is, numerical convex optimization methods \cite{dover-book}).
Specifically, the optimal recovery operation to maximize the entanglement
fidelity \cite{schumy} for a given encoding and noise process was uncovered.
Unfortunately, numerically computed recovery maps are difficult to describe
and understand analytically. Furthermore, recovery operations generated
through convex optimization methods suffer two significant drawbacks. First,
the dimensions of the optimization problem grow exponentially with the length
of the code, limiting the technique to short codes. Second, the optimal
recovery operation may be quite difficult to implement, despite being
physically legitimate. However, this exponential growth can be mitigated in
two manners. First, it is possible to reduce the high dimensionality of the
convex optimization procedures for generating recovery operations in QEC by
transforming the problem in a suitable manner. Consider a quantum noisy
channel $\Lambda:\mathcal{H}_{1}\rightarrow\mathcal{H}_{2}$ with
$\dim\mathcal{H}_{i}=d_{i}$ for $i=1$, $2$. By embedding the encoding into the
noise process and redefining $\Lambda$ as a quantum spreading channel with
$d_{1}$ strictly less than $d_{2}$, the dimensionality of the convex
optimization decreases from $d_{1}^{2}d_{2}^{2}$ to $d_{1}^{4}$. For instance,
for the $\left[  \left[  5\text{, }1\text{, }3\right]  \right]  $ five-qubit
code, the dimensionality reduces from $2^{20}$ to $2^{12}$ in terms of the
optimization variables. For more details, we refer to \cite{andy2}. Second, if
the focus is on near-optimal rather than on optimal recovery schemes, it turns
out that it is possible to compute such recovery operations with less
computationally intensive numerical algorithms that are more scalable than
those involved in semidefinite programs. Briefly speaking, regarding the
algorithm in terms of eigenanalysis, it is possible to significantly reduce
the size of the eigenvector problem and this has a significant effect on the
computational cost of these new algorithms. For illustrative purposes,
consider the AD channel and an $\left[  \left[  n\text{, }k\text{, }d\right]
\right]  $ quantum code. In this case, the dimensionality of the full optimal
semidefinite program grows as $4^{n}$. Instead, if one considers the
semidefinite programs for the first- and second-order subspaces, the
dimensionality of the approximate programs only grows as $n^{2}$ and $n^{4}$,
respectively. For instance, using the $\left[  \left[  7\text{, }1\text{,
}3\right]  \right]  $ CSS seven-qubit code, the full optimal semidefinite
program requires $65536$ optimization variables. However, the first-order
semidefinite program requires $1024$ variables and the second-order
semidefinite program has $7056$ optimization variables. For further details,
we refer to \cite{andy3}. In particular, in \cite{andy3}, to mitigate such
drawbacks of the optimal recovery, a structured near-optimal channel-adapted
recovery procedure was determined and applied to the AD noise model. In
\cite{jmp}, an analytical approach to channel-adapted recovery based on the
pretty-good measurement and the average entanglement fidelity appeared.
Following \cite{jmp}, a simple analytical approach to approximate QEC based on
the transpose channel was derived and used for the AD noise model in
\cite{ngm}. In particular, it was shown in \cite{ngm} that the transpose
channel is a recovery map that coincides with the perfect recovery map for
codes satisfying the perfect Knill-Laflamme\textbf{ }QEC conditions \cite{kl}.
Very recently, it was also proved that the transpose channel works nearly as
well as the optimal recovery channel, with optimality defined in terms of
worst-case fidelity over all code states \cite{ngm2}. As a side remark, we
underline that no definitive choice for the best figure of merit in quantum
information processing tasks has been made yet \cite{miken} and this fact
becomes especially relevant when quantifying the performance of quantum codes
\cite{sainz, cafaro-osid, ric1, ric2, ric3}. In \cite{sainz}, focusing on the
AD noise model, it was shown that fidelity alone may be not sufficient to
compare the efficiency of different error correction codes. In
\cite{lang-shor}, a numerical search based upon a greedy algorithm
\cite{dover-book} was employed to construct a family of high rate nonadditive
quantum codes adapted to the AD noise model that outperform (in terms of
encoded dimension) the stabilizer codes presented in \cite{andy1}. In
\cite{shor-2011}, families of high performance nonadditive quantum codes of
the codeword stabilized (CWS, \cite{cws}) type for single AD-errors that
outperform, in terms of encoded dimension, the best possible additive codes
were presented. These code families were built from nonlinear error-correcting
codes for classical asymmetric channels or classical codes over $GF(3)$.
Finally, for the sake of completeness, we also point out that a method for the
construction of good multi-error-correcting AD codes that are both degenerate
and additive can be found in \cite{bz}.

Taking into account all these above-mentioned facts, we emphasize that there is:

\begin{itemize}
\item no clear understanding of the role played by degeneracy in the analysis
of the performance of stabilizer quantum codes for AD errors, \cite{andy3};

\item no explicit evidence of the relevance of the role played by
self-complementarity in the analysis of the performance (quantified by means
of the entanglement fidelity) of nonadditive quantum codes, \cite{shor-2011};

\item no explicit analytical computation of the entanglement fidelity of
arbitrary quantum codes for AD errors for simple-to-construct recovery maps,
\cite{ngm2};

\item no explicit performance comparison in terms of the entanglement fidelity
between additive and nonadditive quantum codes for AD errors, \cite{lang-shor}.
\end{itemize}

In this article, following the working methodology advocated by one of the
Authors in \cite{cafaro-pra1, cafaro-pra2}, we seek to advance our
quantitative understanding via simple analytical computations of the role
played by degeneracy, self-complementarity, additivity and nonadditivity of
quantum codes for the GAD error model \cite{nielsen-book}. Specifically, we
present the analysis of the performance of various approximate quantum error
correction schemes for the GAD\ channel. We consider both stabilizer and
nonadditive quantum codes. The performance of such error-correcting schemes is
quantified by means of the entanglement fidelity as a function of the damping
probability and the non-zero environmental temperature. The recovery scheme
employed throughout our work applies, in principle, to arbitrary quantum codes
and is the analogue of the perfect Knill-Laflamme recovery scheme adapted to
the approximate quantum error correction framework for the GAD error model. We
also analytically recover and/or clarify some previously known numerical
results in the limiting case of vanishing temperature of the environment. In
addition, our extended analytical investigation suggests that degenerate
stabilizer codes and self-complementary nonadditive codes are especially
suitable for the error correction of the GAD noise model. Finally, comparing
the properly normalized entanglement fidelities of the best performant
stabilizer and nonadditive codes characterized by the same length, we show
that nonadditive codes outperform stabilizer codes not only in terms of
encoded dimension but also in terms of fidelity.

This article is organized as follows. In Section II, we describe the
GAD\ noise model. In Section III, we present some preliminary material
concerning exact and approximate \ QEC conditions, recovery maps, and
entanglement fidelity. In Section IV, we analyze the performances of various
stabilizer codes, both degenerate and nondegenerate. In Section V, we quantify
the performances of various nonadditive codes, both self-complementary and
non-self-complementary. Finally, our conclusions are presented in Section VI.
A number of appendices with technical details of calculations are also provided.

\section{The GAD\ noise model}

The AD quantum operation can characterize the behavior of different types of
dissipative open quantum systems \cite{nielsen-book}: the spontaneous emission
of a single atom coupled to a single mode of the electromagnetic radiation,
the gradual loss of energy from a principal system to the environment where
both systems are modeled by simple harmonic oscillators or the scattering of a
photon via a beam splitter represent physical processes modeled by an AD channel.

It can be shown that the GAD qubit channel can be realized by considering the
evolution of a two-level quantum system (that is, a qubit) in a dissipative
interaction, in the Born-Markov rotating-wave approximation \cite{scully},
with a bath of harmonic oscillators taken to be initially in a thermal state
\cite{epj, jee}.

The Lindblad form of the master equation that describes the evolution that
generates the GAD channel reads \cite{jee},%
\begin{equation}
\frac{d\rho^{S}\left(  t\right)  }{dt}=\sum_{j=1}^{2}\left(  2R_{j}\rho
^{S}R_{j}^{\dagger}-R_{j}^{\dagger}R_{j}\rho^{S}-\rho^{S}R_{j}^{\dagger}%
R_{j}\right)  \text{,} \label{le}%
\end{equation}
where the operators $R_{1}$, $R_{2}$ and $R$ are given by,%
\begin{equation}
R_{1}\overset{\text{def}}{=}\left[  \frac{\gamma_{0}}{2}\left(  N_{\text{th}%
}+1\right)  \right]  ^{\frac{1}{2}}R\text{, }R_{2}\overset{\text{def}}%
{=}\left(  \frac{\gamma_{0}N_{\text{th}}}{2}\right)  ^{\frac{1}{2}}R^{\dagger
}\text{, }R\overset{\text{def}}{=}\sigma_{-}\text{,}%
\end{equation}
with $\gamma_{0}$, $N_{\text{th}}$ and $\sigma_{-}$ defined as,%
\begin{equation}
\gamma_{0}\overset{\text{def}}{=}\frac{4\omega^{3}\left\vert \mathbf{d}%
\right\vert ^{2}}{3\hbar c^{3}}\text{, }N_{\text{th}}\overset{\text{def}}%
{=}\frac{1}{e^{\frac{\hbar\omega}{k_{B}T}}-1}\text{, }\sigma_{-}%
\overset{\text{def}}{=}\frac{\sigma_{x}-i\sigma_{y}}{2}=\left\vert
1\right\rangle \left\langle 0\right\vert \text{.} \label{q}%
\end{equation}
In Eq. (\ref{q}), $\gamma_{0}$ denotes the spontaneous emission rate, $\omega$
is the photonic frequency, $\hbar$ is the Planck constant divided by $2\pi$,
$c$ is the speed of light, $\mathbf{d}$ is the transition matrix of the atomic
dipole operator describing the interaction between the two-level quantum
system with the bath of harmonic oscillators, $k_{B}$ is the Boltzmann
constant, $\sigma_{-}$ is the lowering operator, $N_{\text{th}}$ is the Planck
distribution that gives the number of thermal photons at frequency $\omega$
and, finally, $T$ denotes the temperature of the environment. The operator
$\rho^{S}$ denotes the reduced density matrix operator of the two-level
quantum system interacting with a thermal bath in the weak Born-Markov
rotating-wave approximation \cite{jee}. We remark that when $T=0$, then
$N_{\text{th}}=0$ and $R_{2}=0$. Therefore, when the temperature of the
environment is zero, a single Lindblad operator is sufficient to describe the
master equation.

The evolution of the density operator $\rho^{S}$ in Eq. (\ref{le}) can be
given a Kraus operator-sum decomposition. The Kraus representation is useful,
because it provides an intrinsic description of the principal system, without
explicitly considering the detailed properties of the environment. The
essential features of the problem are contained in the Kraus error operators
$A_{k}$. This not only simplifies calculations, but often provides theoretical insight.

Following \cite{jee}, it turns out that the Kraus decomposition of the GAD
channel becomes,%
\begin{equation}
\Lambda_{\text{GAD}}\left(  \rho\right)  \overset{\text{def}}{=}\sum_{k=0}%
^{3}A_{k}\rho A_{k}^{\dagger}\text{,}%
\end{equation}
where the Kraus error operators $A_{k}$ read,
\begin{align}
&  A_{0}\overset{\text{def}}{=}\frac{\sqrt{p}}{2}\left[  \left(
1+\sqrt{1-\gamma}\right)  I+\left(  1-\sqrt{1-\gamma}\right)  \sigma
_{z}\right]  \text{, }A_{1}\overset{\text{def}}{=}\frac{\sqrt{p}\sqrt{\gamma}%
}{2}\left(  \sigma_{x}+i\sigma_{y}\right)  \text{,}\nonumber\\
& \nonumber\\
&  A_{2}\overset{\text{def}}{=}\frac{\sqrt{1-p}}{2}\left[  \left(
1+\sqrt{1-\gamma}\right)  I-\left(  1-\sqrt{1-\gamma}\right)  \sigma
_{z}\right]  \text{, }A_{3}\overset{\text{def}}{=}\frac{\sqrt{1-p}\sqrt
{\gamma}}{2}\left(  \sigma_{x}-i\sigma_{y}\right)  \text{,} \label{ak}%
\end{align}
where $\gamma$ is the damping parameter and $0\leq p\leq1$ \cite{nielsen-book}%
. The $\left(  2\times2\right)  $-matrix representation of the operators
$A_{k}$ in Eq. (\ref{ak}) is given by,%
\begin{equation}
A_{0}=\sqrt{p}\left(
\begin{array}
[c]{cc}%
1 & 0\\
0 & \sqrt{1-\gamma}%
\end{array}
\right)  \text{, }A_{1}=\sqrt{p}\left(
\begin{array}
[c]{cc}%
0 & \sqrt{\gamma}\\
0 & 0
\end{array}
\right)  \text{, }A_{2}=\sqrt{1-p}\left(
\begin{array}
[c]{cc}%
\sqrt{1-\gamma} & 0\\
0 & 1
\end{array}
\right)  \text{, }A_{3}=\sqrt{1-p}\left(
\begin{array}
[c]{cc}%
0 & 0\\
\sqrt{\gamma} & 0
\end{array}
\right)  \text{,} \label{ko}%
\end{equation}
and their action on the computational basis vectors $\left\vert 0\right\rangle
$ and $\left\vert 1\right\rangle $ of the \emph{complex} Hilbert space
$\mathcal{H}_{2}^{1}$ (the Hilbert space of $1$-qubit quantum states) reads,%
\begin{align}
A_{0}\left\vert 0\right\rangle  &  =\sqrt{p}\left\vert 0\right\rangle \text{,
}A_{0}\left\vert 1\right\rangle =\sqrt{p}\sqrt{1-\gamma}\left\vert
1\right\rangle \text{, }A_{1}\left\vert 0\right\rangle \equiv0\text{, }%
A_{1}\left\vert 1\right\rangle =\sqrt{p}\sqrt{\gamma}\left\vert 0\right\rangle
\text{,}\nonumber\\
& \nonumber\\
A_{2}\left\vert 0\right\rangle  &  =\sqrt{1-p}\sqrt{1-\gamma}\left\vert
0\right\rangle \text{, }A_{2}\left\vert 1\right\rangle =\sqrt{1-p}\left\vert
1\right\rangle \text{ }A_{3}\left\vert 0\right\rangle =\sqrt{1-p}\sqrt{\gamma
}\left\vert 1\right\rangle \text{, }A_{3}\left\vert 1\right\rangle
\equiv0\text{.}%
\end{align}
Notice that for $p=1$, the Kraus operator-sum decomposition of the AD channel
can be recovered. The GAD channel generalizes the AD channel in that it allows
transitions from $\left\vert 0\right\rangle \rightarrow$ $\left\vert
1\right\rangle $ as well as from $\left\vert 1\right\rangle \rightarrow$
$\left\vert 0\right\rangle $. For an alternative and explicit derivation of
the operator-sum decomposition of the GAD channel in the context of scattering
of a photon via a beam splitter, we refer to Appendix A.

We emphasize that the GAD channel is particularly noisy and, unlike the AD
channel, is characterized by a two-dimensional parametric region $\left(
\gamma\text{, }p\left(  \gamma\right)  \right)  $ where it exhibits
entanglement breaking features (for details, see Appendix B). From \cite{jee},
it also follows that the two GAD channel parameters $\gamma$ and $p$ are
formally given by,
\begin{equation}
\gamma\left(  t\right)  \overset{\text{def}}{=}1-e^{-\gamma_{0}\left(
2N_{\text{th}}+1\right)  t}\text{ and, }p\overset{\text{def}}{=}%
\frac{N_{\text{th}}+1}{2N_{\text{th}}+1}\text{.} \label{gamma}%
\end{equation}
Observe that $p=1$ when $T=0$ and $p=\frac{1}{2}$ when $T$ approaches
infinity. For the sake of future convenience, we also introduce a new
additional parameter $\varepsilon$ defined as,%
\begin{equation}
\varepsilon\left(  T\right)  \overset{\text{def}}{=}1-p\left(  T\right)
=\frac{e^{-\frac{\hbar\omega}{k_{B}T}}}{1+e^{-\frac{\hbar\omega}{k_{B}T}}%
}\overset{T\ll1}{\approx}e^{-\frac{\hbar\omega}{k_{B}T}}\text{.}
\label{epsilon}%
\end{equation}
Combining Eqs. (\ref{gamma}) and (\ref{epsilon}), we get%
\begin{equation}
\gamma\left(  t\right)  \equiv\gamma_{\varepsilon}\left(  t\right)
\overset{\text{def}}{=}1-\exp\left[  -\left(  \frac{1}{1-2\varepsilon}\right)
\gamma_{0}t\right]  \overset{\varepsilon\ll1}{\approx}1-e^{-\gamma_{0}\left(
1+2\varepsilon\right)  t}\text{.} \label{gamma-epsilon}%
\end{equation}
From Eq. (\ref{gamma-epsilon}), we conclude that $\gamma_{\varepsilon}$ is a
monotonic increasing function of $\varepsilon$ for $\varepsilon\ll1$ and fixed
values of $\gamma_{0}$ and $t$.

\section{QEC conditions, recovery maps and entanglement fidelity}

\subsection{QEC conditions}

\subsubsection{Exact QEC}

Sufficient conditions for approximate QEC were introduced by Leung et
\textit{al}. in \cite{debbie}. They showed that quantum codes can be effective
in the error correction procedure even though they violate\textbf{ }the
traditional (exact) Knill-Laflamme QEC conditions \cite{kl}. However, these
violations, characterized by small deviations from the standard
error-correction conditions are allowed provided that they do not affect the
desired fidelity order.

For the sake of reasoning, let us consider a binary quantum stabilizer code
$\mathcal{C}$ with code parameters $\left[  \left[  n,k,d\right]  \right]  $
encoding $k$-logical qubits in the Hilbert space $\mathcal{H}_{2}^{k}$ into
$n$-physical qubits in the Hilbert space $\mathcal{H}_{2}^{n}$ with distance
$d$. Assume that the noise model after the encoding procedure is
$\Lambda\left(  \rho\right)  $ and can be described by an operator-sum
representation,%
\begin{equation}
\Lambda\left(  \rho\right)  \overset{\text{def}}{=}\sum_{k\in\mathcal{K}}%
A_{k}\rho A_{k}^{\dagger}\text{,}%
\end{equation}
where $\mathcal{K}$ is the index set of all the enlarged Kraus operators
$A_{k}$ that appear in the sum. The noise channel\textbf{ }$\Lambda$\textbf{
}is a CPTP (completely positive and trace preserving) map. The codespace of
$\mathcal{C}$ is a $2^{k}$-dimensional subspace of $\mathcal{H}_{2}^{n}$ where
some error operators that characterize the error model $\Lambda$ being
considered can be reversed. Denote with $\mathcal{A}_{\text{reversible}%
}\subset$ $\mathcal{A}\overset{\text{def}}{=}\left\{  A_{k}\right\}  $ with
$k\in\mathcal{K}$ the set of reversible enlarged errors $A_{k}$ on
$\mathcal{C}$ such that $\mathcal{K}_{\text{reversible}}\overset{\text{def}%
}{=}\left\{  k:A_{k}\in\mathcal{A}_{\text{reversible}}\right\}  $ is the index
set of $\mathcal{A}_{\text{reversible}}$. Therefore, the noise model
$\Lambda^{\prime}\left(  \rho\right)  $ given by,%
\begin{equation}
\Lambda^{\prime}\left(  \rho\right)  \overset{\text{def}}{=}\sum
_{k\in\mathcal{K}_{\text{reversible}}}A_{k}\rho A_{k}^{\dagger}\text{,}
\label{g1}%
\end{equation}
is reversible on $\emph{C}\subset\mathcal{H}_{2}^{n}$. The noise
channel\textbf{ }$\Lambda^{\prime}$\textbf{ }denotes a CP but non-TP map. The
enlarged error operators $A_{k}$ in $\mathcal{A}_{\text{reversible}}$ satisfy
the standard QEC conditions \cite{nielsen-book},%
\begin{equation}
P_{\mathcal{C}}A_{l}^{\dagger}A_{m}P_{\mathcal{C}}=\alpha_{lm}P_{\mathcal{C}%
}\text{,} \label{vai}%
\end{equation}
or, equivalently,%
\begin{equation}
\left\langle i_{L}\left\vert A_{l}^{\dagger}A_{m}\right\vert j_{L}%
\right\rangle =\alpha_{lm}\delta_{ij}\text{,} \label{kl-may}%
\end{equation}
for any $l$, $m\in\mathcal{K}_{\text{reversible}}$, $P_{\mathcal{C}}$ denotes
the projector on the codespace and $\alpha_{lm}$ are entries of a positive
Hermitian matrix. \ It is helpful to regard the Knill-Laflamme condition in
Eq. (\ref{kl-may}) as embodying two conditions \cite{looi}: the obvious
off-diagonal condition saying that the matrix elements of $A_{l}^{\dagger
}A_{m}$ must vanish when $i\neq j$ (orthogonality condition);\textbf{ }and the
diagonal condition which, since $\alpha_{lm}$ are entries of a positive
Hermitian \emph{complex} matrix, is nothing but the requirement that all
diagonal elements of\textbf{ }$A_{l}^{\dagger}A_{m}$ (inside the coding space)
be identical (non-deformability condition).\textbf{ }The fulfillment of Eq.
(\ref{vai}) for some subset of enlarged error operators $A_{k}$ that
characterize the operator-sum representation of the noise model $\Lambda$
implies that there exists a new operator-sum decomposition of $\Lambda$ such
that $\Lambda^{\prime}\left(  \rho\right)  $ in Eq. (\ref{g1}) becomes,%
\begin{equation}
\Lambda^{\prime}\left(  \rho\right)  \overset{\text{def}}{=}\sum
_{k\in\mathcal{K}_{\text{reversible}}^{\prime}}A_{k}^{\prime}\rho
A_{k}^{\prime\dagger}\text{,}%
\end{equation}
where Eq. (\ref{vai}) is replaced by%
\begin{equation}
P_{\mathcal{C}}A_{l}^{\prime\dagger}A_{m}^{\prime}P_{\mathcal{C}}=p_{m}%
\delta_{lm}P_{\mathcal{C}}\text{,} \label{vai2}%
\end{equation}
for any $l$, $m\in\mathcal{K}_{\text{reversible}}^{\prime}$ with the error
detection probabilities $p_{m}$ non-negative $c$-numbers. We remark that Eq.
(\ref{vai2}) is equivalent to the traditional orthogonality and
non-deformation conditions (see Eq. (\ref{kl-may})) for a nondegenerate code,%
\begin{equation}
\left\langle i_{L}\left\vert A_{l}^{\dagger}A_{m}\right\vert j_{L}%
\right\rangle =\delta_{ij}\delta_{lm}p_{m} \label{cacchio}%
\end{equation}
for any $i$, $j$ labelling the logical states and $l$, $m\in\mathcal{K}%
_{\text{reversible}}$. Observe that for any linear operator $A_{k}^{\prime}$
on a vector space $V$ there exists a unitary $U_{k}$ and a positive operator
$J\overset{\text{def}}{=}\sqrt{A_{k}^{\prime\dagger}A_{k}^{\prime}}$ such that
\cite{nielsen-book},%
\begin{equation}
A_{k}^{\prime}=U_{k}J=U_{k}\sqrt{A_{k}^{\prime\dagger}A_{k}^{\prime}}\text{.}
\label{pd}%
\end{equation}
We stress that $J$ is the unique positive operator that satisfies Eq.
(\ref{pd}). As a matter of fact, multiplying $A_{k}^{\prime}=U_{k}J$ on the
left by the adjoint equation $A_{k}^{\prime\dagger}=JU_{k}^{\dagger}$ gives%
\begin{equation}
A_{k}^{\prime\dagger}A_{k}^{\prime}=JU_{k}^{\dagger}U_{k}J=J^{2}\Rightarrow
J=\sqrt{A_{k}^{\prime\dagger}A_{k}^{\prime}}\text{.}%
\end{equation}
Furthermore, if $A_{k}^{\prime}$ is invertible (that is, $\det A_{k}^{\prime
}\neq0$), $U_{k}$ is unique and reads\textbf{,}%
\begin{equation}
U_{k}\overset{\text{def}}{=}A_{k}^{\prime}J^{-1}=A_{k}^{\prime}\left(
\sqrt{A_{k}^{\prime\dagger}A_{k}^{\prime}}\right)  ^{-1}\text{.} \label{ai1}%
\end{equation}
How do we choose the unitary $U_{k}$ when $A_{k}^{\prime}$ is not invertible?
The operator $J$ is a positive operator and belongs to a special subclass of
Hermitian operators such that for any vector $\left\vert v\right\rangle \in
V$, $\left(  \left\vert v\right\rangle \text{, }J\left\vert v\right\rangle
\right)  $ is a \emph{real} and non-negative number. Therefore, $J$ has a
spectral decomposition%
\begin{equation}
J\overset{\text{def}}{=}\sqrt{A_{k}^{\prime\dagger}A_{k}^{\prime}}=%
%TCIMACRO{\dsum \limits_{l}}%
%BeginExpansion
{\displaystyle\sum\limits_{l}}
%EndExpansion
\lambda_{l}\left\vert l\right\rangle \left\langle l\right\vert \text{,}%
\end{equation}
where $\lambda_{l}\geq0$ and $\left\{  \left\vert l\right\rangle \right\}  $
denotes an orthonormal basis for the vector space $V$. Define the vectors
$\left\vert \psi_{l}\right\rangle \overset{\text{def}}{=}A_{k}^{\prime
}\left\vert l\right\rangle $ and notice that,%
\begin{equation}
\left\langle \psi_{l}\left\vert \psi_{l}\right.  \right\rangle =\left\langle
l\left\vert A_{k}^{\prime\dagger}A_{k}^{\prime}\right\vert l\right\rangle
=\left\langle l\left\vert J^{2}\right\vert l\right\rangle =\lambda_{l}%
^{2}\text{.}%
\end{equation}
For the time being, consider only those $l$ for which $\lambda_{l}\neq0$. For
those $l$, consider the vectors $\left\vert e_{l}\right\rangle $ defined as%
\begin{equation}
\left\vert e_{l}\right\rangle \overset{\text{def}}{=}\frac{\left\vert \psi
_{l}\right\rangle }{\lambda_{l}}=\frac{A_{k}^{\prime}\left\vert l\right\rangle
}{\lambda_{l}}\text{,}%
\end{equation}
with $\left\langle e_{l}\left\vert e_{l^{\prime}}\right.  \right\rangle
=\delta_{ll^{\prime}}$. For those $l$ for which $\lambda_{l}=0$, extend the
orthonormal set $\left\{  \left\vert e_{l}\right\rangle \right\}  $ in such a
manner that it forms an orthonormal basis $\left\{  \left\vert E_{l}%
\right\rangle \right\}  $. Then, a suitable choice for the unitary operator
$U_{k}$ such that%
\begin{equation}
A_{k}^{\prime}\left\vert l\right\rangle =U_{k}J\left\vert l\right\rangle
\text{,}%
\end{equation}
with $\left\{  \left\vert l\right\rangle \right\}  $ an orthonormal basis for
$V$ \ reads,%
\begin{equation}
U_{k}\overset{\text{def}}{=}%
%TCIMACRO{\dsum \limits_{l}}%
%BeginExpansion
{\displaystyle\sum\limits_{l}}
%EndExpansion
\left\vert E_{l}\right\rangle \left\langle l\right\vert \text{.} \label{ani}%
\end{equation}
In summary, the unitary $U_{k}$ is uniquely determined by Eq. (\ref{ai1}) when
$A_{k}^{\prime}$ is invertible or Eq. (\ref{ani}) when $A_{k}^{\prime}$ is not
necessarily invertible. We finally stress that the non-uniqueness of $U_{k}$
when $\det A_{k}^{\prime}=0$ is due to the freedom in choosing the orthonormal
basis $\left\{  \left\vert l\right\rangle \right\}  $ for the vector space $V$.

In the scenario being considered, when Eq. (\ref{vai2}) is satisfied, the
enlarged error operators $A_{m}^{\prime}$ admit polar decompositions,%
\begin{equation}
A_{m}^{\prime}P_{\mathcal{C}}=\sqrt{p_{m}}U_{m}P_{\mathcal{C}}\text{,}
\label{pd1}%
\end{equation}
with $m\in\mathcal{K}_{\text{reversible}}$. From Eqs. (\ref{vai2}) and
(\ref{pd1}), we get%
\begin{equation}
p_{m}\delta_{lm}P_{\mathcal{C}}=P_{\mathcal{C}}A_{l}^{\prime\dagger}%
A_{m}^{\prime}P_{\mathcal{C}}=\sqrt{p_{l}p_{m}}P_{\mathcal{C}}U_{l}^{\dagger
}U_{m}P_{\mathcal{C}}\text{,}%
\end{equation}
that is,%
\begin{equation}
P_{\mathcal{C}}U_{l}^{\dagger}U_{m}P_{\mathcal{C}}=\delta_{lm}P_{\mathcal{C}%
}\text{.} \label{condo}%
\end{equation}
We stress that Eq. (\ref{condo}) is needed for an unambiguous syndrome
detection since, as a consequence of the orthogonality of different
$R_{m}^{\dagger}\overset{\text{def}}{=}U_{m}P_{\mathcal{C}}$, the recovery
operation $\mathcal{R}$ is trace preserving. This can be shown as follows.

Let $\mathcal{V}^{i_{L}}$ be the subspace of $\mathcal{H}_{2}^{n}$ spanned by
the corrupted images $\left\{  A_{k}^{\prime}\left\vert i_{L}\right\rangle
\right\}  $ of the codewords $\left\vert i_{L}\right\rangle $. Let $\left\{
\left\vert v_{r}^{i_{L}}\right\rangle \right\}  $ be an orthonormal basis for
$\mathcal{V}^{i_{L}}$. We define such a subspace $\mathcal{V}^{i_{L}}$ for
each of the codewords. Because of the traditional Knill-Laflamme QEC
conditions \cite{kl},%
\begin{align}
\left\langle i_{L}|A_{k}^{\dagger}A_{k^{\prime}}|i_{L}\right\rangle  &
=\left\langle j_{L}|A_{k}^{\dagger}A_{k^{\prime}}|j_{L}\right\rangle \text{,
}\forall i\text{, }j\nonumber\\
& \nonumber\\
\left\langle i_{L}|A_{k}^{\dagger}A_{k^{\prime}}|j_{L}\right\rangle  &
=0\text{, }\forall\text{ }i\neq j\text{,}%
\end{align}
the subspaces $\mathcal{V}^{i_{L}}$ and $\mathcal{V}^{j_{L}}$ with $i\neq j$
are orthogonal subspaces. If $\mathcal{V}^{i_{L}}\oplus\mathcal{V}^{j_{L}}$ is
a proper subset of $\mathcal{H}_{2}^{n}$ with $\mathcal{V}^{i_{L}}%
\oplus\mathcal{V}^{j_{L}}\neq\mathcal{H}_{2}^{n}$, we denote its orthogonal
complement by $\mathcal{O}$,%
\begin{equation}
\mathcal{H}_{2}^{n}\overset{\text{def}}{=}\left(  \mathcal{V}^{i_{L}}%
\oplus\mathcal{V}^{j_{L}}\right)  \oplus\mathcal{O}\text{,}%
\end{equation}
where,%
\begin{equation}
\mathcal{O}\overset{\text{def}}{=}\left(  \mathcal{V}^{i_{L}}\oplus
\mathcal{V}^{j_{L}}\right)  ^{\perp}\text{.} \label{deo}%
\end{equation}
Let $\left\{  \left\vert o_{k}\right\rangle \right\}  $ be an orthonormal
basis for $\mathcal{O}$. Then, the set of states $\left\{  \left\vert
v_{r}^{i_{L}}\right\rangle \text{, }\left\vert o_{k}\right\rangle \right\}  $
constitutes an orthonormal basis for $\mathcal{H}_{2}^{n}$. Notice that, since
$\left\vert v_{r}^{i_{L}}\right\rangle $ are mutually orthogonal, there exist
unitary $V_{r}$ such that $V_{r}\left\vert v_{r}^{i_{L}}\right\rangle
=\left\vert i_{L}\right\rangle $ ($V_{r}$ is an isometry which returns
$\left\vert v_{r}^{i_{L}}\right\rangle $ to the corresponding $\left\vert
i_{L}\right\rangle $). We introduce the quantum recovery operation
$\mathcal{R}$ with operation elements%
\begin{equation}
\mathcal{R}\overset{\text{def}}{=}\left\{  R_{1}\text{,..., }R_{r}\text{,...,
}\hat{O}\right\}  \text{,} \label{ecce}%
\end{equation}
with,%
\begin{equation}
\mathcal{R}\left(  \rho\right)  =\sum_{k\in\mathcal{K}_{\text{reversible}%
}^{\prime}}R_{k}\rho R_{k}^{\dagger}+\hat{O}\rho\hat{O}^{\dagger}\text{,}%
\end{equation}
where \cite{kl},%
\begin{equation}
R_{r}\overset{\text{def}}{=}V_{r}\sum_{i}\left\vert v_{r}^{i_{L}}\right\rangle
\left\langle v_{r}^{i_{L}}\right\vert =\sum_{i}\left\vert i_{L}\right\rangle
\left\langle v_{r}^{i_{L}}\right\vert \text{,} \label{the recovery}%
\end{equation}
and $\hat{O}$ (with $\hat{O}=\hat{O}^{\dagger}=\hat{O}^{\dagger}\hat{O}$) is a
projector onto the subspace $\mathcal{O}$ in Eq. (\ref{deo}),%
\begin{equation}
\hat{O}\overset{\text{def}}{=}\sum_{k}\left\vert o_{k}\right\rangle
\left\langle o_{k}\right\vert \text{.}%
\end{equation}
We remark that the recovery operation $\mathcal{R}$ is a trace preserving
quantum operation since,%
\begin{align}
\sum_{r}R_{r}^{\dagger}R_{r}+\hat{O}^{\dagger}\hat{O}  &  =\sum_{r}\left[
\left(  \sum_{i}\left\vert i_{L}\right\rangle \left\langle v_{r}^{i_{L}%
}\right\vert \right)  ^{\dagger}\left(  \sum_{j}\left\vert j_{L}\right\rangle
\left\langle v_{r}^{j_{L}}\right\vert \right)  \right]  +\left(  \sum
_{k}\left\vert o_{k}\right\rangle \left\langle o_{k}\right\vert \right)
^{\dagger}\left(  \sum_{k^{\prime}}\left\vert o_{k^{\prime}}\right\rangle
\left\langle o_{k^{\prime}}\right\vert \right) \nonumber\\
& \nonumber\\
&  =\sum_{r\text{, }i\text{, }j}\left\vert v_{r}^{i_{L}}\right\rangle
\left\langle i_{L}|j_{L}\right\rangle \left\langle v_{r}^{j_{L}}\right\vert
+\sum_{k\text{, }k^{\prime}}\left\vert o_{k}\right\rangle \left\langle
o_{k}|o_{k^{\prime}}\right\rangle \left\langle o_{k^{\prime}}\right\vert
\nonumber\\
& \nonumber\\
&  =\sum_{r\text{, }i\text{, }j}\left\vert v_{r}^{i_{L}}\right\rangle
\left\langle v_{r}^{j_{L}}\right\vert \delta_{ij}+\sum_{k\text{, }k^{\prime}%
}\left\vert o_{k}\right\rangle \left\langle o_{k^{\prime}}\right\vert
\delta_{kk^{\prime}}\nonumber\\
& \nonumber\\
&  =\sum_{r\text{, }i}\left\vert v_{r}^{i_{L}}\right\rangle \left\langle
v_{r}^{i_{L}}\right\vert +\sum_{k}\left\vert o_{k}\right\rangle \left\langle
o_{k}\right\vert \nonumber\\
& \nonumber\\
&  =\mathcal{I}_{2^{n}\times2^{n}}\text{,}%
\end{align}
because $\mathcal{B}_{\mathcal{H}_{2}^{n}}\overset{\text{def}}{=}\left\{
\left\vert v_{r}^{j_{L}}\right\rangle \text{, }\left\vert o_{k}\right\rangle
\right\}  $ is an orthonormal basis for $\mathcal{H}_{2}^{n}$. We emphasize
that $\mathcal{R}$ is indeed a CPTP superoperator (whose recovery operators
$R_{k}$ can be regarded as projective measurements followed by unitary
rotations), since it is a sum of orthogonal projections followed by unitary
operators where the projections span the Hilbert space $\mathcal{H}_{2}^{n}$.
Furthermore, we point out that the recovery scheme $\mathcal{R}$ in Eq.
(\ref{ecce}) applies, in principle, to any quantum code satisfying the QEC
conditions independent of the stabilizer formalism \cite{kl}. For more
technical details, we refer to \cite{kl} and \cite{gaitan}.

\subsubsection{Approximate QEC}

In general, approximate QEC becomes useful when the operator-sum
representation of the noise model is defined by errors parametrized by a
certain number of small parameters such as the coupling strength between the
environment and the quantum system. For the sake of simplicity, suppose the
error model is characterized by a single small parameter $\delta$ and assume
the goal is to uncover a quantum code for the noise model $\Lambda^{\prime}$
with fidelity,%
\begin{equation}
\mathcal{F}\geq1-O\left(  \delta^{\beta+1}\right)  \text{.} \label{preserve}%
\end{equation}
How strong can be the violation of the traditional perfect Knill-Laflamme QEC
conditions in order to preserve the desired fidelity order in Eq.
(\ref{preserve})?\ In other words, how relaxed can the approximate error
correction conditions be so that the inequality in Eq. (\ref{preserve}) is
satisfied? The answer to this important question was provided by Leung et
\textit{al}. in \cite{debbie}.

It turns out that for both exact and approximate QEC conditions, it is
necessary that%
\begin{equation}
P_{\text{detection}}\overset{\text{def}}{=}\sum_{k\in\mathcal{K}%
_{\text{reversible}}^{\prime}}p_{k}\geq\mathcal{F}\text{,} \label{detection}%
\end{equation}
where $P_{\text{detection}}$ denotes the total error detection probability.
Eq. (\ref{detection}) requires that all the enlarged error operators
$A_{l}^{\prime}$ with maximum detection probability must be included in
$\mathcal{A}_{\text{reversible}}^{\prime}$,%
\begin{equation}
\max_{\left\vert \psi_{in}\right\rangle \in\mathcal{C}}\text{Tr}\left(
\left\vert \psi_{in}\right\rangle \left\langle \psi_{in}\right\vert
A_{l}^{\prime\dagger}A_{l}^{\prime}\right)  \approx O\left(  \delta^{\alpha
}\right)  \text{ with }\alpha\leq\beta\text{.}%
\end{equation}
The important point is that a good overlap between the input and output states
is needed while it is not necessary to recover the exact input state
$\left\vert \psi_{in}\right\rangle \left\langle \psi_{in}\right\vert $, since
we do not require $\mathcal{F}=1$. In terms of the enlarged error operators
restricted to the codespace, this means that such errors need to be only
approximately unitary and mutually orthogonal. These considerations lead to
the relaxed sufficient QEC conditions.

In analogy to Eq. (\ref{pd1}), assume that the polar decomposition for
$A_{l}^{\prime}$ is given by,%
\begin{equation}
A_{l}^{\prime}P_{\mathcal{C}}=U_{l}\sqrt{P_{\mathcal{C}}A_{l}^{\prime\dagger
}A_{l}^{\prime}P_{\mathcal{C}}}\text{.} \label{sup}%
\end{equation}
Since $P_{\mathcal{C}}A_{l}^{\prime\dagger}A_{l}^{\prime}P_{\mathcal{C}}$
restricted to the codespace $\mathcal{C}$ have different eigenvalues, the
exact error correction conditions are not fulfilled. Let us say that
$\lambda_{l}^{\text{(max)}}\overset{\text{def}}{=}p_{l}$ and $\lambda
_{l}^{\text{(min)}}\overset{\text{def}}{=}\lambda_{l}p_{l}$ are the largest
and the smallest eigenvalues, respectively, where both $p_{l}$ and
$\lambda_{l}$ are $c$-numbers. Furthermore, let us define the so-called
residue operator $\pi_{l}$ as \cite{debbie},%
\begin{equation}
\pi_{l}\overset{\text{def}}{=}\sqrt{P_{\mathcal{C}}A_{l}^{\prime\dagger}%
A_{l}^{\prime}P_{\mathcal{C}}}-\sqrt{\lambda_{l}p_{l}}P_{\mathcal{C}}\text{,}
\label{pi}%
\end{equation}
where,%
\begin{equation}
0\leq\left\vert \pi_{l}\right\vert \overset{\text{def}}{=}\left(  \pi
_{l}^{\dagger}\pi_{l}\right)  ^{\frac{1}{2}}\leq\sqrt{p_{l}}-\sqrt{\lambda
_{l}p_{l}}\text{.}%
\end{equation}
Substituting Eq. (\ref{pi}) into Eq. (\ref{sup}), we obtain%
\begin{equation}
A_{l}^{\prime}P_{\mathcal{C}}=U_{l}\left(  \sqrt{\lambda_{l}p_{l}}I+\pi
_{l}\right)  P_{\mathcal{C}}\text{.} \label{qs}%
\end{equation}
From Eq. (\ref{qs}) and imposing that $P_{\mathcal{C}}U_{l}^{\dagger}%
U_{m}P_{\mathcal{C}}=\delta_{lm}P_{\mathcal{C}}$, the analog of Eq.
(\ref{vai2}) becomes%
\begin{equation}
P_{\mathcal{C}}A_{l}^{\prime\dagger}A_{m}^{\prime}P_{\mathcal{C}}=\left(
\sqrt{\lambda_{l}p_{l}}I+\pi_{l}^{\dagger}\right)  \left(  \sqrt{\lambda
_{m}p_{m}}I+\pi_{m}\right)  P_{\mathcal{C}}\delta_{lm}\text{,} \label{vai3}%
\end{equation}
where,%
\begin{equation}
\lambda_{l}^{\text{(max)}}-\lambda_{l}^{\text{(min)}}\equiv p_{l}\left(
1-\lambda_{l}\right)  \leq O\left(  \delta^{\beta+1}\right)  \text{, }\forall
l\in\mathcal{K}_{\text{reversible}}^{\prime}\text{.}%
\end{equation}
We point out that when the exact QEC conditions are satisfied, $\lambda_{l}=1$
and $\pi_{l}=0$, thus, Eqs. (\ref{vai2}) and (\ref{vai3}) coincide. Finally,
we point out that an approximate recovery operation $\mathcal{R}%
\overset{\text{def}}{=}\left\{  R_{1}\text{,..., }R_{r}\text{,..., }\hat
{O}\right\}  $ with $R_{k}$ defined in Eq. (\ref{the recovery}) and $\hat{O}$
formally defined just as in the exact case can be employed in this new
scenario as well. However, extra care in the explicit computation of the
unitary operators $U_{k}$ is needed in view of the fact that the polar
decomposition in Eq. (\ref{pd1}) is replaced by the one in Eq. (\ref{sup}).
More details can be found in \cite{debbie}.

\subsection{Recovery maps}

In general, numerically constructed recovery maps do not exhibit any practical
implementation structure while the perfect Knill-Laflamme recovery map can be
implemented simply using syndrome measurements and conditional unitary gates
\cite{kl}. For these reasons, our intention here is to pursue an analytical
approach to the recovery scheme that reduces to the perfect Knill-Laflamme
recovery scheme in the limiting case of small deviations from the exact QEC conditions.

We stress that one of the main points advocated in \cite{kl} includes treating
a code solely in terms of its subspace in a larger Hilbert space and defining
decoding operations in terms of general recovery superoperator. Basically, the
focus is on the construction of the recovery superoperator rather than on the
encoding and decoding operators. This allows studying the codes and their
properties for arbitrary interaction superoperator and avoids explicitly
dealing with decoding and encoding issues when studying the fidelity of a code
given its recovery operator. We also emphasize that the approximate QEC
conditions, like the perfect QEC conditions, provide a way to check if a code
is approximately correctable, without requiring knowledge of the optimal
recovery.\textbf{ }Once again, we emphasize here that the perfect
Knill-Laflamme recovery scheme\textbf{ }$\mathcal{R}$\textbf{ }in Eq.
(\ref{ecce}) applies, in principle, to any quantum code satisfying the QEC
conditions and no mention to the stabilizer formalism appeared in \cite{kl}.
Furthermore, for the sake of completeness, we also remark that the traditional
recovery operation for stabilizer codes can be summarized as follows: first,
measure all the eigenvalues of the stabilizer generators. This is the
so-called syndrome measurement; second, given the measured syndrome, compute
the minimum Hamming weight error (that is, the most probable error) that could
have caused the syndrome; third, apply the Pauli matrices that correct this
error\textbf{. }For the sake of completeness, we point out that for various
codes and error models, the minimum Hamming weight error cannot be efficiently
determined. More generally, rather than the most likely single error, the most
likely equivalence class of errors is determined.

One of the first important theoretical approach to near-optimal recovery
schemes was presented in \cite{jmp} where reversal recovery operations that
are near-optimal for the average entanglement fidelity $\mathcal{\bar{F}%
}\left(  E\text{, }\Lambda\right)  $,%
\begin{equation}
\mathcal{\bar{F}}\left(  E\text{, }\Lambda\right)  \overset{\text{def}}{=}%
%TCIMACRO{\dsum \limits_{i}}%
%BeginExpansion
{\displaystyle\sum\limits_{i}}
%EndExpansion
p_{i}\mathcal{F}\left(  \rho_{i}\text{, }\Lambda\right)  \text{,}%
\end{equation}
with $E\overset{\text{def}}{=}\left\{  p_{i}\text{, }\rho_{i}\right\}  $
denoting an ensemble with states $\rho_{i}$ that occur with probability
$p_{i}$ (where $\mathcal{F}$ denotes the entanglement fidelity, see Eq.
(\ref{defef})) were constructed analytically. The near-optimal reversal
operation reads \cite{jmp},%
\begin{equation}
\mathcal{R}_{\Lambda\text{, }\rho}^{\text{(Barnum-Knill)}}\sim\left\{
\rho^{\frac{1}{2}}A_{k}^{\dagger}\Lambda\left(  \rho\right)  ^{-\frac{1}{2}%
}\right\}  \text{,} \label{bk}%
\end{equation}
where $\Lambda\sim\left\{  A_{k}\right\}  $. In \cite{tyson, beny2010},
generalizing the traditional Knill-Laflamme QEC conditions \cite{kl},
necessary and sufficient conditions for approximate correctability of a
quantum code were derived. In particular, a class of near-optimal recovery
channels for the worst-case entanglement fidelity (that is, entanglement
fidelity minimized over all input states) was also provided. Following
\cite{jmp} and assuming $\rho=P_{\mathcal{C}}/d$ where $d$ is the dimension of
the codespace and $P_{\mathcal{C}}$ the projector on the codespace, a special
case of the near-optimal reversal operation in Eq. (\ref{bk}) was introduced
in \cite{ngm, ngm2}. Such reversal operation is the so-called transpose
channel recovery map,%

\begin{equation}
\mathcal{R}_{\text{TC}}\overset{\text{def}}{=}\mathcal{R}_{\Lambda\text{,
}P_{\mathcal{C}}/d}^{\text{(Barnum-Knill)}}\sim\left\{  P_{\mathcal{C}}%
A_{k}^{\dagger}\Lambda\left(  P_{\mathcal{C}}\right)  ^{-\frac{1}{2}}\right\}
\text{.} \label{TC}%
\end{equation}
The transpose channel recovery map $\mathcal{R}_{\text{TC}}$ is a
simple-to-construct recovery map built from the noise channel $\Lambda$ and
the code $\mathcal{C}$. In particular it works nearly as well as the optimal
recovery channel, with optimality defined in terms of worst-case fidelity over
all input states (the fidelity between any two states $\rho$ and $\sigma$ is
given by $f\left(  \rho\text{, }\sigma\right)  \overset{\text{def}}{=}%
$Tr$\sqrt{\rho^{\frac{1}{2}}\sigma\rho^{\frac{1}{2}}}$). We point out, as
mentioned in \cite{jmp} and explicitly shown in \cite{ngm}, that the transpose
channel $\mathcal{R}_{\text{TC}}$ in Eq. (\ref{TC}) reduces to the perfect
Knill-Laflamme recovery operation $\mathcal{R}_{\text{perfect}}%
^{\text{(Knill-Laflamme)}}$ when the traditional exact QEC conditions are satisfied.

The recovery scheme that we choose to use in this work is formally defined in
terms of recovery operators that are just like the operators $R_{k}$ in Eq.
(\ref{the recovery}),%
\begin{equation}
R_{k}\overset{\text{def}}{=}V_{k}P_{k}\equiv V_{k}\sum_{i}\left\vert
v_{k}^{i_{L}}\right\rangle \left\langle v_{k}^{i_{L}}\right\vert =\sum
_{i}\left\vert i_{L}\right\rangle \left\langle v_{k}^{i_{L}}\right\vert
=\frac{\left\vert 0_{L}\right\rangle \left\langle 0_{L}\right\vert
A_{k}^{\dagger}}{\sqrt{\left\langle 0_{L}\left\vert A_{k}^{\dagger}%
A_{k}\right\vert 0_{L}\right\rangle }}+\frac{\left\vert 1_{L}\right\rangle
\left\langle 1_{L}\right\vert A_{k}^{\dagger}}{\sqrt{\left\langle
1_{L}\left\vert A_{k}^{\dagger}A_{k}\right\vert 1_{L}\right\rangle }}\text{,}
\label{rk}%
\end{equation}
where, however, we must now take into account that $\left\langle
0_{L}\left\vert A_{k}^{\dagger}A_{k}\right\vert 0_{L}\right\rangle $ may only
be approximately equal to $\left\langle 1_{L}\left\vert A_{k}^{\dagger}%
A_{k}\right\vert 1_{L}\right\rangle $ in the approximate QEC framework. Thus,
the set of errors $\left\{  A_{k}\right\}  $ that appear in Eq. (\ref{rk}) has
to be considered correctable in the approximate sense specified in the
previous subsection. For this reason, the recovery operators $R_{k}$ in Eq.
(\ref{rk}) cannot assume the simple expression they exhibit in the case of
exact fulfillment of the QEC conditions. In the optimal (exact) case, the
superoperator $\mathcal{R}$ with elements $R_{k}$ in Eq. (\ref{rk}) becomes%
\begin{equation}
\mathcal{R}_{\text{perfect}}^{\text{(Knill-Laflamme)}}\sim\left\{
\frac{P_{\mathcal{C}}A_{k}^{\dagger}}{\sqrt{p_{k}}}\right\}  \text{,}
\label{perfectKL}%
\end{equation}
with $p_{k}\overset{\text{def}}{=}\left\langle 0_{L}\left\vert A_{k}^{\dagger
}A_{k}\right\vert 0_{L}\right\rangle \equiv\left\langle 1_{L}\left\vert
A_{k}^{\dagger}A_{k}\right\vert 1_{L}\right\rangle $. Explicit analytical
investigations in the framework of exact QEC where the superoperator in Eq.
(\ref{perfectKL}) was employed can be found in \cite{cafaro-osid, cafaro-pra1,
cafaro-pra2}.

Before describing the concept of entanglement fidelity, we wish to hint at
what happens with our recovery scheme in the traditional AD noise model when
error correction is performed via the Leung et \textit{al}. four-qubit code
\cite{debbie}. For the sake of clarity, we only consider the recovery operator
for the enlarged error $A_{0000}\overset{\text{def}}{=}A_{0}\otimes
A_{0}\otimes A_{0}\otimes A_{0}$ with $A_{0}$ defined as in Eq. (\ref{a0}). In
this case, we have
\begin{equation}
R_{A_{0000}}A_{0000}\left\vert \psi\right\rangle =\alpha\sqrt{1-2\gamma}%
\sqrt{1+\frac{3\gamma^{2}-2\gamma^{3}+\frac{1}{2}\gamma^{4}}{1-2\gamma}%
}\left\vert 0_{L}\right\rangle +\beta\sqrt{1-2\gamma}\sqrt{1+\frac{\gamma^{2}%
}{1-2\gamma}}\left\vert 1_{L}\right\rangle =\sqrt{1-2\gamma}\left\vert
\psi\right\rangle +\mathcal{O}\left(  \gamma^{2}\right)  \text{,} \label{x}%
\end{equation}
with $\left\vert \psi\right\rangle \overset{\text{def}}{=}\alpha\left\vert
0_{L}\right\rangle +\beta\left\vert 1_{L}\right\rangle $ where $\alpha$,
$\beta\in%
%TCIMACRO{\U{2102} }%
%BeginExpansion
\mathbb{C}
%EndExpansion
$ and $\left\vert \alpha\right\vert ^{2}+\left\vert \beta\right\vert ^{2}=1$
and $\left\{  \left\vert 0_{L}\right\rangle \text{, }\left\vert 1_{L}%
\right\rangle \right\}  $ span the codespace of the four-qubit code. The
approximate nature of Eq. (\ref{x}) is in agreement with the modified version
of the Knill-Laflamme QEC conditions in Eq. (\ref{vai}),%
\begin{equation}
P_{\mathcal{C}}A_{0000}^{\dagger}A_{0000}P_{\mathcal{C}}=\lambda_{00}\left(
\gamma\right)  P_{\mathcal{C}}+P_{\mathcal{C}}\hat{B}_{00}\left(
\gamma\right)  P_{\mathcal{C}}\text{,}%
\end{equation}
with,%
\begin{equation}
\lambda_{00}\left(  \gamma\right)  \overset{\text{def}}{=}1-2\gamma\text{ and,
}\hat{B}_{00}\left(  \gamma\right)  \overset{\text{def}}{=}\left(  3\gamma
^{2}-2\gamma^{3}+\frac{1}{2}\gamma^{4}\right)  \left\vert 0_{L}\right\rangle
\left\langle 0_{L}\right\vert +\gamma^{2}\left\vert 1_{L}\right\rangle
\left\langle 1_{L}\right\vert \text{,}%
\end{equation}
while $P_{\mathcal{C}}$ is the projector on the codespace of the code
$\mathcal{C}$. For more details on this point, we refer to \cite{beny2010}.

\subsection{Entanglement fidelity}

Entanglement fidelity is a useful performance measure of the efficiency of
quantum error correcting codes. It is a quantity that keeps track of how well
the state and entanglement of a subsystem of a larger system are stored,
without requiring the knowledge of the complete state or dynamics of the
larger system. More precisely, the entanglement fidelity is defined for a
mixed state%
\begin{equation}
\rho\overset{\text{def}}{=}\sum_{i}p_{i}\rho_{i}=\text{Tr}_{\mathcal{H}_{R}%
}\left\vert \psi\right\rangle \left\langle \psi\right\vert \text{,}%
\end{equation}
in terms of a purification $\left\vert \psi\right\rangle \in\mathcal{H}%
\otimes\mathcal{H}_{R}$ to a reference system $\mathcal{H}_{R}$. The
purification $\left\vert \psi\right\rangle $ encodes all of the information in
$\rho$. Entanglement fidelity is a measure of how well the channel $\Lambda$
preserves the entanglement of the state $\mathcal{H}$ with its reference
system $\mathcal{H}_{R}$. The entanglement fidelity is defined as follows
\cite{schumy},%
\begin{equation}
\mathcal{F}\left(  \rho\text{, }\Lambda\right)  \overset{\text{def}}%
{=}\left\langle \psi|\left(  \Lambda\otimes I_{\mathcal{H}_{R}}\right)
\left(  \left\vert \psi\right\rangle \left\langle \psi\right\vert \right)
|\psi\right\rangle \text{,}%
\end{equation}
where $\left\vert \psi\right\rangle $ is any purification of $\rho$,
$I_{\mathcal{H}_{R}}$ is the identity map on $\mathcal{M}\left(
\mathcal{H}_{R}\right)  $ (the space of all linear operators on the Hilbert
space $\mathcal{H}_{R}$) and $\Lambda\otimes I_{\mathcal{H}_{R}}$ is the
evolution operator extended to the space $\mathcal{H}\otimes\mathcal{H}_{R}$,
the space on which $\rho$ has been purified. If the quantum operation
$\Lambda$ is written in terms of its Kraus error operators $\left\{
A_{k}\right\}  $ as $\Lambda\left(  \rho\right)  \overset{\text{def}}{=}%
\sum_{k}A_{k}\rho A_{k}^{\dagger}$, then it can be shown that \cite{mike},
\begin{equation}
\mathcal{F}\left(  \rho\text{, }\Lambda\right)  =\sum_{k}\text{Tr}\left(
A_{k}\rho\right)  \text{Tr}\left(  A_{k}^{\dagger}\rho\right)  =\sum
_{k}\left\vert \text{Tr}\left(  \rho A_{k}\right)  \right\vert ^{2}\text{.}
\label{defef}%
\end{equation}
This expression for the entanglement fidelity is very useful for explicit
calculations. Finally, assuming that%
\begin{equation}
\Lambda:\mathcal{M}\left(  \mathcal{H}\right)  \ni\rho\mapsto\Lambda\left(
\rho\right)  =\sum_{k}A_{k}\rho A_{k}^{\dagger}\in\mathcal{M}\left(
\mathcal{H}\right)  \text{, dim}_{%
%TCIMACRO{\U{2102} }%
%BeginExpansion
\mathbb{C}
%EndExpansion
}\mathcal{H}=N \label{pla1}%
\end{equation}
and choosing a purification described by a maximally entangled unit vector for
the mixed state $\rho=I_{\mathcal{H}}/$dim$_{%
%TCIMACRO{\U{2102} }%
%BeginExpansion
\mathbb{C}
%EndExpansion
}\mathcal{H}$ , we obtain%
\begin{equation}
\mathcal{F}\left(  \frac{1}{N}I_{\mathcal{H}}\text{, }\Lambda\right)
=\frac{1}{N^{2}}\sum_{k}\left\vert \text{Tr}A_{k}\right\vert ^{2}\text{.}
\label{nfi}%
\end{equation}
The expression\ in Eq. (\ref{nfi}) represents the entanglement fidelity when
no error correction is performed on the noisy channel $\Lambda$\textbf{
}defined in Eq. in (\ref{pla1}).

Finally, for the sake of completeness, we point out that there exists a
relation between the fidelity of a recovery operation $\mathcal{R}$ and the
worst-case error probability parameter $p$ \cite{mosca},%
\begin{equation}
p\overset{\text{def}}{=}1-\mathcal{F}\left(  \mathcal{R}\text{, }%
\mathcal{C}\text{, }\mathcal{E}\right)  \text{,}%
\end{equation}
where $\mathcal{C}$ and $\mathcal{E}$ denote the code and the noise channel,
respectively. The meaning of the above relation can be described as follows.
Consider a quantum state $\left\vert \psi\right\rangle $ encoded into the
state $U_{\text{enc}}\left\vert \psi\right\rangle \left\vert 00\text{...}%
0\right\rangle $, then subjected to some noise (corresponding to the
$\mathcal{E}_{i}$ operators), then subjected to a recovery operation
(corresponding to the $R_{j}$ operators). Finally, the ancilla work-space is
discarded giving back some state $\rho_{\psi}$ on the original Hilbert space,%
\begin{equation}
\rho_{\psi}=\text{Tr}_{\text{ancilla}}\left[  \sum\limits_{j}R_{j}%
U_{\text{enc}}^{\dagger}\left(  \sum\limits_{i}\mathcal{E}_{i}U_{\text{enc}%
}\left\vert \psi\right\rangle \left\vert 00\text{...}0\right\rangle
\left\langle 0\text{...}00\right\vert \left\langle \psi\right\vert
U_{\text{enc}}^{\dagger}\mathcal{E}_{i}^{\dagger}\right)  U_{\text{enc}}%
R_{j}^{\dagger}\right]  \text{.}%
\end{equation}
We are interested in how close $\rho_{\psi}$ is to the original state
$\left\vert \psi\right\rangle \left\langle \psi\right\vert $. The probability
$p_{\psi}=\left\langle \psi\left\vert \rho_{\psi}\right\vert \psi\right\rangle
$ can be regarded as the probability of no error on the encoded state and the
fidelity of a recovery operation $\mathcal{R}$ is defined as,%
\begin{equation}
\mathcal{F}\left(  \mathcal{R}\text{, }\mathcal{C}\text{, }\mathcal{E}\right)
\overset{\text{def}}{=}\min_{\left\vert \psi\right\rangle }p_{\psi}\text{,}%
\end{equation}
the minimum of all such probabilities $p_{\psi}$ over all encoded states
$\left\vert \psi\right\rangle $. Thus, the probability parameter $p$ gives an
upper bound on the probability with which a generic encoded state will end up
in the wrong state.

\section{Additive codes}

We denote by $\left[  \left[  n,k,d\right]  \right]  $ a stabilizer (or,
additive) code that encodes $k$ logical qubits into $n$ physical qubits
correcting $\left\lfloor \frac{d-1}{2}\right\rfloor $-qubit errors where $d$
is the distance of the code and $\left\lfloor x\right\rfloor $ denotes the
largest integer less than $x$. Additive quantum codes are characterized by a
codespace, the space spanned by the so-called codewords, which is a
simultaneous eigenspace of an Abelian subgroup of the Pauli group. For more
details on the stabilizer formalism, we refer to \cite{daniel-phd}.

\subsection{Nondegenerate codes}

Formally speaking, a quantum code for which the positive Hermitian matrix
$\alpha$ in Eq. (\ref{vai}) is non-singular are called nondegenerate codes.
Instead, codes for which $\alpha$ is singular are called degenerate. For
nondegenerate codes, each error is individually identifiable and, for a given
choice of error operators, the quantum code is transformed into a set of
distinct orthogonal subspaces by applying the errors. In short, for
nondegenerate codes, all the errors acting on the codewords produce linearly
independent quantum states.

\subsubsection{The five-qubit code}

The $\left[  \left[  5\text{, }1\text{, }3\right]  \right]  $ code is the
smallest single-error correcting quantum code \cite{laflamme1, bennett1}. Of
all QECCs that encode $1$ qubit of data and correct all single-qubit errors,
the $\left[  \left[  5\text{, }1\text{, }3\right]  \right]  $ is the most
efficient, saturating the quantum Hamming bound. It encodes $k=1$ qubit in
$n=5$ qubits. The cardinality of its stabilizer group $\mathcal{S}$ is
$\left\vert \mathcal{S}\right\vert =2^{n-k}=16$ and the set $\mathcal{B}%
_{\mathcal{S}}^{\left[  \left[  5,1,3\right]  \right]  }$ of $n-k=4$
stabilizer group generators is given by \cite{gaitan},%
\begin{equation}
\mathcal{B}_{\mathcal{S}}^{\left[  \left[  5,1,3\right]  \right]  }%
\overset{\text{def}}{=}\left\{  X^{1}Z^{2}Z^{3}X^{4}\text{, }X^{2}Z^{3}%
Z^{4}X^{5}\text{, }X^{1}X^{3}Z^{4}Z^{5}\text{, }Z^{1}X^{2}X^{4}Z^{5}\right\}
\text{,} \label{61}%
\end{equation}
with $X\overset{\text{def}}{=}\sigma_{x}$, $Y\overset{\text{def}}{=}\sigma
_{y}$ and $Z\overset{\text{def}}{=}\sigma_{z}$ and $\left\{  \sigma_{x}\text{,
}\sigma_{y}\text{, }\sigma_{z}\right\}  $ as given in Eq. (\ref{paolino}). For
the sake of notational clarity, we emphasize that when describing stabilizer
generators as tensor products of Pauli operators, we may omit to use the
symbol $\otimes$. In addition, the superscripts on the RHS of Eq. (\ref{61})
label the qubits $1$,..., $5$. The distance of the code is $d=3$ and therefore
the weight of the smallest enlarged error operators of the form $A_{l}%
^{\dagger}A_{k}$ \ that cannot be detected by the code is $3$. Finally, we
recall that it is a nondegenerate code, since the smallest weight for elements
of $\mathcal{S}$ (other than identity) is $4$ and therefore it is greater than
the distance $d=3$. The encoding for the $\left[  \left[  5\text{, }1\text{,
}3\right]  \right]  $ code is given by \cite{laflamme1},%
\begin{align}
&  \left\vert 0_{L}\right\rangle \overset{\text{def}}{=}\frac{1}{\sqrt{8}%
}\left[  -\left\vert 00000\right\rangle +\left\vert 01111\right\rangle
-\left\vert 10011\right\rangle +\left\vert 11100\right\rangle +\left\vert
00110\right\rangle +\left\vert 01001\right\rangle +\left\vert
10101\right\rangle +\left\vert 11010\right\rangle \right]  \text{,}\nonumber\\
& \nonumber\\
&  \left\vert 1_{L}\right\rangle \overset{\text{def}}{=}\frac{1}{\sqrt{8}%
}\left[  -\left\vert 11111\right\rangle +\left\vert 10000\right\rangle
+\left\vert 01100\right\rangle -\left\vert 00011\right\rangle +\left\vert
11001\right\rangle +\left\vert 10110\right\rangle -\left\vert
01010\right\rangle -\left\vert 00101\right\rangle \right]  \text{.}
\label{513}%
\end{align}
The enlarged GAD quantum channel after performing the encoding defined by
means of Eq. (\ref{513}) reads,%
\begin{equation}
\Lambda_{\text{GAD}}^{\left[  \left[  5,1,3\right]  \right]  }\left(
\rho\right)  \overset{\text{def}}{=}%
%TCIMACRO{\dsum \limits_{r=0}^{2^{10}-1}}%
%BeginExpansion
{\displaystyle\sum\limits_{r=0}^{2^{10}-1}}
%EndExpansion
A_{r}^{\prime}\rho A_{r}^{\prime\dagger}=%
%TCIMACRO{\dsum \limits_{i\text{, }j\text{, }k\text{, }l\text{, }m=0}^{3}}%
%BeginExpansion
{\displaystyle\sum\limits_{i\text{, }j\text{, }k\text{, }l\text{, }m=0}^{3}}
%EndExpansion
A_{ijklm}\rho A_{ijklm}^{\dagger}\text{,}%
\end{equation}
where to any of the $2^{10}$ values of $r$ we can associate a set of indices
$\left(  i\text{, }j\text{, }k\text{, }l\text{, }m\right)  $ (and vice-versa)
such that,%
\begin{equation}
A_{r}^{\prime}\leftrightarrow A_{ijklm}\overset{\text{def}}{=}A_{i}\otimes
A_{j}\otimes A_{k}\otimes A_{l}\otimes A_{m}\equiv A_{i}A_{j}A_{k}A_{l}%
A_{m}\text{.}%
\end{equation}
The errors $A_{i}$ with $i\in\left\{  0\text{, }1\text{, }2\text{, }3\right\}
$ are defined in Eq. (\ref{ko}) and $\rho\in\mathcal{M}\left(  \mathcal{C}%
\right)  $ with $\mathcal{C}\subset\mathcal{H}_{2}^{5}$. In particular, the
number of weight-$q$ enlarged error operators $A_{r}^{\prime}$ is given by
$3^{q}\binom{5}{q}$ and,%
\begin{equation}
2^{10}=%
%TCIMACRO{\dsum \limits_{q=0}^{5}}%
%BeginExpansion
{\displaystyle\sum\limits_{q=0}^{5}}
%EndExpansion
3^{q}\binom{5}{q}\text{.}%
\end{equation}
We point out that consistency requires that the sum of the probabilities
$P\left(  A_{ijklm}\right)  $ that an $A_{r}^{\prime}=A_{ijklm}$ error occurs
must sum up to unity. For clarity of exposition, consider the limiting case
with $\varepsilon=0$ where only $2^{5}=32$ enlarged errors $A_{r}^{\prime}$
are present. In this case, we have%
\begin{equation}
\sum_{i\text{, }j\text{, }k\text{, }l\text{, }m=0}^{1}P\left(  A_{ijklm}%
\right)  =\sum_{r=0}^{2^{5}-1}P\left(  A_{r}^{\prime}\right)  =\frac{1}{2}%
\sum_{a=0}^{2^{5}-1}\text{Tr}\left(  A_{a}^{\prime}P_{\mathcal{C}}%
A_{a}^{\prime\dagger}\right)  =\frac{1}{2}\sum_{a=0}^{2^{5}-1}\left[
\left\langle 0_{L}\left\vert A_{a}^{\prime\dagger}A_{a}^{\prime}\right\vert
0_{L}\right\rangle +\left\langle 1_{L}\left\vert A_{a}^{\prime\dagger}%
A_{a}^{\prime}\right\vert 1_{L}\right\rangle \right]  =1\text{,}%
\end{equation}
that is,%
\begin{equation}
\sum_{i\text{, }j\text{, }k\text{, }l\text{, }m=0}^{1}P\left(  A_{ijklm}%
\right)  =P_{\text{weight-}0}+P_{\text{weight-}1}+P_{\text{weight-}%
2}+P_{\text{weight-}3}+P_{\text{weight-}4}+P_{\text{weight-}5}=1\text{,}%
\end{equation}
where,%
\begin{align}
P_{\text{weight-}0}  &  =1-\frac{5}{2}\gamma+\frac{5}{2}\gamma^{2}-\frac{5}%
{4}\gamma^{3}+\frac{3}{8}\gamma^{4}-\frac{1}{16}\gamma^{5}\text{,
}P_{\text{weight-}1}=\frac{5}{2}\gamma-5\gamma^{2}+\frac{15}{4}\gamma
^{3}-\frac{3}{2}\gamma^{4}+\frac{5}{16}\gamma^{5}\text{,}\nonumber\\
& \nonumber\\
P_{\text{weight-}2}  &  =\frac{5}{2}\gamma^{2}-\frac{15}{4}\gamma^{3}+\frac
{9}{4}\gamma^{4}-\frac{5}{8}\gamma^{5}\text{, }P_{\text{weight-}3}=\frac{5}%
{4}\gamma^{3}-\frac{3}{2}\gamma^{4}+\frac{5}{8}\gamma^{5}\text{,}\nonumber\\
& \nonumber\\
\text{ }P_{\text{weight-}4}  &  =\frac{3}{8}\gamma^{4}-\frac{5}{16}\gamma
^{5}\text{, }P_{\text{weight-}5}=\frac{1}{16}\gamma^{5}\text{.}%
\end{align}
Using the brute-force approach, it would be fairly straightforward, though
very tedious, to check the approximate QEC conditions for all the $2^{10}$
enlarged errors. Fortunately, this is not necessary. Indeed, we aim at finding
an analytical estimate of the entanglement fidelity of the code such that,%
\begin{equation}
\mathcal{F}^{\left[  \left[  5,1,3\right]  \right]  }\left(  \gamma\text{,
}\varepsilon\right)  \geq1-\mathcal{O}\left(  2\right)  \text{,}%
\end{equation}
where $\mathcal{O}\left(  2\right)  \sim\mathcal{O}\left(  \gamma^{n_{1}%
}\varepsilon^{n_{2}}\right)  $; that is, the pair $\left(  n_{1}\text{, }%
n_{2}\right)  $ is such that,%
\begin{equation}
\lim_{\gamma\text{, }\varepsilon\rightarrow0}\frac{\mathcal{O}\left(
2\right)  }{\mathcal{O}\left(  \gamma^{n_{1}}\varepsilon^{n_{2}}\right)
}=\text{constant.}%
\end{equation}
For instance, we may have $\left\{  \left(  n_{1}\text{, }n_{2}\right)
\right\}  =\left\{  \left(  2\text{, }0\right)  \text{, }\left(  0\text{,
}2\right)  \text{, }\left(  1\text{, }1\right)  \right\}  $. Observe that the
codewords that span the code belong to the $2^{5}=32$-dimensional
\emph{complex} Hilbert space $\mathcal{H}_{2}^{5}$. Thus, a basis of
orthonormal vectors for $\mathcal{H}_{2}^{5}$ requires $32$ elements. For
$\varepsilon=0$, it turns out that none of the $\binom{5}{2}=10$ weight-$2$
errors is correctable. Specifically, errors $A_{01100}$, $A_{00011}$,
$A_{01010}$ and $A_{00101}$ are not compatible with $A_{00000}$; $A_{00110}$
and $A_{01001}$ are not compatible with $A_{10000}$, $A_{11000}$ is not
compatible with $A_{01000}$, $A_{10100}$ is not compatible with $A_{00100}$,
$A_{10010}$ is not compatible with $A_{00010}$ and, finally, $A_{10001}$ is
incompatible with $A_{00001}$. For the general case, it can be shown that all
weight-$0$ ($1$ error) and weight-$1$ ($15$ errors) enlarged error operators
satisfy the approximate QEC conditions up to the sought order. However, the
action on the codewords of five weight-$1$ enlarged errors (specifically,
$A_{20000}$, $A_{02000}$, $A_{00200}$, $A_{00020}$ and $A_{00002}$) leads to
vectors that are not orthogonal to those obtained from the action of the
weight-$0$ error $A_{00000}$ on the codewords. Thus, we omit them from the
construction of our recovery scheme. In view of these considerations, we
construct our recovery operation $\mathcal{R}$ as follows,%
\begin{equation}
\mathcal{R}\overset{\text{def}}{=}\left\{  R_{0}\text{, }R_{1}\text{, }%
R_{2}\text{, }R_{3}\text{, }R_{4}\text{, }R_{5}\text{ }R_{6}\text{, }%
R_{7}\text{, }R_{8}\text{, }R_{9}\text{, }R_{10}\text{, }\hat{O}\right\}
\text{,} \label{five-recovery}%
\end{equation}
where,%
\begin{equation}
R_{r}\overset{\text{def}}{=}\left\vert 0_{L}\right\rangle \left\langle
v_{r}^{0_{L}}\right\vert +\left\vert 1_{L}\right\rangle \left\langle
v_{r}^{1_{L}}\right\vert \text{ with, }\left\vert v_{r}^{i_{L}}\right\rangle
\overset{\text{def}}{=}\frac{A_{k}^{\prime}\left\vert i_{L}\right\rangle
}{\sqrt{\left\langle i_{L}\left\vert A_{k}^{\prime\dagger}A_{k}\right\vert
i_{L}\right\rangle }}\text{,}%
\end{equation}
for $i\in\left\{  0\text{, }1\right\}  $ and $\left\langle v_{r}^{i_{L}%
}\left\vert v_{r^{\prime}}^{j_{L}}\right.  \right\rangle =\delta_{rr^{\prime}%
}\delta_{ij}$. To be clear $R_{0}$ is associated with the weight-$0$ error
$A_{00000}$; $R_{k}$ with $k=1$, ..., $5$ are associated with the five
weight-$1$ errors where single-qubit errors of type $A_{1}$ occur; finally,
$R_{k}$ with $k=6$, ..., $10$ are associated with the five weight-$1$ errors
where single-qubit errors of type $A_{3}$ occur. The construction of these
$11$ recovery operators $R_{k}$ is described in terms of $22$ orthonormal
vectors in $\mathcal{H}_{2}^{5}$. The missing $10$ orthonormal vectors can be
uncovered using the rank-nullity (dimension) theorem together with the
Gram-Schmidt orthonormalization procedure (for more details, see Appendix
C\textbf{)}. They define the operator $\hat{O}$ in Eq. (\ref{five-recovery}),%
\begin{equation}
\hat{O}=%
%TCIMACRO{\dsum \limits_{j=1}^{10}}%
%BeginExpansion
{\displaystyle\sum\limits_{j=1}^{10}}
%EndExpansion
\left\vert o_{j}\right\rangle \left\langle o_{j}\right\vert \text{.}%
\end{equation}
For the sake of convenience, we put $R_{11}\overset{\text{def}}{=}\hat{O}$.
Finally, the estimate of the entanglement fidelity of the five-qubit code,
when the recovery operation $\mathcal{R}$ in Eq. (\ref{five-recovery}) is
employed, becomes (for more details, see Appendix D),%
\begin{equation}
\mathcal{F}^{\left[  \left[  5,1,3\right]  \right]  }\left(  \gamma\text{,
}\varepsilon\right)  \overset{\text{def}}{=}\frac{1}{\left(  \dim_{%
%TCIMACRO{\U{2102} }%
%BeginExpansion
\mathbb{C}
%EndExpansion
}\mathcal{C}\right)  ^{2}}\sum_{k=0}^{2^{10}-1}\sum_{l=0}^{11}\left\vert
\text{Tr}\left(  R_{l}A_{k}^{\prime}\right)  _{\left\vert \mathcal{C}\right.
}\right\vert ^{2}\approx1-\frac{5}{2}\gamma^{2}-10\varepsilon^{2}\left(
1+\frac{\gamma}{\varepsilon}\right)  +\mathcal{O}\left(  3\right)  \text{.}
\label{estimate1}%
\end{equation}
We point out that in the limiting case of $\varepsilon=0$, the AD\ noise model
is recovered and our analytical estimate in Eq. (\ref{estimate1}) reduces to
the numerically obtained truncated series expansion appeared in \cite{andy2}
and \cite{fphd} (specifically, see page $48$ in \cite{fphd}). The effect of
the non-zero environmental temperature on the five-qubit code is illustrated
in Fig\textbf{. }$1$\textbf{.}

\begin{figure}[ptb]
\centering
\includegraphics[width=0.5\textwidth]{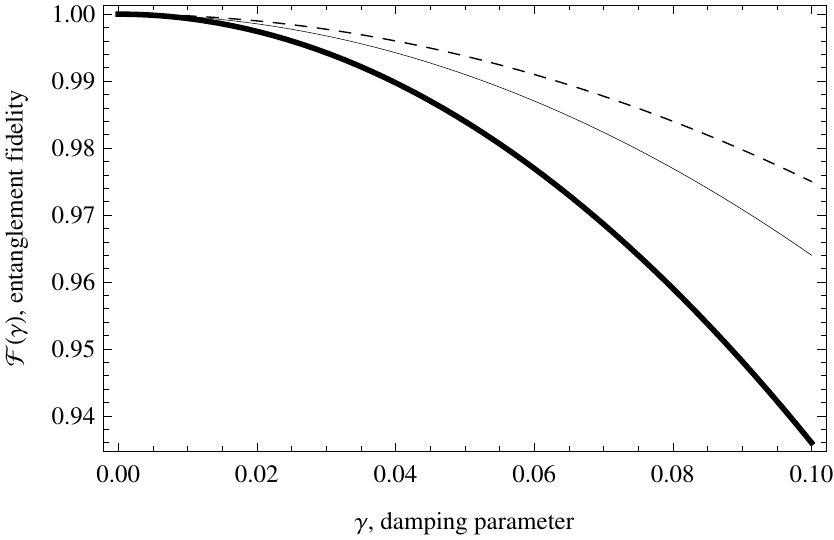}\caption{\textit{Effect of the
non-zero environmental temperature on quantum coding. }The truncated series
expansion of the entanglement fidelity $\mathcal{F}\left(  \gamma\right)  $
vs. the amplitude damping parameter $\gamma$ with $0\leq\gamma\leq10^{-1}$ for
the five-qubit code: $\varepsilon\left(  T\right)  =0$ (dashed line);
$\varepsilon\left(  T\right)  =10^{-1}\gamma$ (thin solid line);
$\varepsilon\left(  T\right)  =3\times10^{-1}\gamma$ (thick solid line).}%
\label{figure1}%
\end{figure}

\subsubsection{The CSS\ seven-qubit code}

The Calderbank-Shor-Steane (CSS) codes are constructed from two classical
binary codes $\mathcal{C}$ and $\mathcal{C}^{\prime}$ that have the following
properties \cite{calderbank-shor96, steane96}: 1) $\mathcal{C}$ and
$\mathcal{C}^{\prime}$ are $\left[  n\text{, }k\text{, }d\right]  $ and
$\left[  n\text{, }k^{\prime}\text{, }d^{\prime}\right]  $ codes,
respectively; 2) $\mathcal{C}^{\prime}\subset\mathcal{C}$; 3) $\mathcal{C}$
and $\mathcal{C}_{\perp}^{\prime}$ (the dual code of $\mathcal{C}^{\prime}$)
are both $t$-error correcting codes. For instance, in case of the seven-qubit
code, the two classical codes are the $\left[  7\text{, }4\text{, }3\right]  $
binary Hamming code $\left(  \mathcal{C}\right)  $ and the $\left[  7\text{,
}3\text{, }4\right]  $ binary simplex code $\left(  \mathcal{C}^{\prime
}\right)  $. The dual code $\mathcal{C}_{\perp}^{\prime}$ is the $\left[
7\text{, }4\text{, }3\right]  $ binary Hamming code. Thus $\mathcal{C}$ and
$\mathcal{C}_{\perp}^{\prime}$ are both $1$-error correcting codes. In this
case, $n=7$, $k=4$, $k^{\prime}=3$, $k-k^{\prime}=1$ so that $1$ qubit is
mapped into $7$ qubits. The seven-qubit code is the simplest example of a CSS
code. Although the seven-qubit code is ostensibly more complicated that the
five-qubit code, it is actually more useful in certain situations by virtue of
being a CSS\ code. The CSS codes are a particularly interesting class of codes
for two reasons. First, they are built using classical codes which have been
more heavily studied than quantum codes, so it is fairly easy to construct
useful quantum codes simply by looking at lists of classical codes. Second,
because of the form of generators, the CSS\ codes are precisely those for
which a CNOT applied between every pair of corresponding qubits in two blocks
performs a valid fault-tolerant operation. This makes them particularly good
candidates in fault-tolerant computation.

The CSS seven-qubit code encodes $k=1$ qubit in $n=7$ qubits. The cardinality
of its stabilizer group $\mathcal{S}$ is $\left\vert \mathcal{S}\right\vert
=2^{n-k}=64$ and the set $\mathcal{B}_{\mathcal{S}}^{\left[  \left[
7,1,3\right]  \right]  }$ of $n-k=6$ stabilizer group generators reads
\cite{gaitan},%
\begin{equation}
\mathcal{B}_{\mathcal{S}}^{\left[  \left[  7,1,3\right]  \right]  }%
\overset{\text{def}}{=}\left\{  X^{4}X^{5}X^{6}X^{7}\text{, }X^{2}X^{3}%
X^{6}X^{7}\text{, }X^{1}X^{3}X^{5}X^{7}\text{, }Z^{4}Z^{5}Z^{6}Z^{7}\text{,
}Z^{2}Z^{3}Z^{6}Z^{7}\text{, }Z^{1}Z^{3}Z^{5}Z^{7}\right\}  \text{.}%
\end{equation}
The distance of the code is $d=3$ and therefore the weight of the smallest
error $A_{l}^{\prime\dagger}A_{k}^{\prime}$ \ that cannot be detected by the
code is $3$. Finally, we recall that it is a nondegenerate code, since the
smallest weight for elements of $\mathcal{S}$ (other than identity) is $4$ and
therefore it is greater than the distance $d=3$. The encoding for the $\left[
\left[  7\text{, }1\text{, }3\right]  \right]  $ code is given by
\cite{gaitan},%
\begin{equation}
\left\vert 0\right\rangle \rightarrow\left\vert 0_{L}\right\rangle
\overset{\text{def}}{=}\frac{1}{\left(  \sqrt{2}\right)  ^{3}}\left[
\begin{array}
[c]{c}%
\left\vert 0000000\right\rangle +\left\vert 0110011\right\rangle +\left\vert
1010101\right\rangle +\left\vert 1100110\right\rangle +\\
\\
+\left\vert 0001111\right\rangle +\left\vert 0111100\right\rangle +\left\vert
1011010\right\rangle +\left\vert 1101001\right\rangle
\end{array}
\right]  \text{,} \label{code7a}%
\end{equation}
and,%
\begin{equation}
\left\vert 1\right\rangle \rightarrow\left\vert 1_{L}\right\rangle
\overset{\text{def}}{=}\frac{1}{\left(  \sqrt{2}\right)  ^{3}}\left[
\begin{array}
[c]{c}%
\left\vert 1111111\right\rangle +\left\vert 1001100\right\rangle +\left\vert
0101010\right\rangle +\left\vert 0011001\right\rangle +\\
\\
+\left\vert 1110000\right\rangle +\left\vert 1000011\right\rangle +\left\vert
0100101\right\rangle +\left\vert 0010110\right\rangle
\end{array}
\right]  \text{.} \label{code7b}%
\end{equation}
The enlarged GAD quantum channel after performing the encoding defined by
means of Eqs. (\ref{code7a}) and (\ref{code7b}) reads,%
\begin{equation}
\Lambda_{\text{GAD}}^{\left[  \left[  7,1,3\right]  \right]  }\left(
\rho\right)  \overset{\text{def}}{=}%
%TCIMACRO{\dsum \limits_{r=0}^{2^{14}-1}}%
%BeginExpansion
{\displaystyle\sum\limits_{r=0}^{2^{14}-1}}
%EndExpansion
A_{r}^{\prime}\rho A_{r}^{\prime\dagger}=%
%TCIMACRO{\dsum \limits_{i\text{, }j\text{, }k\text{, }l\text{, }m\text{,
%}n\text{, }s=0}^{3}}%
%BeginExpansion
{\displaystyle\sum\limits_{i\text{, }j\text{, }k\text{, }l\text{, }m\text{,
}n\text{, }s=0}^{3}}
%EndExpansion
A_{ijklmns}\rho A_{ijklmns}^{\dagger}\text{,}%
\end{equation}
where to any of the $2^{14}$ values of $r$ we can associate a set of indices
$\left(  i\text{, }j\text{, }k\text{, }l\text{, }m\text{, }n\text{, }s\right)
$ (and vice-versa) such that,%
\begin{equation}
A_{r}^{\prime}\leftrightarrow A_{ijklmns}\overset{\text{def}}{=}A_{i}\otimes
A_{j}\otimes A_{k}\otimes A_{l}\otimes A_{m}\otimes A_{n}\otimes A_{s}\equiv
A_{i}A_{j}A_{k}A_{l}A_{m}A_{n}A_{s}\text{.}%
\end{equation}
The errors $A_{i}$ with $i\in\left\{  0\text{, }1\text{, }2\text{, }3\right\}
$ are defined in Eq. (\ref{ko}) and $\rho\in\mathcal{M}\left(  \mathcal{C}%
\right)  $ with $\mathcal{C}\subset\mathcal{H}_{2}^{7}$. In particular, the
number of weight-$q$ enlarged error operators $A_{r}^{\prime}$ is given by
$3^{q}\binom{7}{q}$ and,%
\begin{equation}
2^{14}=%
%TCIMACRO{\dsum \limits_{q=0}^{7}}%
%BeginExpansion
{\displaystyle\sum\limits_{q=0}^{7}}
%EndExpansion
3^{q}\binom{7}{q}\text{.}%
\end{equation}
For $\varepsilon=0$, it can be shown that for any of the $\binom{7}{2}=21$
weight-$2$ errors, there exists at least one of the $7$ weight-$1$ errors for
which the correctability conditions are not satisfied. For the general case,
it can be shown that all weight-$0$ ($1$ error) and weight-$1$ ($21$ errors)
enlarged error operators satisfy the approximate QEC conditions up to the
sought order. In this case, the recovery scheme $\mathcal{R}$ that we use can
be described as follows: $R_{0}$ is associated with the weight-$0$ error
$A_{0000000}$; $R_{k}$ with $k=1$, ..., $7$ are associated with the seven
weight-$1$ errors where single-qubit errors of type $A_{1}$ occur; finally,
$R_{k}$ with $k=8$, ..., $14$ are associated with the five weight-$1$ errors
where single-qubit errors of type $A_{3}$ occur. The construction of these
$15$ recovery operators $R_{k}$ is described in terms of $30$ orthonormal
vectors in $\mathcal{H}_{2}^{7}$. In analogy to the case of the five-qubit
code, the action on the codewords of seven weight-$1$ enlarged errors
(specifically, $A_{2000000}$, $A_{0200000}$, $A_{0020000}$, $A_{0002000}$,
$A_{0000200}$, $A_{0000020}$ and $A_{0000002}$) leads to vectors that are not
orthogonal to those obtained from the action of the weight-$0$ error
$A_{0000000}$ on the codewords. Thus, we omit them from the construction of
our recovery scheme. The missing $98$ orthonormal vectors (needed to obtain an
orthonormal basis of $\mathcal{H}_{2}^{7}$ and to construct $R_{15}%
\overset{\text{def}}{=}\hat{O}$) can be formally computed by using the
rank-nullity theorem together with the Gram-Schmidt orthonormalization
procedure. Omitting further technical details (for more details, see
Appendix\ D) but using the very same line of reasoning presented for the
five-qubit code, our analytical estimate of the entanglement fidelity of the
CSS seven-qubit code reads,%
\begin{equation}
\mathcal{F}^{\left[  \left[  7,1,3\right]  \right]  }\left(  \gamma\text{,
}\varepsilon\right)  \overset{\text{def}}{=}\frac{1}{\left(  \dim_{%
%TCIMACRO{\U{2102} }%
%BeginExpansion
\mathbb{C}
%EndExpansion
}\mathcal{C}\right)  ^{2}}\sum_{k=0}^{2^{14}-1}\sum_{l=0}^{15}\left\vert
\text{Tr}\left(  R_{l}A_{k}^{\prime}\right)  _{\left\vert \mathcal{C}\right.
}\right\vert ^{2}\approx1-\frac{21}{4}\gamma^{2}-21\varepsilon^{2}\left(
1+\frac{\gamma}{\varepsilon}\right)  +\mathcal{O}\left(  3\right)  \text{.}
\label{estimate2}%
\end{equation}
To the best of our knowledge and unlike the case of the five-qubit code, no
truncated series expansion of $\mathcal{F}^{\left[  \left[  7,1,3\right]
\right]  }\left(  \gamma\text{, }\varepsilon\right)  $ with $\varepsilon=0$ is
available in the literature. However, we emphasize that in the special case of
$\varepsilon=0$, our analytical estimate in Eq. (\ref{estimate2}) appears to
exhibit a fairly good agreement with the numerical plot presented in
\cite{andy3} (specifically, see Figure $9$ in \cite{andy3}). For
$\varepsilon=0$, we compared our non truncated analytical estimate in Eq.
(\ref{estimate2}) to the single-qubit baseline performance (entanglement
fidelity when no QEC is performed) given by,%
\begin{equation}
\mathcal{F}_{\text{baseline}}^{1\text{-qubit}}\left(  \gamma\right)
\overset{\text{def}}{=}2^{-2}\left(  1+\sqrt{1-\gamma}\right)  ^{2}\text{.}
\label{baseline}%
\end{equation}
Then, we checked the good overlap between our results (non-truncated fidelity
expressions with and without error correction) and those plotted in
\cite{andy3}. For some more details, see Appendix E. The robustness against
non-zero environmental temperature of the CSS seven-qubit code is compared to
that of the Shor nine-qubit code in Fig.\textbf{ }$2$\textbf{.}

\begin{figure}[ptb]
\centering
\includegraphics[width=0.5\textwidth]{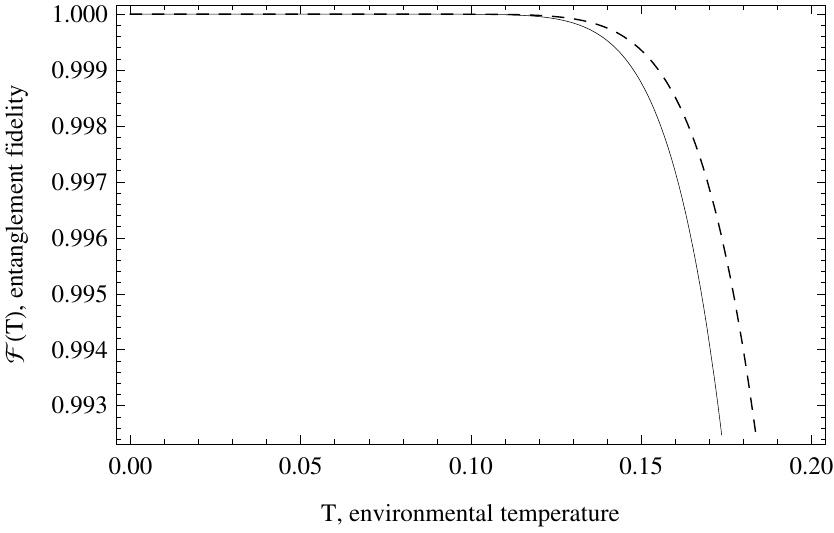}\caption{\textit{Robustness
against non-zero environmental temperature of degenerate and nondegenerate
codes}. The truncated series expansions of the entanglement fidelity
$\mathcal{F}\left(  T\right)  $ vs. the environmental temperature $T$ for
$\hbar=\omega=k_{B}=1$ and $\gamma=10\varepsilon$: the Shor nine-qubit code
(dashed line) and the CSS\ seven-qubit code (thin solid line).}%
\label{figure2}%
\end{figure}

\subsubsection{The eight-qubit concatenated code}

The notion of correctability depends on all the errors in the error set that
one is considering and, unlike detectability, cannot be applied to individual
errors. However, for a given code $\mathcal{C}$, both sets of detectable and
correctable errors are closed under linear combinations. Within the stabilizer
formalism, the error correction conditions can be described as follows
\cite{nielsen-book, daniel-phd}: an $\left[  \left[  n,k,d\right]  \right]  $
quantum code with stabilizer $\mathcal{S}$ and generators $g_{j}$ where
$j=1$,..., $n-k$, corrects an error set $\mathcal{A}$ if every error pair
$A_{l}^{\dagger}A_{m}\in\mathcal{A}$ either anticommutes with at least one
stabilizer generator,%
\[
\exists\text{ }g_{j}\in\mathcal{S}:\left\{  g_{j}\text{, }A_{l}^{\dagger}%
A_{m}\right\}  =0\text{,}%
\]
or is in the stabilizer, $A_{l}^{\dagger}A_{m}\in\mathcal{S}$.

The two Kraus operators for the AD noise model are given by $A_{0}%
=I-\mathcal{O}\left(  \gamma\right)  $ and $A_{1}\propto\sigma_{x}+i\sigma_{y}
$. In the GAD noise model, the error $A_{3}\propto\sigma_{x}-i\sigma_{y}$
appears as well. The linear span of $A_{1}$ and $A_{3}$ equals the linear span
of $\sigma_{x}$ and $\sigma_{y}$. Thus, if a code is capable of correcting $t$
$\sigma_{x}$- and $t$ $\sigma_{y}$-errors, it can also correct $t$ $A_{1}$ and
$t$ $A_{3}$ errors. The stabilizer of the Leung et \textit{al}. four-qubit
code,%
\begin{equation}
\left\vert 0_{L}\right\rangle \overset{\text{def}}{=}\frac{\left\vert
0000\right\rangle +\left\vert 1111\right\rangle }{\sqrt{2}}\text{ ,
}\left\vert 1_{L}\right\rangle \overset{\text{def}}{=}\frac{\left\vert
0011\right\rangle +\left\vert 1100\right\rangle }{\sqrt{2}}\text{,}%
\end{equation}
is given by $\mathcal{S}\overset{\text{def}}{\mathcal{=}}\left\langle
\sigma_{x}^{1}\sigma_{x}^{2}\sigma_{x}^{3}\sigma_{x}^{4}\text{, }\sigma
_{z}^{1}\sigma_{z}^{2}\text{, }\sigma_{z}^{3}\sigma_{z}^{4}\right\rangle $.
According to the above-mentioned considerations, it follows that the error set
$\left\{  I\text{, }\sigma_{x}^{i}\text{, }\sigma_{y}^{i}\right\}  $ with
$i=1$, $2$, $3$, $4$ is not correctable. For instance, the set $\left\{
\sigma_{x}^{1}\text{, }\sigma_{x}^{2}\right\}  $ is not correctable because
$\sigma_{x}^{1}\sigma_{x}^{2}$ commutes will all the stabilizer generators.

To construct a quantum code capable of error-correcting the set $\left\{
I\text{, }\sigma_{x}^{i}\text{, }\sigma_{y}^{i}\right\}  $ with $i=1$, $2$,
$3$, $4$, we concatenate the quantum dual rail code $\mathcal{C}_{\text{QDR}}$
(inner code) with the perfect $1$-erasure correcting code $\mathcal{C}%
_{\text{erasure}}$ (outer code) given by,%
\begin{equation}
\left\vert 0_{L}\right\rangle \overset{\text{def}}{=}\left\vert
01\right\rangle \text{, }\left\vert 1_{L}\right\rangle \overset{\text{def}}%
{=}\left\vert 10\right\rangle \text{,} \label{qdr}%
\end{equation}
and \cite{markus},%
\begin{equation}
\left\vert 0_{L}\right\rangle \overset{\text{def}}{=}\frac{\left\vert
0000\right\rangle +\left\vert 1111\right\rangle }{\sqrt{2}}\text{ ,
}\left\vert 1_{L}\right\rangle \overset{\text{def}}{=}\frac{\left\vert
0110\right\rangle +\left\vert 1001\right\rangle }{\sqrt{2}}\text{,}
\label{erasure}%
\end{equation}
respectively. Both $\mathcal{C}_{\text{QDR}}$ and $\mathcal{C}_{\text{erasure}%
}$ are stabilizer codes with stabilizer groups given by%
\begin{equation}
\mathcal{S}\overset{\text{def}}{=}\left\langle -\sigma_{z}^{1}\sigma_{z}%
^{2}\right\rangle \text{,}%
\end{equation}
and,%
\begin{equation}
\mathcal{S}\overset{\text{def}}{=}\left\langle \sigma_{x}^{1}\sigma_{x}%
^{2}\sigma_{x}^{3}\sigma_{x}^{4}\text{, }\sigma_{z}^{1}\sigma_{z}^{4}\text{,
}\sigma_{z}^{2}\sigma_{z}^{3}\right\rangle \text{,}%
\end{equation}
respectively. We recall that, as pointed out in \cite{preskill}, minus signs
do not really matter when the stabilizers are specified. Erasures are errors
at known positions and a $t$-error correcting code is a $2t$-erasure
correcting code. It can be shown that the perfect $1$-erasure correcting code
is also a single AD-error correcting codes and is local permutation equivalent
to the Leung et \textit{al}. four-qubit code. Using Eqs. (\ref{qdr}) and
(\ref{erasure}), the concatenated code $\mathcal{C}_{\text{conc.}}%
\overset{\text{def}}{=}\mathcal{C}_{\text{QDR}}\circ\mathcal{C}%
_{\text{erasure}}$ is spanned by the following codewords,%
\begin{align}
&  \left\vert 0_{L}\right\rangle \overset{\text{def}}{=}\frac{\left\vert
00000110\right\rangle +\left\vert 00001001\right\rangle +\left\vert
11110110\right\rangle +\left\vert 11111001\right\rangle }{\sqrt{4}}%
\text{,}\nonumber\\
& \nonumber\\
&  \left\vert 1_{L}\right\rangle \overset{\text{def}}{=}\frac{\left\vert
01100000\right\rangle +\left\vert 01101111\right\rangle +\left\vert
10010000\right\rangle +\left\vert 10011111\right\rangle }{\sqrt{4}}\text{.}
\label{conc}%
\end{align}
The stabilizer generators of the concatenated code can be obtained as follows.
The concatenated code uses $8$ qubits that parse two blocks, each containing
$4$ qubits. Qubits $1$-$4$ belong to block $1$; qubits $5$-$8$ belong to block
$2$. To each block we associate a copy of the generators of $\mathcal{C}%
_{\text{erasure}}$. This gives the following six generators,%
\begin{equation}
g_{1}\overset{\text{def}}{=}\sigma_{x}^{1}\sigma_{x}^{2}\sigma_{x}^{3}%
\sigma_{x}^{4}\text{, }g_{2}\overset{\text{def}}{=}\sigma_{x}^{5}\sigma
_{x}^{6}\sigma_{x}^{7}\sigma_{x}^{8}\text{, }g_{3}\overset{\text{def}}%
{=}\sigma_{z}^{1}\sigma_{z}^{4}\text{, }g_{4}\overset{\text{def}}{=}\sigma
_{z}^{5}\sigma_{z}^{8}\text{, }g_{5}\overset{\text{def}}{=}\sigma_{z}%
^{2}\sigma_{z}^{3}\text{, }g_{6}\overset{\text{def}}{=}\sigma_{z}^{6}%
\sigma_{z}^{7}\text{.}%
\end{equation}
The remaining generator $g_{7}$ is the encoded version of $-\sigma_{z}%
^{1}\sigma_{z}^{2}$, that is $g_{7}\overset{\text{def}}{=}-\sigma_{z}%
^{1}\sigma_{z}^{2}\sigma_{z}^{5}\sigma_{z}^{6}$. Summing up, the stabilizer
group for the concatenated code reads,%
\begin{equation}
\mathcal{S}_{\mathcal{C}_{\text{conc.}}}\overset{\text{def}}{=}\left\langle
\sigma_{x}^{1}\sigma_{x}^{2}\sigma_{x}^{3}\sigma_{x}^{4}\text{, }\sigma
_{x}^{5}\sigma_{x}^{6}\sigma_{x}^{7}\sigma_{x}^{8}\text{, }\sigma_{z}%
^{1}\sigma_{z}^{4}\text{, }\sigma_{z}^{5}\sigma_{z}^{8}\text{, }\sigma_{z}%
^{2}\sigma_{z}^{3}\text{, }\sigma_{z}^{6}\sigma_{z}^{7}\text{, }-\sigma
_{z}^{1}\sigma_{z}^{2}\sigma_{z}^{5}\sigma_{z}^{6}\text{ }\right\rangle
\text{.} \label{stabilizer structure}%
\end{equation}
It turns out that the concatenated code with stabilizer structure defined in
Eq. (\ref{stabilizer structure}) is a nondegenerate code of distance $2$.
Furthermore, it can be explicitly checked that the error set $\left\{
I\text{, }\sigma_{x}^{i}\text{, }\sigma_{y}^{i}\right\}  $ with $i=1$, ...,
$8$ is a set of linearly independent errors with unequal error syndromes, a
property of correctable errors by means of nondegenerate codes \cite{gaitan}.
We recall that the syndrome $\emph{s}\left(  E\right)  $ for an error $E$ in
the Pauli group $\mathcal{P}_{\mathcal{H}_{2}^{n}}$ is the bit string
$l=l_{1}$...$l_{n-k}$ where the component bits $l_{i}$ are given by,%
\begin{equation}
l_{i}\overset{\text{def}}{=}\left\{
\begin{array}
[c]{c}%
0\text{, if }\left[  E\text{, }g_{i}\right]  =0\\
1\text{, if }\left\{  E\text{, }g_{i}\right\}  =0
\end{array}
\right.  \text{,}%
\end{equation}
with $i=1$,..., $n-k$ and $\mathcal{S}\overset{\text{def}}{=}\left\langle
g_{i}\right\rangle $ the stabilizer group of the quantum code.

Before discussing the computation of the entanglement fidelity, recall that
the quantum Hamming bound places an upper bound on the number of errors $t$
that an $\left[  \left[  n,k,d\right]  \right]  $ nondegenerate code can
correct for given $n$ and $k$,%
\begin{equation}
2^{k}%
%TCIMACRO{\dsum \limits_{j=0}^{t}}%
%BeginExpansion
{\displaystyle\sum\limits_{j=0}^{t}}
%EndExpansion
3^{j}\binom{n}{j}\leq2^{n}\text{.}%
\end{equation}
Since the code distance $d$ equals $2t+1$, it also places an upper bound on
the code distance. At present it is unknown whether a degenerate code might
allow a violation of the Hamming bound \cite{sarvepalli}. For the sake of
completeness, we also point out that the Hamming bound for $q$-dimensional
systems reads%
\begin{equation}
K%
%TCIMACRO{\dsum \limits_{j=0}^{t}}%
%BeginExpansion
{\displaystyle\sum\limits_{j=0}^{t}}
%EndExpansion
\left(  q^{2}-1\right)  ^{j}\binom{n}{j}\leq q^{n}\text{,}%
\end{equation}
where $k=\log_{q}K$ and $\dim_{%
%TCIMACRO{\U{2102} }%
%BeginExpansion
\mathbb{C}
%EndExpansion
}\mathcal{H}=q^{n}$ with $\mathcal{H}=\left(  \mathcal{%
%TCIMACRO{\U{2102} }%
%BeginExpansion
\mathbb{C}
%EndExpansion
}^{q}\right)  ^{\otimes n}$.

The enlarged GAD quantum channel after performing the encoding defined by
means of Eq. (\ref{conc}) reads,%
\begin{equation}
\Lambda_{\text{GAD}}^{\left[  \left[  8,1\right]  \right]  }\left(
\rho\right)  \overset{\text{def}}{=}%
%TCIMACRO{\dsum \limits_{r=0}^{2^{16}-1}}%
%BeginExpansion
{\displaystyle\sum\limits_{r=0}^{2^{16}-1}}
%EndExpansion
A_{r}^{\prime}\rho A_{r}^{\prime\dagger}=%
%TCIMACRO{\dsum \limits_{i\text{, }j\text{, }k\text{, }l\text{, }m\text{,
%}n\text{, }s\text{, }t=0}^{3}}%
%BeginExpansion
{\displaystyle\sum\limits_{i\text{, }j\text{, }k\text{, }l\text{, }m\text{,
}n\text{, }s\text{, }t=0}^{3}}
%EndExpansion
A_{ijklmnst}\rho A_{ijklmnst}^{\dagger}\text{,}%
\end{equation}
where to any of the $2^{16}$ values of $r$ we can associate a set of indices
$\left(  i\text{, }j\text{, }k\text{, }l\text{, }m\text{, }n\text{, }s\text{,
}t\right)  $ (and vice-versa) such that,%
\begin{equation}
A_{r}^{\prime}\leftrightarrow A_{ijklmnst}\overset{\text{def}}{=}A_{i}\otimes
A_{j}\otimes A_{k}\otimes A_{l}\otimes A_{m}\otimes A_{n}\otimes A_{s}\otimes
A_{t}\equiv A_{i}A_{j}A_{k}A_{l}A_{m}A_{n}A_{s}A_{t}\text{.}%
\end{equation}
The errors $A_{i}$ with $i\in\left\{  0\text{, }1\text{, }2\text{, }3\right\}
$ are defined in Eq. (\ref{ko}) and $\rho\in\mathcal{M}\left(  \mathcal{C}%
\right)  $ with $\mathcal{C}\subset\mathcal{H}_{2}^{8}$. In particular, the
number of weight-$q$ enlarged error operators $A_{r}^{\prime}$ is given by
$3^{q}\binom{8}{q}$ and,%
\begin{equation}
2^{16}=%
%TCIMACRO{\dsum \limits_{q=0}^{8}}%
%BeginExpansion
{\displaystyle\sum\limits_{q=0}^{8}}
%EndExpansion
3^{q}\binom{8}{q}\text{.}%
\end{equation}
In this case, the recovery scheme $\mathcal{R}$ that we use can be described
as follows: $R_{0}$ is associated with the weight-$0$ error $A_{00000000}$;
$R_{k}$ with $k=1$, ..., $8$ are associated with the eight weight-$1$ errors
where single-qubit errors of type $A_{1}$ occur; $R_{k}$ with $k=9$, ..., $16$
are associated with the eight weight-$1$ errors where single-qubit errors of
type $A_{3}$ occur. In analogy to the case of the CSS seven-qubit code, the
action on the codewords of seven weight-$1$ enlarged errors (specifically,
$A_{20000000}$, $A_{02000000}$, $A_{00200000}$, $A_{00020000}$, $A_{00002000}%
$, $A_{00000200}$, $A_{00000020}$ and $A_{00000002}$) leads to vectors that
are not orthogonal to those obtained from the action of the weight-$0$ error
$A_{00000000}$ on the codewords. Thus, we omit them from the construction of
our recovery scheme. We also choose to recover the weight-$2$ errors that are
more likely to occur where the likelihood can be expressed in terms of the
perturbation parameters $\gamma$ and $\epsilon$. For instance, in the limiting
case of $\varepsilon=0$, it can be shown that the sum of all the probabilities
for errors of weight-$k$ to occur reads,
\begin{equation}
P_{weight\text{-}k}\overset{\text{def}}{=}\frac{1}{2}\left[  \binom{2}%
{k}\gamma^{k}\left(  1-\gamma\right)  ^{2-k}+\binom{6}{k}\gamma^{k}\left(
1-\gamma\right)  ^{6-k}\right]  \text{,}%
\end{equation}
with the normalization constraint,%
\begin{equation}
\sum\limits_{k=0}^{8}P_{weight\text{-}k}=1\text{.}%
\end{equation}
Therefore, to the $17$ recovery operators $R_{k}$ constructed so far, we add
the additional $20$ out of the possible $28=\binom{8}{2}$ recovery operators
constructed by means of weight-$2$ errors where errors of type $A_{1}$ occur.
We point out that we only consider $20$ recovery operators, since the
following eight errors are not correctable,%
\begin{equation}
A_{11000000}\text{, }A_{10100000}\text{, }A_{01010000}\text{, }A_{00110000}%
\text{, }A_{00001100}\text{, }A_{00001010}\text{, }A_{00000101}\text{,
}A_{00000011}\text{.}%
\end{equation}
We finally arrive at the construction of $37$ recovery operators $R_{k}$ with
$k=0$,..., $36$ described in terms of $74$ orthonormal vectors in
$\mathcal{H}_{2}^{8}$. The missing $182$ orthonormal vectors (needed to obtain
an orthonormal basis of $\mathcal{H}_{2}^{8}$ and to construct $R_{37}%
\overset{\text{def}}{=}\hat{O}$) can be formally computed using the
rank-nullity theorem together with the Gram-Schmidt orthonormalization.
Omitting further technical details but using the very same line of reasoning
presented for the CSS seven-qubit code, our analytical estimate of the
entanglement fidelity of the eight-qubit concatenated code reads,%
\begin{equation}
\mathcal{F}^{\left[  \left[  8,1\right]  \right]  }\left(  \gamma\text{,
}\varepsilon\right)  \overset{\text{def}}{=}\frac{1}{\left(  \dim_{%
%TCIMACRO{\U{2102} }%
%BeginExpansion
\mathbb{C}
%EndExpansion
}\mathcal{C}\right)  ^{2}}\sum_{k=0}^{2^{16}-1}\sum_{l=0}^{37}\left\vert
\text{Tr}\left(  R_{l}A_{k}^{\prime}\right)  _{\left\vert \mathcal{C}\right.
}\right\vert ^{2}\approx1-2\gamma^{2}-28\varepsilon^{2}\left(  1+\frac{\gamma
}{\varepsilon}\right)  +\mathcal{O}\left(  3\right)  \text{.}
\label{estimate3}%
\end{equation}
We remark that in the limiting case of $\varepsilon=0$, the GAD\ noise model
reduces to the traditional AD model and the concatenated code $\mathcal{C}%
_{\text{conc.}}$ applied to AD\ errors \ works as follows: the inner code
$\mathcal{C}_{\text{QDR}}$ transforms AD errors into erasures which are then
corrected by the outer code $\mathcal{C}_{\text{erasure}}$ \cite{bz}. We also
point out that in the limit of $\varepsilon=0$ our estimated series expansion
of the entanglement fidelity of the eight-qubit concatenated code coincides
with that obtained by means of the traditional Leung et \textit{al}.
four-qubit code applied to AD errors,%
\begin{equation}
\mathcal{F}^{\left[  \left[  8,1\right]  \right]  }\left(  \gamma\text{,
}\varepsilon=0\right)  \approx1-2\gamma^{2}+\mathcal{O}\left(  3\right)
\approx\mathcal{F}_{\text{Leung}}^{\left[  \left[  4,1\right]  \right]
}\left(  \gamma\right)  \text{,} \label{finding-ok}%
\end{equation}
where we assume to use recovery schemes with the same structure as in Eq.
(\ref{the recovery}). This finding is not unexpected and is in agreement with
the fact that, as pointed out earlier, $\mathcal{C}_{\text{erasure}}$ and
$\mathcal{C}_{\text{Leung}}$ are local permutation equivalent quantum codes.

Finally, we compare the performances of the additive codes employed in our
error correction schemes in Fig\textbf{. }$3$\textbf{.}

\begin{figure}[ptb]
\centering
\includegraphics[width=0.5\textwidth]{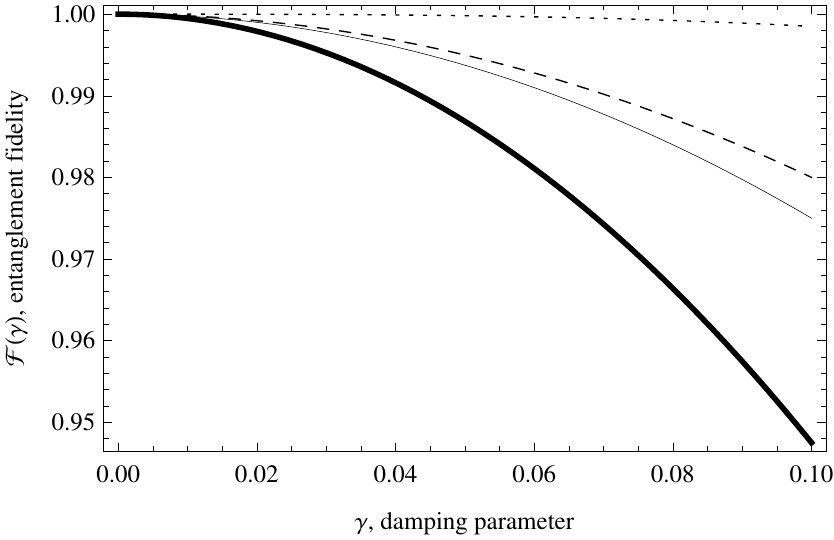}\caption{\textit{Ranking additive
codes}. The truncated series expansions of the entanglement fidelity
$\mathcal{F}\left(  \gamma\right)  $ vs. the amplitude damping parameter
$\gamma$ for $\varepsilon=0$ and $0\leq\gamma\leq10^{-1}$: the Shor nine-qubit
code (dotted line), the degenerate six-qubit code (dashed line), the
five-qubit code (thin solid line), and the CSS seven-qubit code (thick solid
line).}%
\label{figure3}%
\end{figure}

\subsection{Degenerate codes}

Degeneracy is a property of quantum codes which has no analog for classical
error correcting codes and it arises from the fact that two different error
patterns can have indistinguishable effects on a coded quantum state
\cite{vincenzo}. A degenerate code has linearly independent matrices that act
in a linearly dependent way on the codewords, while in a nondegenerate code,
all the errors acting on the codewords produce linearly independent quantum
states. For instance, the Shor nine-qubit code is a degenerate code, since
phase errors within a group of $3$ qubits act the same way. A striking feature
of degenerate quantum codes is that they can be used to correct more errors
than they can uniquely identify \cite{GS}. Also, strictly speaking, degeneracy
is not a property of a code, but a property of a code together with a family
of errors it is designed to correct \cite{gaitan, GS}.

\subsubsection{The six-qubit code}

Calderbank et \textit{al}. discovered two distinct six-qubit degenerate
quantum codes encoding one logical qubit into six physical qubits
\cite{robert}. The first of these codes was discovered by trivially extending
the five-qubit code while the other one through an exhaustive search of the
encoding space. In particular, in \cite{robert} it was argued that this second
example is unique up to equivalence. The example that we consider here was
originally introduced by Bilal et \textit{al}. in \cite{bilal}. They argue
that, since their example is not reducible to the trivial six-qubit code
because every one of its qubits is entangled with the others, their code is
equivalent to the second nontrivial six-qubit code according to the arguments
of Calderbank et \textit{al}. The codespace of this nontrivial $\left[
\left[  6,1,3\right]  \right]  $ six-qubit code is spanned by the codewords
$\left\vert 0_{L}\right\rangle $ and $\left\vert 1_{L}\right\rangle $ defined
as \cite{bilal},%
\begin{equation}
\left\vert 0_{L}\right\rangle \overset{\text{def}}{=}\frac{1}{\sqrt{8}}\left[
\left\vert 000000\right\rangle -\left\vert 100111\right\rangle +\left\vert
001111\right\rangle -\left\vert 101000\right\rangle -\left\vert
010010\right\rangle +\left\vert 110101\right\rangle +\left\vert
011101\right\rangle -\left\vert 111010\right\rangle \right]  \text{,}
\label{6ca}%
\end{equation}
and,%
\begin{equation}
\left\vert 1_{L}\right\rangle \overset{\text{def}}{=}\frac{1}{\sqrt{8}}\left[
\left\vert 001010\right\rangle +\left\vert 101101\right\rangle +\left\vert
000101\right\rangle +\left\vert 1000010\right\rangle -\left\vert
011000\right\rangle -\left\vert 111111\right\rangle +\left\vert
010111\right\rangle +\left\vert 110000\right\rangle \right]  \text{,}
\label{6cb}%
\end{equation}
respectively. The five stabilizer generators for this code can be written as,%
\begin{equation}
g_{1}\overset{\text{def}}{=}Y^{1}Z^{3}X^{4}X^{5}Y^{6}\text{, }g_{2}%
\overset{\text{def}}{=}Z^{1}X^{2}X^{5}Z^{6}\text{, }g_{3}\overset{\text{def}%
}{=}Z^{2}X^{3}X^{4}X^{5}X^{6}\text{, }g_{4}\overset{\text{def}}{=}Z^{4}%
Z^{6}\text{, }g_{5}\overset{\text{def}}{=}Z^{1}Z^{2}Z^{3}Z^{5}\text{.}
\label{stabilizer613}%
\end{equation}
Within the quantum stabilizer formalism, an $\left[  \left[  n,k,d\right]
\right]  $ code is degenerate if the stabilizer group contains elements of
weight less than $d$ (other than the identity) \cite{gaitan}. Thus, it appears
evident from Eq. (\ref{stabilizer613}) that this distance $d=3$ code is degenerate.

The enlarged GAD quantum channel after performing the encoding defined by
means of Eqs. (\ref{6ca}) and (\ref{6cb}) reads,%
\begin{equation}
\Lambda_{\text{GAD}}^{\left[  \left[  6,1,3\right]  \right]  }\left(
\rho\right)  \overset{\text{def}}{=}%
%TCIMACRO{\dsum \limits_{r=0}^{2^{12}-1}}%
%BeginExpansion
{\displaystyle\sum\limits_{r=0}^{2^{12}-1}}
%EndExpansion
A_{r}^{\prime}\rho A_{r}^{\prime\dagger}=%
%TCIMACRO{\dsum \limits_{i\text{, }j\text{, }k\text{, }l\text{, }m\text{, }%
%n=0}^{3}}%
%BeginExpansion
{\displaystyle\sum\limits_{i\text{, }j\text{, }k\text{, }l\text{, }m\text{,
}n=0}^{3}}
%EndExpansion
A_{ijklmn}\rho A_{ijklmn}^{\dagger}\text{,}%
\end{equation}
where to any of the $2^{12}$ values of $r$ we can associate a set of indices
$\left(  i\text{, }j\text{, }k\text{, }l\text{, }m\text{, }n\right)  $ (and
vice-versa) such that,%
\begin{equation}
A_{r}^{\prime}\leftrightarrow A_{ijklmn}\overset{\text{def}}{=}A_{i}\otimes
A_{j}\otimes A_{k}\otimes A_{l}\otimes A_{m}\otimes A_{n}\equiv A_{i}%
A_{j}A_{k}A_{l}A_{m}A_{n}\text{.}%
\end{equation}
The errors $A_{i}$ with $i\in\left\{  0\text{, }1\text{, }2\text{, }3\right\}
$ are defined in Eq. (\ref{ko}) and $\rho\in\mathcal{M}\left(  \mathcal{C}%
\right)  $ with $\mathcal{C}\subset\mathcal{H}_{2}^{6}$. In particular, the
number of weight-$q$ enlarged error operators $A_{r}^{\prime}$ is given by
$3^{q}\binom{6}{q}$ and,%
\begin{equation}
2^{12}=%
%TCIMACRO{\dsum \limits_{q=0}^{6}}%
%BeginExpansion
{\displaystyle\sum\limits_{q=0}^{6}}
%EndExpansion
3^{q}\binom{6}{q}\text{.}%
\end{equation}
For $\varepsilon=0$, it can be shown that among the $\binom{6}{2}=15$
weight-$2$ errors, the set of five weight-$2$ errors given by $\left\{
A_{110000}\text{, }A_{100010}\text{, }A_{011000}\text{, }A_{001010}\text{,
}A_{000101}\right\}  $ is not correctable, since they are not compatible with
$A_{000000}$. In addition, the action of the two weight-$2$ errors
$A_{101000}$ and $A_{010010}$ on the codewords leads to state vectors that are
not orthogonal to $A_{000000}\left\vert i_{L}\right\rangle $ with
$i\in\left\{  0\text{, }1\right\}  $. All weight-$1$ errors are correctable,
of course. In view of these considerations, we construct the recovery scheme
$\mathcal{R}$ for the general case as follows: $R_{0}$ is associated with the
weight-$0$ error $A_{000000}$; $R_{k}$ with $k=1$, ..., $6$ are associated
with the six weight-$1$ errors where single-qubit errors of type $A_{1}$
occur; $R_{k}$ with $k=7$, ..., $12$ are associated with the six weight-$1$
errors where single-qubit errors of type $A_{3}$ occur; finally, $R_{k}$ with
$k=13$, ..., $20$ are associated with the eight correctable weight-$2$ errors
where errors of type $A_{1}$ occur. The construction of these $21$ recovery
operators $R_{k}$ is described in terms of $42$ orthonormal vectors in
$\mathcal{H}_{2}^{6}$. As pointed out earlier, the missing $22$ orthonormal
vectors (needed to obtain an orthonormal basis of $\mathcal{H}_{2}^{6}$ and to
construct $R_{21}\overset{\text{def}}{=}\hat{O}$) can be formally computed by
using the rank-nullity theorem together with the Gram-Schmidt
orthonormalization procedure. Our analytical estimate of the entanglement
fidelity of the nontrivial six-qubit degenerate code reads,%
\begin{equation}
\mathcal{F}^{\left[  \left[  6,1,3\right]  \right]  }\left(  \gamma\text{,
}\varepsilon\right)  \overset{\text{def}}{=}\frac{1}{\left(  \dim_{%
%TCIMACRO{\U{2102} }%
%BeginExpansion
\mathbb{C}
%EndExpansion
}\mathcal{C}\right)  ^{2}}\sum_{k=0}^{2^{12}-1}\sum_{l=0}^{21}\left\vert
\text{Tr}\left(  R_{l}A_{k}^{\prime}\right)  _{\left\vert \mathcal{C}\right.
}\right\vert ^{2}\approx1-2\gamma^{2}-15\varepsilon^{2}\left(  1+\frac{\gamma
}{\varepsilon}\right)  +\mathcal{O}\left(  3\right)  \text{.}
\label{estimate4}%
\end{equation}
Comparing Eqs. (\ref{estimate1}), (\ref{estimate2}) and (\ref{estimate3}) to
Eq. (\ref{estimate4}), we observe that the six-qubit degenerate code
outperforms the five-, CSS\ seven-, and eight-qubit concatenated nondegenerate
codes. Our findings show that despite the fact that the CSS seven- and
eight-qubit concatenated codes have larger Hilbert spaces for encoding than
that allowed for the six-qubit code, their error-correcting capability is
smaller, given the noise model considered and the recovery schemes employed.
Our finding in Eq. (\ref{estimate4}) strengthens the \emph{suspect} advanced
in \cite{andy3} where it was conjectured that thanks to their degenerate
structure, such codes can outperform nondegenerate codes despite their shorter length.

\subsubsection{The Shor nine-qubit code}

We consider here the $\left[  \left[  9,1,3\right]  \right]  $ Shor nine-qubit
code \cite{shor1995}, the code that gave birth to the subject of quantum error
correcting codes. The codespace of such a code is spanned by the following two
codewords \cite{gaitan},%
\begin{equation}
\left\vert 0_{L}\right\rangle \overset{\text{def}}{=}\frac{1}{\sqrt{8}}\left[
\left\vert 000\right\rangle +\left\vert 111\right\rangle \right]  \left[
\left\vert 000\right\rangle +\left\vert 111\right\rangle \right]  \left[
\left\vert 000\right\rangle +\left\vert 111\right\rangle \right]  \text{,}
\label{code9a}%
\end{equation}
and,%
\begin{equation}
\left\vert 1_{L}\right\rangle \overset{\text{def}}{=}\frac{1}{\sqrt{8}}\left[
\left\vert 000\right\rangle -\left\vert 111\right\rangle \right]  \left[
\left\vert 000\right\rangle -\left\vert 111\right\rangle \right]  \left[
\left\vert 000\right\rangle -\left\vert 111\right\rangle \right]  \text{.}
\label{code9b}%
\end{equation}
This degenerate code can be constructed by concatenating two nondegenerate
$\left[  \left[  3,1,1\right]  \right]  $ codes and its eight stabilizer
generators can be written as,%
\begin{align}
&  g_{1}\overset{\text{def}}{=}Z^{1}Z^{2}\text{, }g_{2}\overset{\text{def}}%
{=}Z^{1}Z^{3}\text{, }g_{3}\overset{\text{def}}{=}Z^{4}Z^{5}\text{, }%
g_{4}\overset{\text{def}}{=}Z^{4}Z^{6}\text{, }g_{5}\overset{\text{def}}%
{=}Z^{7}Z^{8}\text{, }g_{6}\overset{\text{def}}{=}Z^{7}Z^{9}\text{,
}\nonumber\\
& \nonumber\\
&  g_{7}\overset{\text{def}}{=}X^{1}X^{2}X^{3}X^{4}X^{5}X^{6}\text{, }%
g_{8}\overset{\text{def}}{=}X^{1}X^{2}X^{3}X^{7}X^{8}X^{9}\text{.}%
\end{align}
The enlarged GAD quantum channel after performing the encoding defined by
means of Eqs. (\ref{code9a}) and (\ref{code9b}) reads,%
\begin{equation}
\Lambda_{\text{GAD}}^{\left[  \left[  9,1,3\right]  \right]  }\left(
\rho\right)  \overset{\text{def}}{=}%
%TCIMACRO{\dsum \limits_{r=0}^{2^{18}-1}}%
%BeginExpansion
{\displaystyle\sum\limits_{r=0}^{2^{18}-1}}
%EndExpansion
A_{r}^{\prime}\rho A_{r}^{\prime\dagger}=%
%TCIMACRO{\dsum \limits_{i\text{, }j\text{, }k\text{, }l\text{, }m\text{,
%}n\text{, }s\text{, }t\text{, }u=0}^{3}}%
%BeginExpansion
{\displaystyle\sum\limits_{i\text{, }j\text{, }k\text{, }l\text{, }m\text{,
}n\text{, }s\text{, }t\text{, }u=0}^{3}}
%EndExpansion
A_{ijklmnstu}\rho A_{ijklmnstu}^{\dagger}\text{,}%
\end{equation}
where to any of the $2^{18}$ values of $r$ we can associate a set of indices
$\left(  i\text{, }j\text{, }k\text{, }l\text{, }m\text{, }n\text{, }s\text{,
}t\text{, }u\right)  $ (and vice-versa) such that,%
\begin{equation}
A_{r}^{\prime}\leftrightarrow A_{ijklmnstu}\overset{\text{def}}{=}A_{i}\otimes
A_{j}\otimes A_{k}\otimes A_{l}\otimes A_{m}\otimes A_{n}\otimes A_{s}\otimes
A_{t}\otimes A_{u}\equiv A_{i}A_{j}A_{k}A_{l}A_{m}A_{n}A_{s}A_{t}A_{u}\text{.}%
\end{equation}
The errors $A_{i}$ with $i\in\left\{  0\text{, }1\text{, }2\text{, }3\right\}
$ are defined in Eq. (\ref{ko}) and $\rho\in\mathcal{M}\left(  \mathcal{C}%
\right)  $ with $\mathcal{C}\subset\mathcal{H}_{2}^{9}$. In particular, the
number of weight-$q$ enlarged error operators $A_{r}^{\prime}$ is given by
$3^{q}\binom{9}{q}$ and,%
\begin{equation}
2^{18}=%
%TCIMACRO{\dsum \limits_{q=0}^{9}}%
%BeginExpansion
{\displaystyle\sum\limits_{q=0}^{9}}
%EndExpansion
3^{q}\binom{9}{q}\text{.}%
\end{equation}
As a side remark, we recall that a quantum code has distance $d$ if all errors
of weight less than $d$ satisfy the QEC conditions $\left\langle
i_{L}\left\vert A_{l}^{\dagger}A_{m}\right\vert j_{L}\right\rangle
=\alpha_{lm}\delta_{ij}$ and at least one error of weight $d$ exists that
violates it. Otherwise stated, the distance of a code is the weight of the
smallest error $A_{l}^{\dagger}A_{m}$ that cannot be detected by the code. For
instance, using the three-qubit bit-flip repetition code $\left[  \left[
3,1,1\right]  \right]  $ to correct bit-flip errors, it turns out that the
weight-$1$ error $\sigma_{z}^{1}$ cannot be detected. However, this code of
distance $d=1$ also detects errors of weight-$2$ such as, for instance,
$\sigma_{x}^{1}\sigma_{x}^{2}$.

For $\varepsilon=0$, it turns out that all the nine weight-$1$ and thirty-six
weight-$2$ errors are correctable. In addition, all weight-$3$ errors can be
recovered as well, except for $A_{111000000}$, $A_{000111000}$ and
$A_{000000111}$. These three errors could be potentially recovered by means of
the recovery operator constructed with the weight-$0$ error $A_{000000000}$.
However, it can be checked that their contributions is null. For the general
case, the recovery scheme $\mathcal{R}$ that we use can be described as
follows: $R_{0}$ is associated with the weight-$0$ error $A_{000000000}$;
$R_{k}$ with $k=1$, ..., $9$ are associated with the nine weight-$1$ errors
where single-qubit errors of type $A_{1}$ occur; $R_{k}$ with $k=10$, ...,
$18$ are associated with the nine weight-$1$ errors where single-qubit errors
of type $A_{3}$ occur; $R_{k}$ with $k=19$, ..., $54$ are associated with the
thirty-six weight-$2$ errors where errors of type $A_{1}$ occur; finally,
$R_{k}$ with $k=55$, ..., $135$ are associated with the eighty-one weight-$3$
errors where errors of type $A_{1}$ occur. The construction of these $136$
recovery operators $R_{k}$ is described in terms of $272$ orthonormal vectors
in $\mathcal{H}_{2}^{9}$. The missing $240$ orthonormal vectors, needed to
obtain an orthonormal basis of $\mathcal{H}_{2}^{9}$ and to construct
$R_{136}\overset{\text{def}}{=}\hat{O}$, can be formally computed by using the
rank-nullity theorem together with the Gram-Schmidt orthonormalization
procedure. Our analytical estimate of the entanglement fidelity of the Shor
nine-qubit code reads,%
\begin{equation}
\mathcal{F}^{\left[  \left[  9,1,3\right]  \right]  }\left(  \gamma\text{,
}\varepsilon\right)  \overset{\text{def}}{=}\frac{1}{\left(  \dim_{%
%TCIMACRO{\U{2102} }%
%BeginExpansion
\mathbb{C}
%EndExpansion
}\mathcal{C}\right)  ^{2}}\sum_{k=0}^{2^{18}-1}\sum_{l=0}^{136}\left\vert
\text{Tr}\left(  R_{l}A_{k}^{\prime}\right)  _{\left\vert \mathcal{C}\right.
}\right\vert ^{2}\approx1-\frac{3}{2}\gamma^{3}-36\varepsilon^{2}\left(
1+\frac{\gamma}{\varepsilon}\right)  +\mathcal{O}\left(  3\right)  \text{.}
\label{estimate5}%
\end{equation}
To the best of our knowledge and unlike the case of the five-qubit code, no
truncated series expansion of $\mathcal{F}^{\left[  \left[  9,1,3\right]
\right]  }\left(  \gamma\text{, }\varepsilon\right)  $ with $\varepsilon=0$ is
available in the literature. However, we emphasize that in the special case of
$\varepsilon=0$, our analytical estimate in Eq. (\ref{estimate5}) appears to
exhibit a fairly good agreement with the numerical plot presented in
\cite{andy1} (specifically, see Figure $12$ in \cite{andy1}). For
$\varepsilon=0$, we compared our non truncated analytical estimate in Eq.
(\ref{estimate5}) to the baseline performance of a single qubit given by Eq.
(\ref{baseline}). Then, we checked the good overlap between our results (non
truncated fidelity expressions with and without error correction) and the ones
plotted in \cite{andy1}. For some more details, see Appendix F.

\section{Nonadditive codes}

There are codes that do not exhibit stabilizer structures. Such codes are
known as nonadditive quantum codes. A quantum code is nonadditive if it is not
additive, that is if it cannot be constructed within the stabilizer framework.

Examples of nonstabilizer codes can be found when one of the two following
requirements are satisfied \cite{looi}:

\begin{itemize}
\item It exists a state $\left\vert \psi\right\rangle \notin\mathcal{C}$ such
that $g\left\vert \psi\right\rangle =\left\vert \psi\right\rangle $ for any
operator $g$ that belongs to the group stabilizer $\mathcal{S}_{\mathcal{C}}$
of the code $\mathcal{C}$. This happens when $\mathcal{C}$ is not maximal;

\item It exists $g\notin\mathcal{S}_{\mathcal{C}}$ such that $g\left\vert
\psi\right\rangle =\left\vert \psi\right\rangle $ for any state in the
codespace of $\mathcal{C}$. This happens when $\mathcal{S}_{\mathcal{C}}$ is
not maximal.
\end{itemize}

We point out it can also occur that the code $\mathcal{C}$ does not allow any
stabilizer group $\mathcal{S}_{\mathcal{C}}$ at all, neither maximal nor
minimal. For instance, consider the code $\mathcal{C}$ defined as,%
\begin{equation}
\mathcal{C}\overset{\text{def}}{=}\text{Span}\left\{  \left\vert
0_{L}\right\rangle \overset{\text{def}}{=}\frac{\left\vert 01\right\rangle
+\left\vert 10\right\rangle }{\sqrt{2}}\text{, }\left\vert 1_{L}\right\rangle
\overset{\text{def}}{=}\left\vert 11\right\rangle \right\}  \text{.}
\label{example}%
\end{equation}
The code $\mathcal{C}$ in Eq. (\ref{example}) is not the joint $+1$-eigenspace
of any operator in the Pauli group $\mathcal{P}_{\mathcal{H}_{1}^{2}}$.
Therefore, this code is not a stabilizer code, since it does not have the
standard stabilizer structure.

A nonadditive $\left(  \left(  n,K,d\right)  \right)  $ code is a
$K$-dimensional subspace of a $n$-qubit Hilbert space correcting $\left\lfloor
\frac{d-1}{2}\right\rfloor $-qubit errors and $d$ is the distance of the code.
The first example of nonadditive code was a $\left(  \left(  5,6,2\right)
\right)  $ code presented in \cite{rains}. This code was constructed
numerically by building a projector operator with a given weight distribution.
It encodes six logical qubits into five physical qubits and can correct
single-qubit erasure. This code outperforms any known stabilizer code in terms
of encoded dimension ($\log_{2}6/5$). In \cite{smolin}, a family of
nonadditive codes of distance $d=2$ capable of detecting any single-qubit
error (or, equivalently, correct any single-qubit erasure) with high encoded
dimensions was introduced. The simplest example of a nonadditive code in such
a family is represented by a self-complementary $\left(  \left(  5,5,2\right)
\right)  $ nonadditive quantum code that is not a subcode of the $\left(
\left(  5,6,2\right)  \right)  $ code in \cite{rains}. Necessary and
sufficient conditions for the error correction of amplitude damping errors
with self-complementary nonadditive quantum codes were presented in
\cite{lang-shor, shor-2011}. In particular, in \cite{lang-shor} a numerical
investigation of a $\left(  \left(  8,12\right)  \right)  $ self-complementary
nonadditive quantum code with a high encoding rate ($\log_{2}12/8$) was
presented for the correction of amplitude damping errors. The performance of
this code was quantified by means of the entanglement fidelity and evaluated
numerically for a maximum likelihood recovery scheme that corrects all the
first-order amplitude damping errors. We stress that for nonadditive codes,
the decoding procedure does not have the syndrome-diagnosis and the recovery
structure of the stabilizer codes. However, like the additive case, recovery
schemes for nonadditive codes may exhibit a projection nature as well.
Furthermore, thanks to the graph-state formalism \cite{hein}, many nonadditive
codes can be characterized by a stabilizer-like structure \cite{cws}. The
stabilizer-like structure of some classes of nonadditive codes simplify
significantly the encoding and decoding procedures for these codes \cite{yu2}.
Nonadditive codes of distance $d=2$ can detect but cannot correct arbitrary
single-qubit errors. To achieve this task, codes of distance $d=3$ are needed.
In \cite{vatan}, sufficient general conditions for the existence of
nonadditive codes were given. In particular, an example of a strongly
nonadditive $\left(  \left(  11,2,3\right)  \right)  $ quantum code was
presented. However, the question of whether the nonadditive codes correcting
errors beyond erasures are more efficient (in terms of the encoded dimension)
than the corresponding stabilizer codes remained open. The very first example
of a $1$-error correcting nonadditive code capable of outperforming the
optimal stabilizer code (the $\left[  \left[  9,3,3\right]  \right]  $ code)
of the same length was the nondegenerate $\left(  \left(  9,12,3\right)
\right)  $ code in \cite{yu}.

\subsection{Non-self-complementary codes}

\subsubsection{The $\left(  \left(  11,2,3\right)  \right)  $ code}

According to \cite{vatan}, two quantum codes $\mathcal{C}_{1}$ and
$\mathcal{C}_{2}$ in $%
%TCIMACRO{\U{2102} }%
%BeginExpansion
\mathbb{C}
%EndExpansion
^{2^{n}}$ are \emph{locally equivalent} if there is a transversal operator
$U\overset{\text{def}}{=}u_{1}\otimes...\otimes u_{n}$ with $u_{j}\in
SU\left(  2,%
%TCIMACRO{\U{2102} }%
%BeginExpansion
\mathbb{C}
%EndExpansion
\right)  $, mapping $\mathcal{C}_{1}$ into $\mathcal{C}_{2}$. Instead, two
codes are \emph{globally equivalent}, or simply equivalent, if $\mathcal{C}%
_{1}$ is locally equivalent to a code obtained from $\mathcal{C}_{2}$ by a
permutation on qubits. A\ quantum code $\mathcal{C}\subset%
%TCIMACRO{\U{2102} }%
%BeginExpansion
\mathbb{C}
%EndExpansion
^{2^{n}}$ is called nonadditive if it is not equivalent to any additive code;
moreover, $\mathcal{C}$ is strongly nonadditive if the only additive code that
contains any code equivalent to $\mathcal{C}$ is the trivial code $%
%TCIMACRO{\U{2102} }%
%BeginExpansion
\mathbb{C}
%EndExpansion
^{2^{n}}$ (in other words, if $\pm X_{\mathbf{\alpha}}Z_{\mathbf{\beta}}$ with
$\mathbf{\alpha}$, $\mathbf{\beta\in}\left\{  0,1\right\}  ^{n}$ is in the
stabilizer of any code equivalent to a supercode of $\mathcal{C}$ then
$\mathbf{\alpha=}$ $\mathbf{\beta}=\mathbf{0}$). Also, the generalized
stabilizer $\mathcal{GS}_{\mathcal{C}}$ of a code $\mathcal{C}\subset$ $%
%TCIMACRO{\U{2102} }%
%BeginExpansion
\mathbb{C}
%EndExpansion
^{2^{n}}$ is the set of all unitary operators $V$ on $%
%TCIMACRO{\U{2102} }%
%BeginExpansion
\mathbb{C}
%EndExpansion
^{2^{n}}$ such that $V\left\vert c\right\rangle =\left\vert c\right\rangle $
for every $\left\vert c\right\rangle \in\mathcal{C}$. Then, the stabilizer
$\mathcal{S}_{\mathcal{C}}$ of a code $\mathcal{C}$ is $\mathcal{S}%
_{\mathcal{C}}=\mathcal{P}_{%
%TCIMACRO{\U{2102} }%
%BeginExpansion
\mathbb{C}
%EndExpansion
^{2^{n}}}\cap\mathcal{GS}_{\mathcal{C}}$ where $\mathcal{P}_{%
%TCIMACRO{\U{2102} }%
%BeginExpansion
\mathbb{C}
%EndExpansion
^{2^{n}}}$ is the $n$-qubit Pauli group.

The strong nonadditivity is guaranteed by the fulfillment of two conditions
\cite{vatan}: i) the identity operator is the only operator in the stabilizer
of the code; ii) there is no element in $\mathcal{GS}_{\mathcal{C}}$ of the
form $X_{\mathbf{\alpha}}T$ with $\left\{  0,1\right\}  ^{n}\ni\mathbf{\alpha
}\neq\mathbf{0}$ where $T$ is a $Z$-type unitary operator of the form,%
\begin{equation}
T\overset{\text{def}}{=}%
%TCIMACRO{\dbigotimes \limits_{j=1}^{n}}%
%BeginExpansion
{\displaystyle\bigotimes\limits_{j=1}^{n}}
%EndExpansion
\left(
\begin{array}
[c]{cc}%
e^{i\theta_{j}} & 0\\
0 & \pm e^{-i\theta_{j}}%
\end{array}
\right)  \text{,}%
\end{equation}
where $i$ is the \emph{complex} imaginary unit.

The first example of a $1$-error correcting strongly nonadditive code was a
$\left(  \left(  11,2,3\right)  \right)  $ code with codespace spanned by the
following two codewords \cite{vatan},%
\begin{equation}
\left\vert 0_{L}\right\rangle \overset{\text{def}}{=}\frac{1}{\sqrt{12}}%
\sum_{i=1}^{12}\left\vert r_{i}\right\rangle \text{,} \label{code11a}%
\end{equation}
and,%
\begin{equation}
\left\vert 1_{L}\right\rangle \overset{\text{def}}{=}\frac{1}{\sqrt{12}}%
\sum_{i=1}^{12}\left\vert \mathbf{1}+r_{i}\right\rangle \text{.}
\label{code11b}%
\end{equation}
The quantity $\mathbf{1}$ is the all-$1$ vector of length $11$ and $r_{i}$
denotes the $i$th row of the following $\left(  12\times11\right)  $-matrix
$H$ defined as,%
\begin{equation}
H\overset{\text{def}}{=}\left(
\begin{array}
[c]{ccccccccccc}%
0 & 0 & 0 & 0 & 0 & 0 & 0 & 0 & 0 & 0 & 0\\
1 & 0 & 1 & 0 & 0 & 0 & 1 & 1 & 1 & 0 & 1\\
1 & 1 & 0 & 1 & 0 & 0 & 0 & 1 & 1 & 1 & 0\\
0 & 1 & 1 & 0 & 1 & 0 & 0 & 0 & 1 & 1 & 1\\
1 & 0 & 1 & 1 & 0 & 1 & 0 & 0 & 0 & 1 & 1\\
1 & 1 & 0 & 1 & 1 & 0 & 1 & 0 & 0 & 0 & 1\\
1 & 1 & 1 & 0 & 1 & 1 & 0 & 1 & 0 & 0 & 0\\
0 & 1 & 1 & 1 & 0 & 1 & 1 & 0 & 1 & 0 & 0\\
0 & 0 & 1 & 1 & 1 & 0 & 1 & 1 & 0 & 1 & 0\\
0 & 0 & 0 & 1 & 1 & 1 & 0 & 1 & 1 & 0 & 1\\
1 & 0 & 0 & 0 & 1 & 1 & 1 & 0 & 1 & 1 & 0\\
0 & 1 & 0 & 0 & 0 & 1 & 1 & 1 & 0 & 1 & 1
\end{array}
\right)  \text{.}%
\end{equation}
The enlarged GAD quantum channel after performing the encoding defined by
means of Eqs. (\ref{code11a}) and (\ref{code11b}) reads,%
\begin{equation}
\Lambda_{\text{GAD}}^{\left(  \left(  11,2,3\right)  \right)  }\left(
\rho\right)  \overset{\text{def}}{=}%
%TCIMACRO{\dsum \limits_{r=0}^{2^{22}-1}}%
%BeginExpansion
{\displaystyle\sum\limits_{r=0}^{2^{22}-1}}
%EndExpansion
A_{r}^{\prime}\rho A_{r}^{\prime\dagger}=%
%TCIMACRO{\dsum \limits_{a_{1}\text{,..., }a_{11}=0}^{3}}%
%BeginExpansion
{\displaystyle\sum\limits_{a_{1}\text{,..., }a_{11}=0}^{3}}
%EndExpansion
A_{a_{1}a_{2}\text{...}a_{10}a_{11}}\rho A_{a_{1}a_{2}\text{...}a_{10}a_{11}%
}^{\dagger}\text{,}%
\end{equation}
where to any of the $2^{22}$ values of $r$ we can associate a set of indices
$\left(  a_{1}\text{,..., }a_{11}\right)  $ (and vice-versa) such that,%
\begin{equation}
A_{r}^{\prime}\leftrightarrow A_{a_{1}a_{2}\text{...}a_{10}a_{11}}%
\overset{\text{def}}{=}A_{a_{1}}\otimes A_{a_{2}}\otimes\text{...}\otimes
A_{a_{10}}\otimes A_{a_{11}}\equiv A_{a_{1}}A_{a_{2}}\text{...}A_{a_{10}%
}A_{a_{11}}\text{.}%
\end{equation}
The errors $A_{i}$ with $i\in\left\{  0\text{, }1\text{, }2\text{, }3\right\}
$ are defined in Eq. (\ref{ko}) and $\rho\in\mathcal{M}\left(  \mathcal{C}%
\right)  $ with $\mathcal{C}\subset\mathcal{H}_{2}^{11}$. In particular, the
number of weight-$q$ enlarged error operators $A_{r}^{\prime}$ is given by
$3^{q}\binom{11}{q}$ and,%
\begin{equation}
2^{22}=%
%TCIMACRO{\dsum \limits_{q=0}^{11}}%
%BeginExpansion
{\displaystyle\sum\limits_{q=0}^{11}}
%EndExpansion
3^{q}\binom{11}{q}\text{.}%
\end{equation}
We assume to focus on the recovery of weight-$1$ errors only and pay no
attention to the possible recovery of higher-order enlarged errors. For
instance, this working hypothesis is especially plausible for values of the
perturbation parameters $\gamma$ and $\varepsilon$ with $0\leq\varepsilon
\ll\gamma\ll1$. In this case, the recovery scheme $\mathcal{R}$ that we use
can be described as follows: $R_{0}$ is associated with the weight-$0$ error;
$R_{k}$ with $k=1$, ..., $11$ are associated with the eleven weight-$1$ errors
where single-qubit errors of type $A_{1}$ occur; finally, $R_{k}$ with $k=12$,
..., $22$ are associated with the eleven weight-$1$ errors where single-qubit
errors of type $A_{3}$ occur. The construction of these $23$ recovery
operators $R_{k}$ is described in terms of $46$ orthonormal vectors in
$\mathcal{H}_{2}^{11}$. The missing orthonormal vectors needed to obtain an
orthonormal basis of $\mathcal{H}_{2}^{11}$ and to construct $R_{23}%
\overset{\text{def}}{=}\hat{O}$ can be formally computed by using the
rank-nullity theorem together with the Gram-Schmidt orthonormalization
procedure. Finally, our analytical estimate of the entanglement fidelity of
the $\left(  \left(  11,2,3\right)  \right)  $ code reads,%
\begin{equation}
\mathcal{F}_{\text{first-order}}^{\left(  \left(  11,2,3\right)  \right)
}\left(  \gamma\text{, }\varepsilon\right)  \overset{\text{def}}{=}\frac
{1}{\left(  \dim_{%
%TCIMACRO{\U{2102} }%
%BeginExpansion
\mathbb{C}
%EndExpansion
}\mathcal{C}\right)  ^{2}}\sum_{k=0}^{2^{14}-1}\sum_{l=0}^{22}\left\vert
\text{Tr}\left(  R_{l}A_{k}^{\prime}\right)  _{\left\vert \mathcal{C}\right.
}\right\vert ^{2}\approx1-\frac{55}{4}\gamma^{2}-55\varepsilon^{2}\left(
1+\frac{\gamma}{\varepsilon}\right)  +\mathcal{O}\left(  3\right)  \text{,}
\label{estimate6}%
\end{equation}
with,%
\begin{equation}
\mathcal{F}^{\left(  \left(  11,2,3\right)  \right)  }\left(  \gamma\text{,
}\varepsilon\right)  \overset{\text{def}}{=}\frac{1}{\left(  \dim_{%
%TCIMACRO{\U{2102} }%
%BeginExpansion
\mathbb{C}
%EndExpansion
}\mathcal{C}\right)  ^{2}}\sum_{k=0}^{2^{14}-1}\sum_{l=0}^{23}\left\vert
\text{Tr}\left(  R_{l}A_{k}^{\prime}\right)  _{\left\vert \mathcal{C}\right.
}\right\vert ^{2}\text{.}%
\end{equation}
From Eq. (\ref{estimate6}), we conclude that the performance of this strongly
nonadditive code whose encoding rate equals $1/11$ is not especially good for
GAD errors. After all, this code was originally introduced in \cite{vatan} for
conceptual reasons without any claim about its error-correcting capabilities
against any specific noise model. This code lacks two essential features: high
encoding rate and self-complementarity \cite{smolin}. The nonadditive code we
consider next, although not self-complementary, has a very high encoding rate.

\subsubsection{The $\left(  \left(  9,12,3\right)  \right)  $ code}

We consider next the nondegenerate $\left(  \left(  9,12,3\right)  \right)  $
code \cite{yu}, the first example of a $1$-error correcting nonadditive code
capable of outperforming (in terms of encoded dimension) the optimal
stabilizer code with the same length, namely, the $\left[  \left[
9,3,3\right]  \right]  $ code. This is trivially a nonadditive code, since it
encodes a fractional number of qubits, $k=\log_{2}12\approx3.6$. This code was
constructed by means of the graph-state formalism \cite{hein}. Therefore, in
order to justify the structure of the codewords spanning this $K=12$%
-dimensional codespace of this code, a subspace of the $2^{9}$-dimensional
\emph{complex} Hilbert space $\mathcal{H}_{2}^{9}$, we introduce first the
basic ingredients of the graph-state formalism and we refer to \cite{hein} and
\cite{werner} for more details on this specific point.

The starting point in the graph-state formalism is the notion of graph. An
unidirected simple graph $G\overset{\text{def}}{=}G\left(  V\text{, }%
\Gamma\right)  $ with $n=\left\vert V\right\vert $ vertices is characterized
by the so-called adjacency matrix $\Gamma$. This is a $n\times n$ symmetric
matrix with vanishing diagonal elements such that $\Gamma_{ij}=1$ if vertices
$i$ and $j$ are connected and $\Gamma_{ij}=0$ otherwise. The graph-state
$\left\vert G\right\rangle $ associated with the graph $G$ reads,%
\begin{equation}
\left\vert G\right\rangle \overset{\text{def}}{=}\frac{1}{\sqrt{2^{n}}}%
%TCIMACRO{\dsum \limits_{\vec{\mu}=\mathbf{0}}^{\mathbf{1}}}%
%BeginExpansion
{\displaystyle\sum\limits_{\vec{\mu}=\mathbf{0}}^{\mathbf{1}}}
%EndExpansion
\left(  -1\right)  ^{\frac{1}{2}\vec{\mu}\cdot\Gamma\cdot\vec{\mu}}\left\vert
\vec{\mu}\right\rangle _{z}\text{,}%
\end{equation}
where $\left\vert \vec{\mu}\right\rangle _{z}$ are the simultaneous
eigenstates of $\left\{  Z_{j}\right\}  _{j\in V}$ with $\left(  -1\right)
^{\mu_{j}}$ as eigenvalues and $Z_{j}$ the Pauli operator acting on qubit
$j\in V$.

The $\left(  \left(  9,12,3\right)  \right)  $ code is associated with the
so-called loop graph $L_{9}$ whose $9\times9$ adjacency matrix reads,%
\begin{equation}
\Gamma_{L_{9}}\overset{\text{def}}{=}\left(
\begin{array}
[c]{ccccccccc}%
0 & 1 & 0 & 0 & 0 & 0 & 0 & 0 & 1\\
1 & 0 & 1 & 0 & 0 & 0 & 0 & 0 & 0\\
0 & 1 & 0 & 1 & 0 & 0 & 0 & 0 & 0\\
0 & 0 & 1 & 0 & 1 & 0 & 0 & 0 & 0\\
0 & 0 & 0 & 1 & 0 & 1 & 0 & 0 & 0\\
0 & 0 & 0 & 0 & 1 & 0 & 1 & 0 & 0\\
0 & 0 & 0 & 0 & 0 & 1 & 0 & 1 & 0\\
0 & 0 & 0 & 0 & 0 & 0 & 1 & 0 & 1\\
1 & 0 & 0 & 0 & 0 & 0 & 0 & 1 & 0
\end{array}
\right)  \text{,}%
\end{equation}
and its corresponding graph-state is denoted as $\left\vert L_{9}\right\rangle
$. In terms of $\left\vert L_{9}\right\rangle $, the codespace of the code is
spanned by the following states%
\begin{equation}
\left\vert i_{L}\right\rangle \overset{\text{def}}{=}Z_{V_{i}}\left\vert
L_{9}\right\rangle \text{,} \label{codewords1}%
\end{equation}
where $i=1$,..., $12$ and,%
\begin{equation}
Z_{V_{i}}\overset{\text{def}}{=}%
%TCIMACRO{\dprod \limits_{a\in V_{i}}}%
%BeginExpansion
{\displaystyle\prod\limits_{a\in V_{i}}}
%EndExpansion
Z_{a}\text{,}%
\end{equation}
with the set of vertices $V_{i}$ defined as,%
\begin{align}
&  V_{1}\overset{\text{def}}{=}\left\{  \varnothing\right\}  \text{, }%
V_{2}\overset{\text{def}}{=}\left\{  2\text{, }6\text{, }7\right\}  \text{,
}V_{3}\overset{\text{def}}{=}\left\{  4\text{, }5\text{, }9\right\}  \text{,
}V_{4}\overset{\text{def}}{=}\left\{  2\text{, }3\text{, }6\text{, }8\right\}
\text{, }V_{5}\overset{\text{def}}{=}\left\{  3\text{, }5\text{, }8\text{,
}9\right\}  \text{, }V_{6}\overset{\text{def}}{=}\left\{  2\text{, }3\text{,
}4\text{, }5\text{, }6\text{, }7\text{, }8\text{, }9\right\}  \text{,}%
\nonumber\\
& \nonumber\\
&  V_{7}\overset{\text{def}}{=}\left\{  1\text{, }4\text{, }7\right\}  \text{,
}V_{8}\overset{\text{def}}{=}\left\{  1\text{, }2\text{, }4\text{, }6\right\}
\text{, }V_{9}\overset{\text{def}}{=}\left\{  1\text{, }5\text{, }7\text{,
}9\right\}  \text{, }V_{10}\overset{\text{def}}{=}\left\{  1\text{, }2\text{,
}3\text{, }4\text{, }6\text{, }7\text{, }8\right\}  \text{, }V_{11}%
\overset{\text{def}}{=}\left\{  1\text{, }3\text{, }4\text{, }5\text{,
}7\text{, }8\text{, }9\right\}  \text{,}\nonumber\\
& \nonumber\\
&  V_{12}\overset{\text{def}}{=}\left\{  1\text{, }2\text{, }3\text{,
}5\text{, }6\text{, }8\text{, }9\right\}  \text{.}%
\end{align}
To be explicit, the $12$ codewords are given by%
\begin{align}
&  \left\vert 1_{L}\right\rangle \overset{\text{def}}{=}\left\vert
L_{9}\right\rangle \text{, }\left\vert 2_{L}\right\rangle \overset{\text{def}%
}{=}Z^{2}Z^{6}Z^{7}\left\vert L_{9}\right\rangle \text{, }\left\vert
3_{L}\right\rangle \overset{\text{def}}{=}Z^{4}Z^{5}Z^{9}\left\vert
L_{9}\right\rangle \text{, }\left\vert 4_{L}\right\rangle \overset{\text{def}%
}{=}Z^{2}Z^{3}Z^{6}Z^{8}\left\vert L_{9}\right\rangle \text{, }\left\vert
5_{L}\right\rangle \overset{\text{def}}{=}Z^{3}Z^{5}Z^{8}Z^{9}\left\vert
L_{9}\right\rangle \text{,}\nonumber\\
& \nonumber\\
&  \left\vert 6_{L}\right\rangle \overset{\text{def}}{=}Z^{2}Z^{3}Z^{4}%
Z^{5}Z^{6}Z^{7}Z^{8}Z^{9}\left\vert L_{9}\right\rangle \text{, }\left\vert
7_{L}\right\rangle \overset{\text{def}}{=}Z^{1}Z^{4}Z^{7}\left\vert
L_{9}\right\rangle \text{, }\left\vert 8_{L}\right\rangle \overset{\text{def}%
}{=}Z^{1}Z^{2}Z^{4}Z^{6}\left\vert L_{9}\right\rangle \text{, }\left\vert
9_{L}\right\rangle \overset{\text{def}}{=}Z^{1}Z^{5}Z^{7}Z^{9}\left\vert
L_{9}\right\rangle \text{,}\nonumber\\
& \nonumber\\
&  \left\vert 10_{L}\right\rangle \overset{\text{def}}{=}Z^{1}Z^{2}Z^{3}%
Z^{4}Z^{6}Z^{7}Z^{8}\left\vert L_{9}\right\rangle \text{, }\left\vert
11_{L}\right\rangle \overset{\text{def}}{=}Z^{1}Z^{3}Z^{4}Z^{5}Z^{7}Z^{8}%
Z^{9}\left\vert L_{9}\right\rangle \text{, }\left\vert 12_{L}\right\rangle
\overset{\text{def}}{=}Z^{1}Z^{2}Z^{3}Z^{5}Z^{6}Z^{8}Z^{9}\left\vert
L_{9}\right\rangle \text{.}%
\end{align}
We stress that each codeword is the sum of $512$ state vectors,%
\begin{equation}%
%TCIMACRO{\dsum \limits_{k=0}^{9}}%
%BeginExpansion
{\displaystyle\sum\limits_{k=0}^{9}}
%EndExpansion
\binom{9}{k}=2^{9}=512\text{,}\nonumber
\end{equation}
where $\binom{9}{k}$ denotes the number of state vectors of length $9$ in this
sum with $k$-$1$s in their definition. For each codeword, the sign
distribution of these $512$ state vectors changes according to the action of
$Z_{V_{k}}$ with $k\in\left\{  1\text{,..., }12\right\}  $.

The enlarged GAD quantum channel after performing the encoding defined by
means of Eq. (\ref{codewords1}) reads,%
\begin{equation}
\Lambda_{\text{GAD}}^{\left(  \left(  9,12,3\right)  \right)  }\left(
\rho\right)  \overset{\text{def}}{=}%
%TCIMACRO{\dsum \limits_{r=0}^{2^{18}-1}}%
%BeginExpansion
{\displaystyle\sum\limits_{r=0}^{2^{18}-1}}
%EndExpansion
A_{r}^{\prime}\rho A_{r}^{\prime\dagger}=%
%TCIMACRO{\dsum \limits_{a_{1}\text{,..., }a_{9}=0}^{3}}%
%BeginExpansion
{\displaystyle\sum\limits_{a_{1}\text{,..., }a_{9}=0}^{3}}
%EndExpansion
A_{a_{1}a_{2}\text{...}a_{8}a_{9}}\rho A_{a_{1}a_{2}\text{...}a_{8}a_{9}%
}^{\dagger}\text{,}%
\end{equation}
where to any of the $2^{18}$ values of $r$ we can associate a set of indices
$\left(  a_{1}\text{,..., }a_{9}\right)  $ (and vice-versa) such that,%
\begin{equation}
A_{r}^{\prime}\leftrightarrow A_{a_{1}a_{2}\text{...}a_{8}a_{9}}%
\overset{\text{def}}{=}A_{a_{1}}\otimes A_{a_{2}}\otimes\text{...}\otimes
A_{a_{8}}\otimes A_{a_{9}}\equiv A_{a_{1}}A_{a_{2}}\text{...}A_{a_{8}}%
A_{a_{9}}\text{.}%
\end{equation}
The errors $A_{i}$ with $i\in\left\{  0\text{, }1\text{, }2\text{, }3\right\}
$ are defined in Eq. (\ref{ko}) and $\rho\in\mathcal{M}\left(  \mathcal{C}%
\right)  $ with $\mathcal{C}\subset\mathcal{H}_{2}^{9}$. In particular, the
number of weight-$q$ enlarged error operators $A_{r}^{\prime}$ is given by
$3^{q}\binom{9}{q}$ and,%
\begin{equation}%
%TCIMACRO{\dsum \limits_{q=0}^{9}}%
%BeginExpansion
{\displaystyle\sum\limits_{q=0}^{9}}
%EndExpansion
3^{q}\binom{9}{q}=2^{18}\text{.}%
\end{equation}
Before describing our recovery scheme, let us make two remarks. First, the
codespace of this nonadditive code is a $12$-dimensional subspace of
$\mathcal{H}_{2}^{9}$ spanned by twelve codewords. Each codeword \ is the
sum-decomposition of $512$ vector states in $\mathcal{H}_{2}^{9}$. The number
of vector states in such decomposition with $m$ non-zero components is given
by $\binom{9}{m}$. This binomial factor can be regarded as the cardinality of
vector states of Hamming weight $m$ that appear in the sum-decomposition of
the codewords. The normalization condition requires,%
\begin{equation}
2^{9}=%
%TCIMACRO{\dsum \limits_{q=0}^{9}}%
%BeginExpansion
{\displaystyle\sum\limits_{q=0}^{9}}
%EndExpansion
\binom{9}{q}=2^{9}=512\text{.}%
\end{equation}
Second, after some thinking, it can be shown that the action of any weight-$1$
enlarged error operator (where single-qubit errors of type $A_{1}$ or $A_{3}$
may occur) on each of these codewords with the above-mentioned structure leads
to quantum states in $\mathcal{H}_{2}^{9}$ described in terms of a
sum-decomposition of $162$ vector states which give rise to the following
Hamming weight distribution: $1$ vector with Hamming weight $m=1$, $8$ vectors
with $m=2$, $28$ vectors with $m=3$, $56$ vectors with $m=4$, $35$ vectors
with $m=5$, $20$ vectors with $m=6$, $10$ vectors with $m=7$, $3$ vectors with
$m=8$ and, finally, $1$ vector with $m=9$.

As stated earlier, we assume to focus on the recovery of weight-$1$ errors
only and pay no attention to the possible recovery of higher-order enlarged
errors. In this case, the recovery scheme $\mathcal{R}$ that we use can be
described as follows: $R_{0}$ is associated with the weight-$0$ error; $R_{k}$
with $k=1$, ..., $9$ are associated with the eleven weight-$1$ errors where
single-qubit errors of type $A_{1}$ occur; finally, $R_{k}$ with $k=10$, ...,
$18$ are associated with the eleven weight-$1$ errors where single-qubit
errors of type $A_{3}$ occur. The construction of these $19$ recovery
operators $R_{k}$ is described in terms of $19\times12=228$ orthonormal
vectors in $\mathcal{H}_{2}^{9}$. The missing orthonormal vectors needed to
obtain an orthonormal basis of $\mathcal{H}_{2}^{9}$ and to construct
$R_{19}\overset{\text{def}}{=}\hat{O}$ can be formally computed by using the
rank-nullity theorem together with the Gram-Schmidt orthonormalization
procedure. Finally, our analytical estimate of the entanglement fidelity of
the $\left(  \left(  9,12,3\right)  \right)  $ code reads,%
\begin{equation}
\mathcal{F}_{\text{first-order}}^{\left(  \left(  9,12,3\right)  \right)
}\left(  \gamma\text{, }\varepsilon\right)  \overset{\text{def}}{=}\frac
{1}{\left(  \dim_{%
%TCIMACRO{\U{2102} }%
%BeginExpansion
\mathbb{C}
%EndExpansion
}\mathcal{C}\right)  ^{2}}\sum_{k=0}^{2^{18}-1}\sum_{l=0}^{18}\left\vert
\text{Tr}\left(  R_{l}A_{k}^{\prime}\right)  _{\left\vert \mathcal{C}\right.
}\right\vert ^{2}\approx1-9\gamma^{2}-36\varepsilon^{2}\left(  1+\frac{\gamma
}{\varepsilon}\right)  +\mathcal{O}\left(  3\right)  \text{,}
\label{estimate7}%
\end{equation}
with,%
\begin{equation}
\mathcal{F}^{\left(  \left(  9,12,3\right)  \right)  }\left(  \gamma\text{,
}\varepsilon\right)  \overset{\text{def}}{=}\frac{1}{\left(  \dim_{%
%TCIMACRO{\U{2102} }%
%BeginExpansion
\mathbb{C}
%EndExpansion
}\mathcal{C}\right)  ^{2}}\sum_{k=0}^{2^{18}-1}\sum_{l=0}^{19}\left\vert
\text{Tr}\left(  R_{l}A_{k}^{\prime}\right)  _{\left\vert \mathcal{C}\right.
}\right\vert ^{2}\text{.}%
\end{equation}
From Eqs. (\ref{estimate6}) and (\ref{estimate7}), we conclude that the
nondegenerate $\left(  \left(  9,12,3\right)  \right)  $ code not only
outperforms the $\left(  \left(  11,2,3\right)  \right)  $ in terms of encoded
dimension but also in terms of entanglement fidelity with recovery schemes
limited to first-order recovery. Furthermore from Eq. (\ref{estimate5}), it
can be shown that
\begin{equation}
\mathcal{F}_{\text{first-order}}^{\left[  \left[  9,1,3\right]  \right]
}\left(  \gamma\text{, }\varepsilon=0\right)  \approx1-\frac{45}{4}\gamma
^{2}+\mathcal{O}\left(  \gamma^{3}\right)  \lesssim1-\frac{9}{\log_{2}%
12}\gamma^{2}+\mathcal{O}\left(  \gamma^{3}\right)  \approx\mathcal{\tilde{F}%
}_{\text{first-order}}^{\left(  \left(  9,12,3\right)  \right)  }\left(
\gamma\text{, }\varepsilon=0\right)  \text{,} \label{cumpa-fuck}%
\end{equation}
where\textbf{ }$\mathcal{\tilde{F}}$\textbf{ }denotes the entanglement
fidelity normalized with exponentiation by\textbf{ }$1/k$\textbf{
}where\textbf{ }$k\overset{\text{def}}{=}\log_{2}K$\textbf{ }is the number of
encoded\textbf{ }logical qubits\textbf{ (}$k=1$\textbf{ }for the Shor
nine-qubit code\textbf{) }\cite{andy1}\textbf{. }From the comparison of the
first-order entanglement-based performances of these two codes in Eq.
(\ref{cumpa-fuck}), we are lead to the conclusion that the nondegenerate and
nonadditive code\textbf{ }$\left(  \left(  9,12,3\right)  \right)  $\textbf{
}outperforms the degenerate and additive code\textbf{ }$\left[  \left[
9,1,3\right]  \right]  $ not only in terms of encoded dimension. The
comparison between these two codes is shown in Fig.\textbf{ }$4$\textbf{.}

\begin{figure}[ptb]
\centering
\includegraphics[width=0.5\textwidth]{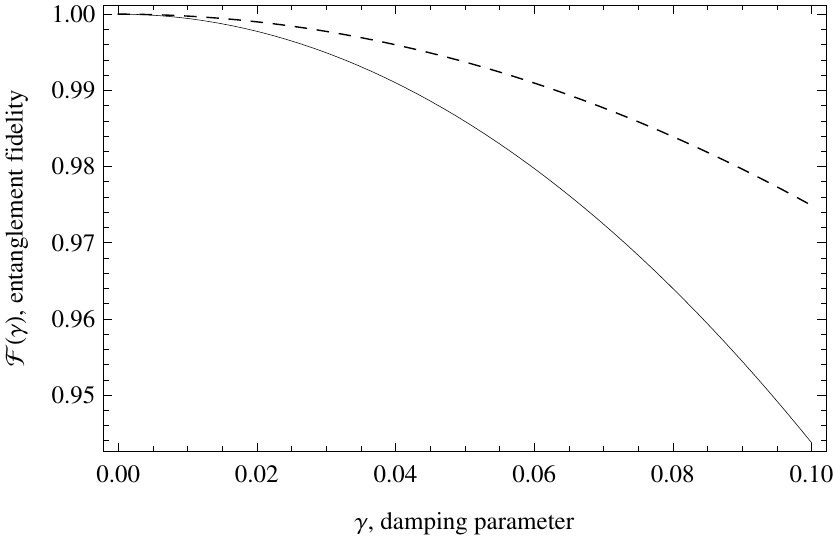}\caption{\textit{Additive vs.
nonadditive codes of length nine}. The truncated series expansions of the
normalized entanglement fidelity $\mathcal{F}\left(  \gamma\right)  $ vs. the
amplitude damping parameter $\gamma$ for $\varepsilon=0$ and $0\leq\gamma
\leq10^{-1}$: the Shor nine-qubit code (thin solid line) and the $\left(
\left(  9,12,3\right)  \right)  $-code (dashed line).}%
\label{figure4}%
\end{figure}

\subsection{Self-complementary codes}

In what follows, we consider single-AD error correcting codes. For this
reason, we set $\varepsilon=0$ and limit our considerations to the approximate
QEC of AD errors by means of self-complementary nonadditive quantum codes. An
$\left(  \left(  n,K,d\right)  \right)  $ code is called self-complementary if
its codespace is spanned by codewords $\left\{  \left\vert c_{a}\right\rangle
\right\}  $ defined as \cite{smolin},%
\begin{equation}
\left\vert c_{a}\right\rangle \overset{\text{def}}{=}\frac{\left\vert
a\right\rangle +\left\vert \bar{a}\right\rangle }{\sqrt{2}}\text{,}%
\end{equation}
where $a$ is a binary string of length $n$ and $\bar{a}\overset{\text{def}}%
{=}\mathbf{1\oplus}a$ is the complement of $a$. The suitability of
self-complemetary nonadditive codes for the error correction of AD errors was
first proposed in \cite{lang-shor, shor-2011}. However, no explicit comparison
of the performances quantified by means of analytical estimates of the
entanglement fidelity of self-complementary and non-self-complementary
nonadditive codes for AD errors was presented. Furthermore, no similar
comparison between nonadditive self-complementary and stabilizer codes for AD
errors exists as well. In view of these considerations, we seek to provide
some useful insights into these unexplored issues.

\subsubsection{The $\left(  \left(  6,5\right)  \right)  $ code}

The first self-complemetary single-AD error-correcting code that we consider
is the $\left(  \left(  6,5\right)  \right)  $ code whose codespace is spanned
by the following five codewords in $\mathcal{H}_{2}^{6}$,%
\begin{align}
&  \left\vert 0_{L}\right\rangle \overset{\text{def}}{=}\frac{\left\vert
000000\right\rangle +\left\vert 111111\right\rangle }{\sqrt{2}}\text{,
}\left\vert 1_{L}\right\rangle \overset{\text{def}}{=}\frac{\left\vert
110000\right\rangle +\left\vert 001111\right\rangle }{\sqrt{2}}\text{,
}\left\vert 2_{L}\right\rangle \overset{\text{def}}{=}\frac{\left\vert
001100\right\rangle +\left\vert 110011\right\rangle }{\sqrt{2}}\text{,}%
\nonumber\\
& \nonumber\\
&  \left\vert 3_{L}\right\rangle \overset{\text{def}}{=}\frac{\left\vert
000011\right\rangle +\left\vert 111100\right\rangle }{\sqrt{2}}\text{,
}\left\vert 4_{L}\right\rangle \overset{\text{def}}{=}\frac{\left\vert
010101\right\rangle +\left\vert 101010\right\rangle }{\sqrt{2}}\text{.}
\label{65}%
\end{align}
We point out that the existence of such a code was originally proposed in
\cite{lang-shor}, although it was neither explicitly shown nor used for error
correction of single-AD errors.

The enlarged AD quantum channel after performing the encoding defined by means
of Eq. (\ref{65}) reads,%
\begin{equation}
\Lambda_{\text{AD}}^{\left(  \left(  6,5\right)  \right)  }\left(
\rho\right)  \overset{\text{def}}{=}%
%TCIMACRO{\dsum \limits_{r=0}^{2^{6}-1}}%
%BeginExpansion
{\displaystyle\sum\limits_{r=0}^{2^{6}-1}}
%EndExpansion
A_{r}^{\prime}\rho A_{r}^{\prime\dagger}=%
%TCIMACRO{\dsum \limits_{a_{1}\text{,..., }a_{6}=0}^{1}}%
%BeginExpansion
{\displaystyle\sum\limits_{a_{1}\text{,..., }a_{6}=0}^{1}}
%EndExpansion
A_{a_{1}a_{2}\text{...}a_{5}a_{6}}\rho A_{a_{1}a_{2}\text{...}a_{5}a_{6}%
}^{\dagger}\text{,}%
\end{equation}
where to any of the $2^{6}$ values of $r$ we can associate a set of indices
$\left(  a_{1}\text{,..., }a_{6}\right)  $ (and vice-versa) such that,%
\begin{equation}
A_{r}^{\prime}\leftrightarrow A_{a_{1}a_{2}\text{...}a_{5}a_{6}}%
\overset{\text{def}}{=}A_{a_{1}}\otimes A_{a_{2}}\otimes\text{...}\otimes
A_{a_{5}}\otimes A_{a_{6}}\equiv A_{a_{1}}A_{a_{2}}\text{...}A_{a_{5}}%
A_{a_{6}}\text{.}%
\end{equation}
The errors $A_{i}$ with $i\in\left\{  0\text{, }1\right\}  $ are defined in
Eq. (\ref{ko}) and $\rho\in\mathcal{M}\left(  \mathcal{C}\right)  $ with
$\mathcal{C}\subset\mathcal{H}_{2}^{6}$. In particular, the number of
weight-$q$ enlarged error operators $A_{r}^{\prime}$ is given by $\binom{6}%
{q}$ and,%
\begin{equation}%
%TCIMACRO{\dsum \limits_{q=0}^{6}}%
%BeginExpansion
{\displaystyle\sum\limits_{q=0}^{6}}
%EndExpansion
\binom{6}{q}=2^{6}\text{.}%
\end{equation}
The recovery scheme $\mathcal{R}$ that we use can be described as follows:
$R_{0}$ is associated with the weight-$0$ error; $R_{k}$ with $k=1$, ..., $6$
are associated with the six weight-$1$ errors where single-qubit errors of
type $A_{1}$ occur. The construction of these $7$ recovery operators $R_{k}$
is described in terms of $7\times5=35$ orthonormal vectors in $\mathcal{H}%
_{2}^{6}$. The missing orthonormal vectors needed to obtain an orthonormal
basis of $\mathcal{H}_{2}^{6}$ and to construct $R_{7}\overset{\text{def}}%
{=}\hat{O}$ can be formally computed by using the rank-nullity theorem
together with the Gram-Schmidt orthonormalization procedure. Finally, our
analytical estimate of the entanglement fidelity of the $\left(  \left(
6,5\right)  \right)  $ code reads,%
\begin{equation}
\mathcal{F}_{\text{first-order}}^{\left(  \left(  6,5\right)  \right)
}\left(  \gamma\right)  \overset{\text{def}}{=}\frac{1}{\left(  \dim_{%
%TCIMACRO{\U{2102} }%
%BeginExpansion
\mathbb{C}
%EndExpansion
}\mathcal{C}\right)  ^{2}}\sum_{k=0}^{2^{6}-1}\sum_{l=0}^{6}\left\vert
\text{Tr}\left(  R_{l}A_{k}^{\prime}\right)  _{\left\vert \mathcal{C}\right.
}\right\vert ^{2}\approx1-\frac{21}{5}\gamma^{2}+\mathcal{O}\left(  \gamma
^{3}\right)  \text{,} \label{estimate8}%
\end{equation}
with,%
\begin{equation}
\mathcal{F}^{\left(  \left(  6,5\right)  \right)  }\left(  \gamma\right)
\overset{\text{def}}{=}\frac{1}{\left(  \dim_{%
%TCIMACRO{\U{2102} }%
%BeginExpansion
\mathbb{C}
%EndExpansion
}\mathcal{C}\right)  ^{2}}\sum_{k=0}^{2^{6}-1}\sum_{l=0}^{7}\left\vert
\text{Tr}\left(  R_{l}A_{k}^{\prime}\right)  _{\left\vert \mathcal{C}\right.
}\right\vert ^{2}\text{.}%
\end{equation}
From Eqs. (\ref{estimate4}) and (\ref{estimate8}), it follows that
\begin{equation}
\mathcal{\tilde{F}}_{\text{first-order}}^{\left(  \left(  6,5\right)  \right)
}\left(  \gamma\right)  \approx1-\frac{21}{5\log_{2}5}\gamma^{2}%
+\mathcal{O}\left(  \gamma^{3}\right)  \geq1-2\gamma^{2}+\mathcal{O}\left(
\gamma^{3}\right)  \approx\mathcal{F}^{\left[  \left[  6,1,3\right]  \right]
}\left(  \gamma\text{, }\varepsilon=0\right)  \geq\mathcal{F}%
_{\text{first-order}}^{\left[  \left[  6,1,3\right]  \right]  }\left(
\gamma\text{, }\varepsilon=0\right)  \text{.} \label{cumpa-fuck2}%
\end{equation}
Eq. (\ref{cumpa-fuck2}) is an explicit manifestation of the superiority, in
terms of both encoded dimension and entanglement fidelity, of nonadditive over
additive codes.

\subsubsection{The $\left(  \left(  8,12\right)  \right)  $ code}

The second self-complemetary single-AD error-correcting code that we consider
is the $\left(  \left(  8,12\right)  \right)  $ code whose codespace is
spanned by the following twelve codewords in $\mathcal{H}_{2}^{8}$
\cite{lang-shor},%
\begin{align}
&  \left\vert 0_{L}\right\rangle \overset{\text{def}}{=}\frac{\left\vert
00000000\right\rangle +\left\vert 11111111\right\rangle }{\sqrt{2}}\text{,
}\left\vert 1_{L}\right\rangle \overset{\text{def}}{=}\frac{\left\vert
00000011\right\rangle +\left\vert 11111100\right\rangle }{\sqrt{2}}\text{,
}\left\vert 2_{L}\right\rangle \overset{\text{def}}{=}\frac{\left\vert
00001100\right\rangle +\left\vert 11110011\right\rangle }{\sqrt{2}}%
\text{,}\nonumber\\
& \nonumber\\
&  \left\vert 3_{L}\right\rangle \overset{\text{def}}{=}\frac{\left\vert
00110000\right\rangle +\left\vert 11001111\right\rangle }{\sqrt{2}}\text{,
}\left\vert 4_{L}\right\rangle \overset{\text{def}}{=}\frac{\left\vert
11000000\right\rangle +\left\vert 00111111\right\rangle }{\sqrt{2}}\text{,
}\left\vert 5_{L}\right\rangle \overset{\text{def}}{=}\frac{\left\vert
10101000\right\rangle +\left\vert 01010111\right\rangle }{\sqrt{2}}%
\text{,}\nonumber\\
& \nonumber\\
&  \left\vert 6_{L}\right\rangle \overset{\text{def}}{=}\frac{\left\vert
01011000\right\rangle +\left\vert 10100111\right\rangle }{\sqrt{2}}\text{,
}\left\vert 7_{L}\right\rangle \overset{\text{def}}{=}\frac{\left\vert
01100100\right\rangle +\left\vert 10011011\right\rangle }{\sqrt{2}}\text{,
}\left\vert 8_{L}\right\rangle \overset{\text{def}}{=}\frac{\left\vert
10010100\right\rangle +\left\vert 01101011\right\rangle }{\sqrt{2}}%
\text{,}\nonumber\\
& \nonumber\\
&  \left\vert 9_{L}\right\rangle \overset{\text{def}}{=}\frac{\left\vert
11110000\right\rangle +\left\vert 00001111\right\rangle }{\sqrt{2}}\text{,
}\left\vert 10_{L}\right\rangle \overset{\text{def}}{=}\frac{\left\vert
11001100\right\rangle +\left\vert 00110011\right\rangle }{\sqrt{2}}\text{,
}\left\vert 11_{L}\right\rangle \overset{\text{def}}{=}\frac{\left\vert
00111100\right\rangle +\left\vert 11000011\right\rangle }{\sqrt{2}}\text{.}
\label{812}%
\end{align}
The enlarged AD quantum channel after performing the encoding defined by means
of Eq. (\ref{812}) reads,%
\begin{equation}
\Lambda_{\text{AD}}^{\left(  \left(  8,12\right)  \right)  }\left(
\rho\right)  \overset{\text{def}}{=}%
%TCIMACRO{\dsum \limits_{r=0}^{2^{8}-1}}%
%BeginExpansion
{\displaystyle\sum\limits_{r=0}^{2^{8}-1}}
%EndExpansion
A_{r}^{\prime}\rho A_{r}^{\prime\dagger}=%
%TCIMACRO{\dsum \limits_{a_{1}\text{,..., }a_{8}=0}^{1}}%
%BeginExpansion
{\displaystyle\sum\limits_{a_{1}\text{,..., }a_{8}=0}^{1}}
%EndExpansion
A_{a_{1}a_{2}\text{...}a_{7}a_{8}}\rho A_{a_{1}a_{2}\text{...}a_{7}a_{8}%
}^{\dagger}\text{,}%
\end{equation}
where to any of the $2^{8}$ values of $r$ we can associate a set of indices
$\left(  a_{1}\text{,..., }a_{8}\right)  $ (and vice-versa) such that,%
\begin{equation}
A_{r}^{\prime}\leftrightarrow A_{a_{1}a_{2}\text{...}a_{7}a_{8}}%
\overset{\text{def}}{=}A_{a_{1}}\otimes A_{a_{2}}\otimes\text{...}\otimes
A_{a_{7}}\otimes A_{a_{8}}\equiv A_{a_{1}}A_{a_{2}}\text{...}A_{a_{7}}%
A_{a_{8}}\text{.}%
\end{equation}
The errors $A_{i}$ with $i\in\left\{  0\text{, }1\right\}  $ are defined in
Eq. (\ref{ko}) and $\rho\in\mathcal{M}\left(  \mathcal{C}\right)  $ with
$\mathcal{C}\subset\mathcal{H}_{2}^{8}$. In particular, the number of
weight-$q$ enlarged error operators $A_{r}^{\prime}$ is given by $\binom{8}%
{q}$ and,%
\begin{equation}%
%TCIMACRO{\dsum \limits_{q=0}^{8}}%
%BeginExpansion
{\displaystyle\sum\limits_{q=0}^{8}}
%EndExpansion
\binom{8}{q}=2^{8}\text{.}%
\end{equation}
The recovery scheme $\mathcal{R}$ that we use can be described as follows:
$R_{0}$ is associated with the weight-$0$ error; $R_{k}$ with $k=1$, ..., $8$
are associated with the eight weight-$1$ errors where single-qubit errors of
type $A_{1}$ occur. The construction of these $9$ recovery operators $R_{k}$
is described in terms of $9\times12=108$ orthonormal vectors in $\mathcal{H}%
_{2}^{8}$. The missing orthonormal vectors needed to obtain an orthonormal
basis of $\mathcal{H}_{2}^{8}$ and to construct $R_{9}\overset{\text{def}}%
{=}\hat{O}$ can be formally computed by using the rank-nullity theorem
together with the Gram-Schmidt orthonormalization procedure. Finally, our
analytical estimate of the entanglement fidelity of the $\left(  \left(
8,12\right)  \right)  $ code reads,%
\begin{equation}
\mathcal{F}_{\text{first-order}}^{\left(  \left(  8,12\right)  \right)
}\left(  \gamma\right)  \overset{\text{def}}{=}\frac{1}{\left(  \dim_{%
%TCIMACRO{\U{2102} }%
%BeginExpansion
\mathbb{C}
%EndExpansion
}\mathcal{C}\right)  ^{2}}\sum_{k=0}^{2^{8}-1}\sum_{l=0}^{8}\left\vert
\text{Tr}\left(  R_{l}A_{k}^{\prime}\right)  _{\left\vert \mathcal{C}\right.
}\right\vert ^{2}\approx1-\frac{15}{2}\gamma^{2}+\mathcal{O}\left(  \gamma
^{3}\right)  \text{,} \label{estimate9}%
\end{equation}
with,%
\begin{equation}
\mathcal{F}^{\left(  \left(  8,12\right)  \right)  }\left(  \gamma\right)
\overset{\text{def}}{=}\frac{1}{\left(  \dim_{%
%TCIMACRO{\U{2102} }%
%BeginExpansion
\mathbb{C}
%EndExpansion
}\mathcal{C}\right)  ^{2}}\sum_{k=0}^{2^{8}-1}\sum_{l=0}^{9}\left\vert
\text{Tr}\left(  R_{l}A_{k}^{\prime}\right)  _{\left\vert \mathcal{C}\right.
}\right\vert ^{2}\text{.}%
\end{equation}
It is convenient to compare the performance of this code with that of a
multi-qubit encoding stabilizer code with the same length. For instance, the
$\left[  \left[  8,3,3\right]  \right]  $ code is a special case of a class of
$\left[  \left[  2^{j},2^{j}-j-2,3\right]  \right]  $ codes \cite{danielpra}.
It encodes three logical qubits into eight physical qubits and corrects all
single-qubit errors. The five stabilizer generators are given by
\cite{gaitan},%
\begin{align}
&  g_{1}\overset{\text{def}}{=}X^{1}X^{2}X^{3}X^{4}X^{5}X^{6}X^{7}X^{8}\text{,
}g_{2}\overset{\text{def}}{=}Z^{1}Z^{2}Z^{3}Z^{4}Z^{5}Z^{6}Z^{7}Z^{8}\text{,
}g_{3}\overset{\text{def}}{=}X^{2}X^{4}Y^{5}Z^{6}Y^{7}Z^{8}\text{,
}\nonumber\\
& \nonumber\\
&  g_{4}\overset{\text{def}}{=}X^{2}Z^{3}Y^{4}X^{6}Z^{7}Y^{8}\text{, }%
g_{5}\overset{\text{def}}{=}Y^{2}X^{3}Z^{4}X^{5}Z^{6}Y^{8}\text{,}%
\end{align}
and a suitable choice for the logical operations $\bar{X}_{i}$ and $\bar
{Z}_{i}$ with $i\in\left\{  1\text{, }2\text{, }3\right\}  $ reads,
\begin{equation}
\bar{X}_{1}\overset{\text{def}}{=}X^{1}X^{2}Z^{6}Z^{8}\text{, }\bar{X}%
_{2}\overset{\text{def}}{=}X^{1}X^{3}Z^{4}Z^{7}\text{, }\bar{X}_{3}%
\overset{\text{def}}{=}X^{1}Z^{4}X^{5}Z^{6}\text{, }\bar{Z}_{1}\overset
{\text{def}}{=}Z^{2}Z^{4}Z^{6}Z^{8}\text{, }\bar{Z}_{2}\overset{\text{def}}%
{=}Z^{3}Z^{4}Z^{7}Z^{8}\text{, }\bar{Z}_{3}\overset{\text{def}}{=}Z^{5}%
Z^{6}Z^{7}Z^{8}\text{.}%
\end{equation}
A convenient choice for the basis codewords reads,%
\begin{equation}
\left\vert \overline{ijk}\right\rangle \overset{\text{def}}{=}\left(  \bar
{X}_{1}\right)  ^{\bar{\imath}}\left(  \bar{X}_{2}\right)  ^{\bar{j}}\left(
\bar{X}_{3}\right)  ^{\bar{k}}%
%TCIMACRO{\dsum \limits_{g\in\mathcal{S}}}%
%BeginExpansion
{\displaystyle\sum\limits_{g\in\mathcal{S}}}
%EndExpansion
g\left\vert 00000000\right\rangle \text{.}%
\end{equation}
For an explicit representations of the codewords of the $\left[  \left[
8,3,3\right]  \right]  $ code, we refer to \cite{danielpra} (specifically, see
Table III in \cite{danielpra}). Following our line of reasoning presented for
the error correction of AD errors by means of stabilizer codes and omitting
technical details, it turns out that the entanglement fidelity of the $\left[
\left[  8,3,3\right]  \right]  $ with our recovery up to first-order errors
becomes,%
\begin{equation}
\mathcal{F}_{\text{first-order}}^{\left[  \left[  8,3,3\right]  \right]
}\left(  \gamma\right)  \approx1-7\gamma^{2}+\mathcal{O}\left(  \gamma
^{3}\right)  \text{.} \label{estimate10}%
\end{equation}
Unlike the case of single-qubit encoding, in this case the entanglement
fidelity when no QEC is performed is represented by the three-qubit baseline
performance given by,%
\begin{equation}
\mathcal{F}_{\text{baseline}}^{3\text{-qubit}}\left(  \gamma\right)
\overset{\text{def}}{=}8^{-2}\left[  1+3\sqrt{1-\gamma}+3\left(
1-\gamma\right)  +\left(  1-\gamma\right)  ^{\frac{3}{2}}\right]  ^{2}\text{.}%
\end{equation}
In addition, from Eqs. (\ref{estimate9}) and (\ref{estimate10}), we get%
\begin{equation}
\mathcal{\tilde{F}}_{\text{first-order}}^{\left(  \left(  8,12\right)
\right)  }\left(  \gamma\right)  \approx1-\frac{15}{2\log_{2}12}\gamma
^{2}+\mathcal{O}\left(  \gamma^{3}\right)  \geq1-\frac{7}{3}\gamma
^{2}+\mathcal{O}\left(  \gamma^{3}\right)  \approx\mathcal{\tilde{F}%
}_{\text{first-order}}^{\left[  \left[  8,3,3\right]  \right]  }\left(
\gamma\right)  \text{.} \label{cumpa-fuck3}%
\end{equation}
Eq. (\ref{cumpa-fuck3}) is yet another clear fingerprint of the superiority,
in terms of both encoded dimension and entanglement fidelity, of nonadditive
over additive codes. The comparison between these two codes is shown in Fig.
$5$. Finally, we compare the performances of the nonadditive codes employed in
our error correction schemes in Fig.\textbf{ }$6$\textbf{.}

\begin{figure}[ptb]
\centering
\includegraphics[width=0.5\textwidth]{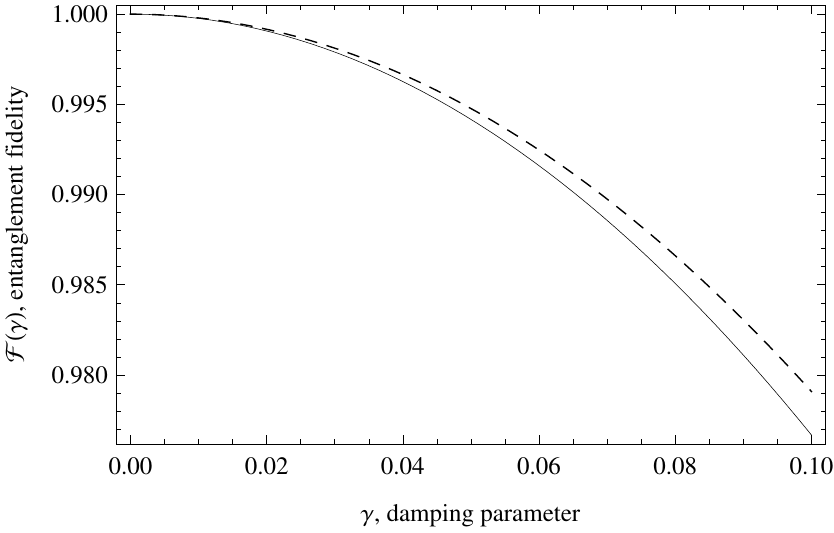}\caption{\textit{Additive
vs. nonadditive codes of length eight}. The truncated series expansions of the
normalized entanglement fidelity $\mathcal{F}\left(  \gamma\right)  $ vs. the
amplitude damping parameter $\gamma$ for $\varepsilon=0$ and $0\leq\gamma
\leq10^{-1}$: Gottesman's $\left[  \left[  8,3,3\right]  \right]  $-code (thin
solid line) and the $\left(  \left(  8,12\right)  \right)  $-code (dashed
line).}%
\label{figure5}%
\end{figure}\begin{figure}[ptb]
\centering
\includegraphics[width=0.5\textwidth]{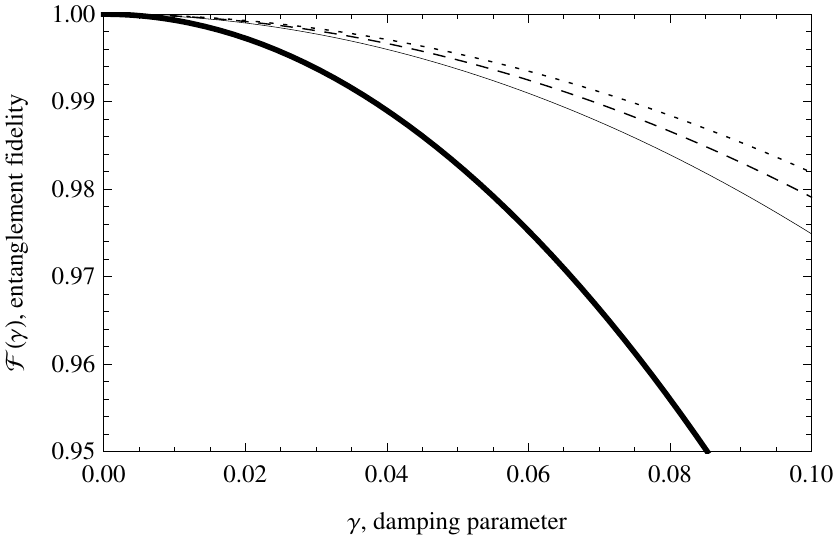}\caption{\textit{Ranking
nonadditive codes}. The truncated series expansions of the normalized
entanglement fidelity $\mathcal{F}\left(  \gamma\right)  $ vs. the amplitude
damping parameter $\gamma$ for $\varepsilon=0$ and $0\leq\gamma\leq10^{-1}$:
the $\left(  \left(  6,5\right)  \right)  $-code (dotted line), the $\left(
\left(  8,12\right)  \right)  $-code (dashed line), the $\left(  \left(
9,12,3\right)  \right)  $-code (thin solid line), and the $\left(  \left(
11,2,3\right)  \right)  $-code (thick solid line).}%
\label{figure6}%
\end{figure}

\section{Final Remarks}

In this article, we presented the first analytic analysis of the performance
of various approximate QEC codes for GAD errors. Specifically, we considered
both stabilizer and nonadditive quantum codes. The performance of such codes
was quantified by means of the entanglement fidelity as a function of the
damping probability and the non-zero environmental temperature. We
analytically recovered and clarified some previously known numerical results
in the limiting case of the AD channel (zero environmental temperature). In
addition, our extended investigation suggested that degenerate stabilizer
codes and self-complementary nonadditive codes are especially suitable for the
error correction of GAD errors. Finally, comparing the properly normalized
entanglement fidelities of the best performant stabilizer and nonadditive
codes characterized by the same length, we showed that, in general,
nonadditive codes outperform stabilizer codes not only in terms of encoded
dimension but also in terms of entanglement fidelity.

\smallskip

Our main findings may be summarized as follows:

\begin{enumerate}
\item We have explicitly shown that in the presence of non-zero environmental
temperature, the performance of both additive and nonadditive quantum codes
decreases (with respect to zero environmental temperature case). In
particular, degenerate stabilizer codes seem to be more robust than the
nondegenerate ones against this effect, as evident from Eqs. (\ref{estimate4})
and (\ref{estimate5});

\item In the limiting case of $\varepsilon=0$ and considering the $\left[
\left[  5,1,3\right]  \right]  $-code, our analytic estimate in Eq.
(\ref{estimate1}) reduces to the numeric ones in \cite{andy2} and \cite{fphd};
in the same limiting case, considering the CSS $\left[  \left[  7,1,3\right]
\right]  $-code, our estimate in Eq. (\ref{estimate2}) reduces to the numeric
one in \cite{andy3}; finally, considering the Shor $\left[  \left[
9,1,3\right]  \right]  $-code, our estimate in Eq. (\ref{estimate5}) reduces
to that numerically obtained in \cite{andy1};

\item We have constructed a nondegenerate eight-qubit concatenated stabilizer
code (see Eq. (\ref{conc})), a natural generalization of the Leung et
\textit{al}. four-qubit code, suitable for the QEC of GAD errors. We have also
checked that in the limit of $\varepsilon=0$, our estimated performance for
such a code reduces to that of the four-qubit code applied to AD errors (see
Eq. (\ref{finding-ok}));

\item We have provided further evidence (see Eq. (\ref{estimate4}) for the
$\left[  \left[  6,1,3\right]  \right]  $-code) in support of the\emph{
suspect} advanced in \cite{andy3} where it was conjectured that thanks to
their degenerate structure, such codes can outperform nondegenerate codes
despite their shorter length;

\item From Eqs. (\ref{estimate6}) and (\ref{estimate7}), we have explicitly
shown that the nonadditive $\left(  \left(  9,12,3\right)  \right)  $-code not
only outperforms the $\left(  \left(  11,2,3\right)  \right)  $-code in terms
of encoded dimension but also in terms of entanglement fidelity with recovery
schemes limited to first-order recovery;

\item We have shown that, to first-order recovery, self-complementary
nonadditive codes can outperform non-self-complementary nonadditive codes
despite exhibiting smaller encoded dimension (see Eqs. (\ref{estimate7}) and
(\ref{estimate8}) for the $\left(  \left(  6,5\right)  \right)  $- and
$\left(  \left(  9,12,3\right)  \right)  $-codes, respectively);

\item From the comparison of the first-order entanglement-based performances
between additive and nonadditive quantum codes with identical lengths (see
Eqs. (\ref{cumpa-fuck}), (\ref{cumpa-fuck2}), and (\ref{cumpa-fuck3})), we
concluded that nonadditive codes outperform, in general, additive codes not
only in terms of encoded dimension. In particular, nonadditivity seems to
matter more than degeneracy in the approximate QEC of both GAD\ and AD errors
(see Eq. (\ref{cumpa-fuck})).
\end{enumerate}

\smallskip

We wish to emphasize three aspects of our analysis:

\begin{itemize}
\item First, despite the great variety of quantum codes employed in this work,
we have limited our attention to qubit channels only. The AD and GAD channels
can, of course, be extended to quantum systems of dimension greater than two,
leading to the so-called qudit channels. For instance, a qutrit is a
three-state quantum system. In such higher-dimensional cases, the physical
processes that we may consider are described by the modeling of atoms as
having more than two states (multilevel atoms) interacting with environments
modeled by a bath of harmonic oscillators which can be initially in the vacuum
state or, more generally, in a thermal state with temperature greater than
zero. The study of the effectiveness of QEC schemes suitable for such
additional realistic scenarios will be the object of our attention in future efforts;

\item Second, the QEC strategy that we have employed for our analytic
estimates may be considered mildly conservative, since we might have slightly
underestimated the quantum codes performances to avoid false over-estimations.
However, we have tried to maintain the same degree of conservativeness in all
our estimates in order to preserve the fairness of the comparisons between
pairs of different quantum codes. After all, exact analytical calculations of
the entanglement fidelities can be quite intimidating. They are conceptually
straightforward but computationally extremely tedious if performed by hands to
avoid the drawbacks of numerical results \cite{ngm, andy3}. We also emphasize
that conservative estimates are not uncommon in QEC since, after all, we are
lead to deal with approximations. For instance, conservativeness appears in
analytic estimates of error thresholds in topological QEC when crude
combinatorics arguments are employed \cite{jimbo} as well as in the numerical
estimation of the error threshold to depolarization of toric codes by means of
Monte Carlo simulations \cite{jimbo2}.

\item Third, despite the variety of quantum codes employed in the approximate
QEC of GAD\ errors presented in this work, we emphasize that we cannot state
without a doubt that for any given arbitrary noise model, nonadditive and
degenerate codes are more efficient, in general, than additive and
nondegenerate codes. In all honesty, we feel uncomfortable in extrapolating a
general statement from a large, but yet limited, number of special cases.
Perhaps, this is the reason why QEC is an art: general statements cannot be
made a priori, each scenario has to be taken into consideration separately. It
is a matter of finding a good matching between the noise model and the quantum
code. Furthermore, we are not able to quantify exactly the effect on the code
performances provided by fractional encoding, a feature of nonadditive codes,
or by the possibility of correcting a number of errors greater than the one
that is uniquely identified, a feature of degenerate codes. Also, we cannot
rule out the possibility that the mathematical structure behind the sign
pattern that characterizes the various codewords spanning the codespaces may
play a relevant role in determining the performance of the error correction
schemes. In conclusion, the merit of achieving higher performances cannot be
ascribed to specific properties of a code, but rather to the properties of a
code \emph{together} with a family of errors it is designed to correct. This
final remark is in excellent agreement with the fact that, for instance,
degeneracy must not be regarded as a property of a quantum code alone, but
rather a property of a code together with a class of errors it is designed to recover.
\end{itemize}

\smallskip

In conclusion, we feel we have truly challenged ourselves in this work from an
analytical stand-point and we tried to advance our understanding of
approximate QEC of GAD errors for various qubit codes, both additive and
nonadditive, as much as we could. We are confident we have succeeded in this
regard. However, in all fairness, it is very likely that quantitative results
in QEC will always require the help of numerical investigations. In future
efforts, we wish to further sharpen our analytic estimations by using, if
possible, both additional analytical and numerical scrutiny.

\begin{acknowledgments}
We thank the ERA-Net CHIST-ERA project HIPERCOM for financial support. CC also
thanks Marcel Bergmann for numerically testing Eq. (\ref{estimate7}) and for
helpful\textbf{ }conversations.
\end{acknowledgments}

\appendix

\section{The Kraus operator-sum decomposition of the GAD\ channel}

In this Appendix, we provide an explicit derivation of the $\left(
2\times2\right)  $-matrix representation of the Kraus error operators for the
GAD channel in Eq. (\ref{ko}).

Let us consider AD from the scattering of a photon via a beam splitter
\cite{nielsen-book}. Consider a single optical mode that contains the quantum
state $\left\vert \psi\right\rangle \overset{\text{def}}{=}\alpha\left\vert
0\right\rangle +\beta\left\vert 1\right\rangle $, where $\left\vert
0\right\rangle $ is the vacuum state while $\left\vert 1\right\rangle $
denotes the single photon state. The scattering of a photon from this mode can
be modeled by inserting a beam splitter in the path of the photon. The beam
splitter acts on two modes: it performs the unitary operation $U\overset
{\text{def}}{=}e^{\chi\left(  a^{\dagger}b-b^{\dagger}a\right)  }$ and allows
the photon to couple to another single optical mode that represents the
environment. The operators\textbf{ }$a$, $a^{\dagger}$ and $b$, $b^{\dagger}$
are the annihilation and creation operators for photons in the two modes.
Assuming the environment starts out with no photons,
\begin{equation}
\rho_{\text{env.}}\overset{\text{def}}{=}\left\vert 0\right\rangle
\left\langle 0\right\vert \text{,}%
\end{equation}
it can be shown that the quantum operation that describes this process reads,%
\begin{equation}
\Lambda_{\text{AD}}\left(  \rho\right)  \overset{\text{def}}{=}A_{0}\rho
A_{0}^{\dagger}+A_{1}\rho A_{1}^{\dagger}\text{,}%
\end{equation}
with\textbf{ }$A_{k}\overset{\text{def}}{=}\left\langle k\left\vert
U\right\vert 0\right\rangle $ and,%
\begin{equation}
A_{0}=\left(
\begin{array}
[c]{cc}%
1 & 0\\
0 & \sqrt{1-\gamma}%
\end{array}
\right)  \text{, }A_{1}=\left(
\begin{array}
[c]{cc}%
0 & \sqrt{\gamma}\\
0 & 0
\end{array}
\right)  \text{,}%
\end{equation}
where $\gamma\overset{\text{def}}{=}\sin^{2}\chi$ denotes the probability of
losing a photon. If we assume that the environment starts out in a linear
superposition of zero and one photons,%
\begin{equation}
\rho_{\text{env.}}\overset{\text{def}}{=}\sum_{j=0}^{1}q_{j}\left\vert
j\right\rangle \left\langle j\right\vert \text{,}%
\end{equation}
where $q_{0}\overset{\text{def}}{=}p$, $q_{1}=1-p$, and $0\leq p\leq1$, it
turns out that the quantum operation that describes this process reads,%
\begin{equation}
\Lambda_{\text{GAD}}\left(  \rho\right)  \overset{\text{def}}{=}A_{00}\rho
A_{00}^{\dagger}+A_{01}\rho A_{01}^{\dagger}+A_{11}\rho A_{11}^{\dagger
}+A_{10}\rho A_{10}^{\dagger}\text{,}%
\end{equation}
with $A_{jk}\overset{\text{def}}{=}\sqrt{q_{j}}\left\langle k\left\vert
U\right\vert j\right\rangle $. Observe that $A_{0k}\overset{\text{def}}%
{=}\sqrt{p}\left\langle k\left\vert U\right\vert 0\right\rangle $ and
$A_{1k}\overset{\text{def}}{=}\sqrt{1-p}\left\langle k\left\vert U\right\vert
1\right\rangle $ with $k=0$, $1$. Let us first focus on $A_{1k}$ which can be
written as,%
\begin{equation}
A_{1k}\overset{\text{def}}{=}\sum_{m\text{, }n}\left(  A_{1k}\right)
_{mn}\left\vert m\right\rangle \left\langle n\right\vert \text{,} \label{a1kk}%
\end{equation}
where the coefficients $\left(  A_{1k}\right)  _{mn}$ are defined as,%
\begin{equation}
\left(  A_{1k}\right)  _{mn}\overset{\text{def}}{=}\sqrt{1-p}\left\langle
m\text{, }k\left\vert U\right\vert n\text{, }1\right\rangle =\sqrt{1-p}\left(
\left\langle m\text{, }k\right\vert \right)  \left(  U\left\vert n\text{,
}1\right\rangle \right)  \text{.} \label{u1}%
\end{equation}
Before computing an explicit expression for $U\left\vert n\text{,
}1\right\rangle $ in Eq. (\ref{u1}), observe that using the
Baker-Campbell-Hausdorff formula,%
\begin{equation}
e^{\chi A}Be^{-\chi A}=B+\chi\left[  A\text{, }B\right]  +\frac{\chi^{2}}%
{2!}\left[  A\text{, }\left[  A\text{, }B\right]  \right]  +\frac{\chi^{3}%
}{3!}\left[  A\text{, }\left[  A\text{, }\left[  A\text{, }B\right]  \right]
\right]  +\mathcal{O}\left(  3\right)  \text{,}%
\end{equation}
we obtain,%
\begin{align}
&  UaU^{\dagger}\overset{\text{def}}{=}e^{\chi\left(  a^{\dagger}b-b^{\dagger
}a\right)  }ae^{-\chi\left(  a^{\dagger}b-b^{\dagger}a\right)  }\nonumber\\
& \nonumber\\
&  =a+\chi\left[  a^{\dagger}b-b^{\dagger}a\text{, }a\right]  +\frac{\chi^{2}%
}{2!}\left[  a^{\dagger}b-b^{\dagger}a\text{, }\left[  a^{\dagger}%
b-b^{\dagger}a\text{, }a\right]  \right]  +\frac{\chi^{3}}{3!}\left[
a^{\dagger}b-b^{\dagger}a\text{, }\left[  a^{\dagger}b-b^{\dagger}a\text{,
}\left[  a^{\dagger}b-b^{\dagger}a\text{, }a\right]  \right]  \right]
+\mathcal{O}\left(  3\right)  \text{.} \label{uau}%
\end{align}
Notice that the commutator $\left[  a^{\dagger}b-b^{\dagger}a\text{,
}a\right]  $ in\ Eq. (\ref{uau}) can be written as,%
\begin{equation}
\left[  a^{\dagger}b-b^{\dagger}a\text{, }a\right]  =-b\text{, }\left[
a^{\dagger}b-b^{\dagger}a\text{, }\left[  a^{\dagger}b-b^{\dagger}a\text{,
}a\right]  \right]  =a\text{, }\left[  a^{\dagger}b-b^{\dagger}a\text{,
}\left[  a^{\dagger}b-b^{\dagger}a\text{, }\left[  a^{\dagger}b-b^{\dagger
}a\text{, }a\right]  \right]  \right]  =-a\text{,}%
\end{equation}
therefore $UaU^{\dagger}$ becomes,%
\begin{equation}
UaU^{\dagger}=a-\chi b-\frac{\chi^{2}}{2!}a+\frac{\chi^{3}}{3!}b+...=a\left(
1-\frac{\chi^{2}}{2!}+...\right)  -b\left(  \chi-\frac{\chi^{3}}%
{3!}+...\right)  =\left(  \cos\chi\right)  a-\left(  \sin\chi\right)
b\text{.}%
\end{equation}
Following this line of reasoning, we can also show that $UbU^{\dagger}$
equals,
\begin{equation}
UbU^{\dagger}=\left(  \cos\chi\right)  b-\left(  \sin\chi\right)  a\text{.}%
\end{equation}
Finally, recalling that $\left(  n+m\right)  ^{l}$ can be written as,%
\begin{equation}
\left(  n+m\right)  ^{l}=\sum_{p=0}^{l}\binom{l}{p}n^{p}m^{l-p}\text{,}%
\end{equation}
we have that $U\left\vert n\text{, }1\right\rangle $ in Eq. (\ref{u1})
becomes,%
\begin{align}
U\left\vert n\text{, }1\right\rangle  &  =U\frac{\left(  a^{\dagger}\right)
^{n}}{\sqrt{n!}}b^{\dagger}\left\vert 00\right\rangle =\left[  U\frac{\left(
a^{\dagger}\right)  ^{n}}{\sqrt{n!}}U^{\dagger}\right]  \left[  Ub^{\dagger
}U^{\dagger}\right]  \left\vert 00\right\rangle \\
& \nonumber\\
&  =\frac{1}{\sqrt{n!}}\left[  \left(  \cos\chi\right)  a^{\dagger}-\left(
\sin\chi\right)  b^{\dagger}\right]  ^{n}\left[  \left(  \cos\chi\right)
b^{\dagger}-\left(  \sin\chi\right)  a^{\dagger}\right]  \left\vert
00\right\rangle \nonumber\\
& \nonumber\\
&  =\frac{1}{\sqrt{n!}}\left[  \sum_{p=0}^{n}\binom{n}{p}\left(  \cos
\chi\right)  ^{p}\left(  -\sin\chi\right)  ^{n-p}\left(  a^{\dagger}\right)
^{p}\left(  b^{\dagger}\right)  ^{n-p}\right]  \left[  \left(  \cos
\chi\right)  \left\vert 01\right\rangle -\left(  \sin\chi\right)  \left\vert
10\right\rangle \right] \nonumber\\
& \nonumber\\
&  =\frac{1}{\sqrt{n!}}\sum_{p=0}^{n}\binom{n}{p}\left[  \left(  \cos
\chi\right)  ^{1+p}\left(  -\sin\chi\right)  ^{n-p}\left(  a^{\dagger}\right)
^{p}\left(  b^{\dagger}\right)  ^{n-p}\left\vert 01\right\rangle +\left(
\cos\chi\right)  ^{p}\left(  -\sin\chi\right)  ^{n-p+1}\left(  a^{\dagger
}\right)  ^{p}\left(  b^{\dagger}\right)  ^{n-p}\left\vert 10\right\rangle
\right] \nonumber\\
& \nonumber\\
&  =\frac{1}{\sqrt{n!}}\sum_{p=0}^{n}\binom{n}{p}\left[  \left(  \cos
\chi\right)  ^{1+p}\left(  -\sin\chi\right)  ^{n-p}\left(  a^{\dagger}\right)
^{p}\left\vert 0\right\rangle \left(  b^{\dagger}\right)  ^{n-p}\left\vert
1\right\rangle +\left(  \cos\chi\right)  ^{p}\left(  -\sin\chi\right)
^{n-p+1}\left(  a^{\dagger}\right)  ^{p}\left\vert 1\right\rangle \left(
b^{\dagger}\right)  ^{n-p}\left\vert 0\right\rangle \right] \nonumber\\
& \nonumber\\
&  =\frac{1}{\sqrt{n!}}\sum_{p=0}^{n}\binom{n}{p}\left[  \left(  \cos
\chi\right)  ^{1+p}\left(  -\sin\chi\right)  ^{n-p}\sqrt{p!}\left\vert
p\right\rangle \left(  b^{\dagger}\right)  ^{n-p+1}\left\vert 0\right\rangle
+\left(  \cos\chi\right)  ^{p}\left(  -\sin\chi\right)  ^{n-p+1}\left(
a^{\dagger}\right)  ^{1+p}\left\vert 0\right\rangle \sqrt{\left(  n-p\right)
!}\left\vert n-p\right\rangle \right]  \text{,}\nonumber
\end{align}
that is, after some more algebra,%
\begin{align}
U\left\vert n\text{, }1\right\rangle  &  =\frac{1}{\sqrt{n!}}\sum_{p=0}%
^{n}\binom{n}{p}\left[
\begin{array}
[c]{c}%
\left(  \cos\chi\right)  ^{1+p}\left(  -\sin\chi\right)  ^{n-p}\sqrt{p!}%
\sqrt{\left(  n-p+1\right)  !}\left\vert p\text{, }n-p+1\right\rangle +\\
\\
+\left(  \cos\chi\right)  ^{p}\left(  -\sin\chi\right)  ^{n-p+1}\sqrt{\left(
1+p\right)  !}\sqrt{\left(  n-p\right)  !}\left\vert 1+p\text{, }%
n-p\right\rangle
\end{array}
\right] \nonumber\\
& \nonumber\\
&  =\sum_{p=0}^{n}\left[
\begin{array}
[c]{c}%
\frac{1}{\sqrt{n!}}\binom{n}{p}\sqrt{p!}\sqrt{\left(  n-p+1\right)  !}\left(
\cos\chi\right)  ^{1+p}\left(  -\sin\chi\right)  ^{n-p}\left\vert p\text{,
}n-p+1\right\rangle +\\
\\
+\frac{1}{\sqrt{n!}}\binom{n}{p}\sqrt{\left(  1+p\right)  !}\sqrt{\left(
n-p\right)  !}\left(  \cos\chi\right)  ^{p}\left(  -\sin\chi\right)
^{n-p+1}\left\vert 1+p\text{, }n-p\right\rangle
\end{array}
\right] \nonumber\\
& \nonumber\\
&  =\sum_{p=0}^{n}\left[
\begin{array}
[c]{c}%
\sqrt{\binom{n}{p}}\sqrt{\left(  n-p+1\right)  }\left(  \cos\chi\right)
^{1+p}\left(  -\sin\chi\right)  ^{n-p}\left\vert p\text{, }n-p+1\right\rangle
+\\
\\
+\sqrt{\binom{n}{p}}\sqrt{\left(  1+p\right)  }\left(  \cos\chi\right)
^{p}\left(  -\sin\chi\right)  ^{n-p+1}\left\vert 1+p\text{, }n-p\right\rangle
\end{array}
\right]  \text{.} \label{un1f}%
\end{align}
Therefore, substituting Eq. (\ref{un1f}) into Eq. (\ref{u1}), we obtain%
\begin{align}
\frac{\left(  A_{1k}\right)  _{mn}}{\sqrt{1-p}}\overset{\text{def}}%
{=}\left\langle m\text{, }k\left\vert U\right\vert n\text{, }1\right\rangle
&  =\sum_{p=0}^{n}\left[
\begin{array}
[c]{c}%
\sqrt{\binom{n}{p}}\sqrt{\left(  n-p+1\right)  }\left(  \cos\chi\right)
^{1+p}\left(  -\sin\chi\right)  ^{n-p}\left\langle m\text{, }k\left\vert
p\text{, }n-p+1\right.  \right\rangle +\\
\\
+\sqrt{\binom{n}{p}}\sqrt{\left(  1+p\right)  }\left(  \cos\chi\right)
^{p}\left(  -\sin\chi\right)  ^{n-p+1}\left\langle m\text{, }k\left\vert
1+p\text{, }n-p\right.  \right\rangle
\end{array}
\right] \nonumber\\
& \nonumber\\
&  =\sum_{p=0}^{n}\left[
\begin{array}
[c]{c}%
\sqrt{\binom{n}{p}}\sqrt{\left(  n-p+1\right)  }\left(  \cos\chi\right)
^{1+p}\left(  -\sin\chi\right)  ^{n-p}\delta_{m\text{, }p}\delta_{k\text{,
}n-p+1}+\\
\\
+\sqrt{\binom{n}{p}}\sqrt{\left(  1+p\right)  }\left(  \cos\chi\right)
^{p}\left(  -\sin\chi\right)  ^{n-p+1}\delta_{m\text{, }1+p}\delta_{k\text{,
}n-p}%
\end{array}
\right] \nonumber\\
& \nonumber\\
&  =\left[
\begin{array}
[c]{c}%
\sqrt{\binom{n}{n-k+1}}\sqrt{k}\left(  \cos\chi\right)  ^{n-k+2}\left(
-\sin\chi\right)  ^{k-1}\delta_{m\text{, }n-k+1}+\\
\\
\sqrt{\binom{n}{n-k}}\sqrt{n-k+1}\left(  \cos\chi\right)  ^{n-k}\left(
-\sin\chi\right)  ^{1+k}\delta_{m\text{, }n-k+1}%
\end{array}
\right]  \text{.} \label{a1k}%
\end{align}
Using Eq. (\ref{a1k}), the Kraus operators $A_{1k}$ in Eq. (\ref{a1kk})
become
\begin{equation}
A_{1k}=\sum_{m\text{, }n}\left(  A_{1k}\right)  _{mn}\left\vert m\right\rangle
\left\langle n\right\vert =\sqrt{1-p}\left[
\begin{array}
[c]{c}%
\sqrt{\binom{n}{n-k+1}}\sqrt{k}\left(  \cos\chi\right)  ^{n-k+2}\left(
-\sin\chi\right)  ^{k-1}\left\vert n-k+1\right\rangle \left\langle
n\right\vert +\\
\\
\sqrt{\binom{n}{n-k}}\sqrt{n-k+1}\left(  \cos\chi\right)  ^{n-k}\left(
-\sin\chi\right)  ^{1+k}\left\vert n-k+1\right\rangle \left\langle
n\right\vert
\end{array}
\right]  \text{.}%
\end{equation}
Therefore, the error operators $A_{11}$ and $A_{10}$ read
\begin{align}
A_{11}  &  =\sqrt{1-p}\sum_{n=0}^{1}\left[  \sqrt{\binom{n}{n}}\left(
\cos\chi\right)  ^{1+n}+\sqrt{\binom{n}{n}}\sqrt{n}\left(  \cos\chi\right)
^{n-1}\left(  -\sin\chi\right)  ^{2}\right]  \left\vert n\right\rangle
\left\langle n\right\vert \nonumber\\
&  =\sqrt{1-p}\left\{  \left(  \cos\chi\right)  \left\vert 0\right\rangle
\left\langle 0\right\vert +\left[  \left(  \cos\chi\right)  ^{2}+\left(
-\sin\chi\right)  ^{2}\right]  \left\vert 1\right\rangle \left\langle
1\right\vert \right\}  =\sqrt{1-p}\left[  \left\vert 1\right\rangle
\left\langle 1\right\vert +\left(  \cos\chi\right)  \left\vert 0\right\rangle
\left\langle 0\right\vert \right] \nonumber\\
&  =\sqrt{1-p}\left(
\begin{array}
[c]{cc}%
\sqrt{1-\gamma} & 0\\
0 & 1
\end{array}
\right)  \text{,}%
\end{align}
where $\gamma\overset{\text{def}}{=}\sin^{2}\chi\equiv1-\cos^{2}\chi$ and,%
\begin{equation}
A_{10}=\sqrt{1-p}\left(  -\sin\chi\right)  \left\vert 1\right\rangle
\left\langle 0\right\vert =\sqrt{1-p}\left(
\begin{array}
[c]{cc}%
0 & 0\\
\sqrt{\gamma} & 0
\end{array}
\right)  \text{,}%
\end{equation}
respectively. Following the same line of reasoning employed for computing
$A_{11}$ and $A_{10}$ and noticing that%
\begin{equation}
U\left\vert 00\right\rangle =\left[  I+\chi\left(  a^{\dagger}b-b^{\dagger
}a\right)  +\frac{\chi^{2}}{2!}\left(  a^{\dagger}b-b^{\dagger}a\right)
^{2}+...\right]  \left\vert 00\right\rangle =\left\vert 00\right\rangle
\text{,}%
\end{equation}
we can also compute the Kraus errors $A_{0k}$, where
\begin{equation}
A_{0k}\overset{\text{def}}{=}\sum_{m\text{, }n}\left(  A_{0k}\right)
_{mn}\left\vert m\right\rangle \left\langle n\right\vert \text{,}%
\end{equation}
and the coefficients $\left(  A_{0k}\right)  _{mn}$ read,%
\begin{equation}
\left(  A_{0k}\right)  _{mn}\overset{\text{def}}{=}\sqrt{p}\left\langle
m\text{, }k\left\vert U\right\vert n\text{, }0\right\rangle =\sqrt{p}\left(
\left\langle m\text{, }k\right\vert \right)  \left(  U\left\vert n\text{,
}0\right\rangle \right)  \text{.}%
\end{equation}
It turns out that the errors $A_{00}$ and $A_{01}$ become,%
\begin{equation}
A_{00}\overset{\text{def}}{=}\sqrt{p}\left[  \left\vert 0\right\rangle
\left\langle 0\right\vert +\sqrt{1-\gamma}\left\vert 1\right\rangle
\left\langle 1\right\vert \right]  =\sqrt{p}\left(
\begin{array}
[c]{cc}%
1 & 0\\
0 & \sqrt{1-\gamma}%
\end{array}
\right)  \text{,}%
\end{equation}
and,
\begin{equation}
A_{01}\overset{\text{def}}{=}\sqrt{p}\left\langle 1\left\vert U\right\vert
0\right\rangle =\sqrt{p}\left[  \sqrt{\gamma}\left\vert 0\right\rangle
\left\langle 1\right\vert \right]  =\sqrt{p}\left(
\begin{array}
[c]{cc}%
0 & \sqrt{\gamma}\\
0 & 0
\end{array}
\right)  \text{,}%
\end{equation}
respectively. In conclusion, the GAD channel $\Lambda_{\text{GAD}}$ is given
by%
\begin{equation}
\Lambda_{\text{GAD}}\left(  \rho\right)  \overset{\text{def}}{=}A_{00}\rho
A_{00}^{\dagger}+A_{01}\rho A_{10}^{\dagger}+A_{10}\rho A_{10}^{\dagger
}+A_{11}\rho A_{11}^{\dagger}\text{,}%
\end{equation}
where,%
\begin{align}
A_{00}\overset{\text{def}}{=}\sqrt{p}\left\langle 0\left\vert U\right\vert
0\right\rangle  &  =\sqrt{p}\left[  \left\vert 0\right\rangle \left\langle
0\right\vert +\sqrt{1-\gamma}\left\vert 1\right\rangle \left\langle
1\right\vert \right]  =\sqrt{p}\left(
\begin{array}
[c]{cc}%
1 & 0\\
0 & \sqrt{1-\gamma}%
\end{array}
\right)  \text{,}\nonumber\\
& \nonumber\\
\text{ }A_{01}\overset{\text{def}}{=}\sqrt{p}\left\langle 1\left\vert
U\right\vert 0\right\rangle  &  =\sqrt{p}\left[  \sqrt{\gamma}\left\vert
0\right\rangle \left\langle 1\right\vert \right]  =\sqrt{p}\left(
\begin{array}
[c]{cc}%
0 & \sqrt{\gamma}\\
0 & 0
\end{array}
\right)  \text{,}\nonumber\\
& \nonumber\\
A_{10}\overset{\text{def}}{=}\sqrt{1-p}\left\langle 0\left\vert U\right\vert
1\right\rangle  &  =\sqrt{1-p}\left[  \sqrt{\gamma}\left\vert 1\right\rangle
\left\langle 0\right\vert \right]  =\sqrt{1-p}\left(
\begin{array}
[c]{cc}%
0 & 0\\
\sqrt{\gamma} & 0
\end{array}
\right)  \text{,}\nonumber\\
& \nonumber\\
\text{ }A_{11}\overset{\text{def}}{=}\sqrt{1-p}\left\langle 1\left\vert
U\right\vert 1\right\rangle  &  =\sqrt{1-p}\left[  \sqrt{1-\gamma}\left\vert
0\right\rangle \left\langle 0\right\vert +\left\vert 1\right\rangle
\left\langle 1\right\vert \right]  =\sqrt{1-p}\left(
\begin{array}
[c]{cc}%
\sqrt{1-\gamma} & 0\\
0 & 1
\end{array}
\right)  \text{.}%
\end{align}
Finally, provided that we relabel $A_{0}\overset{\text{def}}{=}A_{00}$,
$A_{1}\overset{\text{def}}{=}A_{01}$, $A_{2}\overset{\text{def}}{=}A_{11}$,
$A_{3}\overset{\text{def}}{=}A_{10}$, the Kraus operators in Eq. (\ref{ko})
are obtained.

For the sake of completeness, we also observe that the analysis provided for
this model is formally equivalent to that for the model employed in the main
text. For instance, the Hamiltonian of the beam splitter,\textbf{
}$H_{\text{int}}^{\left(  \text{bs}\right)  }\overset{\text{def}}{=}%
$\textbf{\ }$i\chi\left(  ab^{\dagger}-a^{\dagger}b\right)  $\textbf{, }should
be replaced by the interaction Hamiltonian between the two-level atom and the
bath of harmonic oscillators,\textbf{ }$H_{\text{int}}^{\left(  \text{atom-HO}%
\right)  }\overset{\text{def}}{=}g\left(  a^{\dagger}\sigma_{-}+a\sigma
_{+}\right)  $\textbf{. }In the definition of\textbf{ }$H_{\text{int}%
}^{\left(  \text{atom-HO}\right)  }$\textbf{, }the Pauli raising\textbf{
}$\left(  \sigma_{+}\right)  $\textbf{\ }and lowering\textbf{ }$\left(
\sigma_{-}\right)  $\textbf{\ }operators act on the two-level atom. The
creation\textbf{ }$\left(  a^{\dagger}\right)  $\textbf{\ }and
annihilation\textbf{ }$\left(  a\right)  $\textbf{\ }operators are associated
with the harmonic oscillator, instead. Finally,\textbf{ }$g$ is the coupling
constant for the interaction between the atom and the oscillator.

\section{Qubit entanglement-breaking and the GAD channel}

A quantum channel $\Lambda$ is called entanglement-breaking if $\left(
\Lambda\otimes I\right)  \left(  \rho\right)  $ is always separable, i.e., any
entangled density matrix $\rho$ is mapped to a separable one \cite{ruskai}. In
order to check if a channel is entanglement-breaking, it is sufficient to look
at the separability of the output state corresponding just to an input
maximally entangled state. In other words, $\Lambda$ is entanglement-breaking
if and only if $\left(  \Lambda\otimes I\right)  \left(  \left\vert
\beta\right\rangle \left\langle \beta\right\vert \right)  $ is separable for
$\left\vert \beta\right\rangle $ defined as,%
\begin{equation}
\left\vert \beta\right\rangle \overset{\text{def}}{=}\frac{1}{\sqrt{d}}%
%TCIMACRO{\dsum \limits_{j=0}^{d-1}}%
%BeginExpansion
{\displaystyle\sum\limits_{j=0}^{d-1}}
%EndExpansion
\left\vert j\right\rangle \otimes\left\vert j\right\rangle \text{,}%
\end{equation}
$d$ being the dimension of the Hilbert space. A simple way to check the
separability of density matrices is by means of the Peres-Horodecki positive
partial transpose (PPT) criterion \cite{peres, horodecki} which provides a
necessary and sufficient condition for the joint density matrix $\rho$ of two
$d=2$-dimensional systems $A$ and $B$ to be separable. Alternatively, the
entanglement of a mixed state $\rho$ described by a probabilistic mixture of
an ensemble of pure states of quantum systems of dimension $2\times2$ can be
quantified by means of the so-called concurrence \cite{wootters}. For systems
with this dimensionality, the concurrence $\mathcal{C}\left(  \rho\right)  $
of $\rho$ reads%
\begin{equation}
\mathcal{C}\left(  \rho\right)  \overset{\text{def}}{=}\max\left\{  0\text{,
}\sqrt{\lambda_{1}}-\sqrt{\lambda_{2}}-\sqrt{\lambda_{3}}-\sqrt{\lambda_{4}%
}\right\}  \text{,}%
\end{equation}
where $\lambda_{i}$ are the non-negative \emph{real }eigenvalues of the
non-Hermitian matrix $\rho\tilde{\rho}\overset{\text{def}}{=}$ $\rho\left(
\sigma_{y}\otimes\sigma_{y}\right)  \rho^{\ast}\left(  \sigma_{y}\otimes
\sigma_{y}\right)  $ where $\tilde{\rho}$ is the spin-flipped state,
$\lambda_{1}$ is the largest eigenvalue, and the complex conjugation is taken
with respect to the product basis of eigenvectors of $\sigma_{z}$ given by
$\left\{  \left\vert \uparrow\uparrow\right\rangle \text{, }\left\vert
\uparrow\downarrow\right\rangle \text{, }\left\vert \downarrow\uparrow
\right\rangle \text{, }\left\vert \downarrow\downarrow\right\rangle \right\}
$.

For the GAD channel, it can be shown that the concurrence of $\rho
\overset{\text{def}}{=}\left(  \Lambda_{\text{GAD}}\otimes I\right)  \left(
\left\vert \beta\right\rangle \left\langle \beta\right\vert \right)  $ with
$\left\vert \beta\right\rangle \overset{\text{def}}{=}\frac{\left\vert
00\right\rangle +\left\vert 11\right\rangle }{\sqrt{2}}$ reads,%
\begin{equation}
\mathcal{C}_{\text{GAD}}\left(  \gamma\text{, }p\right)  \overset{\text{def}%
}{=}\frac{1}{2}\left[
\begin{array}
[c]{c}%
\sqrt{\left[  \left(  2\left(  1-\gamma\right)  +p\gamma^{2}\left(
1-p\right)  \right)  +2\sqrt{\left(  1-\gamma\right)  \left(  \left(
1-\gamma\right)  +p\gamma^{2}\left(  1-p\right)  \right)  }\right]  }+\\
\\
-\sqrt{\left[  \left(  2\left(  1-\gamma\right)  +p\gamma^{2}\left(
1-p\right)  \right)  -2\sqrt{\left(  1-\gamma\right)  \left(  \left(
1-\gamma\right)  +p\gamma^{2}\left(  1-p\right)  \right)  }\right]  }+\\
\\
-2\sqrt{p\left(  1-p\right)  \gamma^{2}}%
\end{array}
\right]  \text{.}%
\end{equation}
Furthermore, the eigenvalues of the partial transpose of $\rho$ are given by,%
\begin{align}
\lambda_{1}\left(  \gamma\text{, }p\right)   &  =\frac{1}{2}\gamma p+\frac
{1}{2}\left(  1-\gamma\right)  \text{, }\lambda_{2}\left(  \gamma\text{,
}p\right)  =\frac{1}{2}\left(  1-p\gamma\right)  \text{, }\nonumber\\
& \nonumber\\
\lambda_{3}\left(  \gamma\text{, }p\right)   &  =\frac{1}{4}\gamma-\frac{1}%
{2}\sqrt{\frac{1}{4}\gamma^{2}-\gamma-p\gamma^{2}+p^{2}\gamma^{2}+1}\text{,
}\lambda_{4}\left(  \gamma\text{, }p\right)  =\frac{1}{4}\gamma+\frac{1}%
{2}\sqrt{\frac{1}{4}\gamma^{2}-\gamma-p\gamma^{2}+p^{2}\gamma^{2}+1}\text{.}%
\end{align}
The eigenvalues $\lambda_{1}$, $\lambda_{2}$ and $\lambda_{4}$ are positive.
The eigenvalue $\lambda_{3}$ is positive provided that,%
\begin{equation}
p^{2}\gamma^{2}-p\gamma^{2}+1-\gamma\leq0\text{.} \label{inequality}%
\end{equation}
The inequality in Eq. (\ref{inequality}) is fulfilled in the two-dimensional
parametric region $\left(  \gamma\text{, }p\left(  \gamma\right)  \right)  $
where $p_{\text{min}}\left(  \gamma\right)  \leq p\left(  \gamma\right)  \leq
p_{\text{max}}\left(  \gamma\right)  $ with,%
\begin{equation}
p_{\text{min}}\left(  \gamma\right)  \overset{\text{def}}{=}\frac{1}%
{\gamma^{2}}\left(  \frac{1}{2}\gamma^{2}-\frac{1}{2}\sqrt{\gamma^{4}%
+4\gamma^{3}-4\gamma^{2}}\right)  \text{ and, }p_{\text{max}}\left(
\gamma\right)  \overset{\text{def}}{=}\frac{1}{\gamma^{2}}\left(  \frac{1}%
{2}\gamma^{2}+\frac{1}{2}\sqrt{\gamma^{4}+4\gamma^{3}-4\gamma^{2}}\right)
\text{.}%
\end{equation}
It turns out that for pairs $\left(  \gamma\text{, }p\left(  \gamma\right)
\right)  $ within this two-dimensional parametric region, $\mathcal{C}%
_{\text{GAD}}\left(  \gamma\text{, }p\right)  $ equals zero and the
GAD\ channel becomes entanglement-breaking.

\section{Explicit construction of an orthonormal basis}

In general, we can proceed as follows. The rank-nullity theorem allows us with
an algorithmic procedure for enlarging a set of $m$ linearly independent
vectors in $%
%TCIMACRO{\U{2102} }%
%BeginExpansion
\mathbb{C}
%EndExpansion
^{n}$ with $m\leq n$ to a set of $n$ linearly independent vectors in $%
%TCIMACRO{\U{2102} }%
%BeginExpansion
\mathbb{C}
%EndExpansion
^{n}$ \cite{lang}. The Gram-Schmidt orthonormalization procedure, instead, can
be used to construct an orthonormal basis from this set of $n$ linearly
independent vectors.

The rank-nullity theorem states that given a $\left(  m\times n\right)
$-matrix $A$, it turns out that $n=rank\left(  A\right)  +nullity\left(
A\right)  $. The nullity of a $\left(  m\times n\right)  $-matrix $A$
representing a linear map $\hat{A}:%
%TCIMACRO{\U{2102} }%
%BeginExpansion
\mathbb{C}
%EndExpansion
^{n}\rightarrow$ $%
%TCIMACRO{\U{2102} }%
%BeginExpansion
\mathbb{C}
%EndExpansion
^{m}$ is the dimension of its null-space (or $kernel$),%
\begin{equation}
\ker A\overset{\text{def}}{=}\left\{  \vec{x}\in%
%TCIMACRO{\U{2102} }%
%BeginExpansion
\mathbb{C}
%EndExpansion
^{n}:A\vec{x}=\vec{0}\right\}  \text{.}%
\end{equation}
The rank of a matrix is the maximum number of linearly independent columns (or rows).

Let us suppose we wish to construct an orthonormal basis $\left\{  \left\vert
e_{k}\right\rangle \right\}  $ with $k=1$,..., $32$ for the $32$-dimensional
\emph{complex} Hilbert space $\mathcal{H}_{2}^{5}$. The first two orthonormal
basis vectors can be constructed from the action of the weight-$0$ error
operators $A_{00000}$ on the codewords $\left\vert 0_{L}\right\rangle $ and
$\left\vert 1_{L}\right\rangle $. They read,%
\begin{equation}
\left\vert e_{1}\right\rangle \overset{\text{def}}{=}N_{1}\left(
\gamma\right)  \left[
\begin{array}
[c]{c}%
-\left\vert 00000\right\rangle +\left(  1-\gamma\right)  ^{2}\left\vert
01111\right\rangle -\left(  1-\gamma\right)  ^{\frac{3}{2}}\left\vert
10011\right\rangle +\left(  1-\gamma\right)  ^{\frac{3}{2}}\left\vert
11100\right\rangle +\\
+\left(  1-\gamma\right)  \left\vert 00110\right\rangle +\left(
1-\gamma\right)  \left\vert 01001\right\rangle +\left(  1-\gamma\right)
^{\frac{3}{2}}\left\vert 10101\right\rangle +\left(  1-\gamma\right)
^{\frac{3}{2}}\left\vert 11010\right\rangle
\end{array}
\right]  \text{,} \label{e1}%
\end{equation}
and,%
\begin{equation}
\left\vert e_{2}\right\rangle \overset{\text{def}}{=}N_{2}\left(
\gamma\right)  \left[
\begin{array}
[c]{c}%
-\left(  1-\gamma\right)  ^{\frac{5}{2}}\left\vert 11111\right\rangle
+\sqrt{1-\gamma}\left\vert 10000\right\rangle +\left(  1-\gamma\right)
\left\vert 01100\right\rangle -\left(  1-\gamma\right)  \left\vert
00011\right\rangle +\\
+\left(  1-\gamma\right)  ^{\frac{3}{2}}\left\vert 11001\right\rangle +\left(
1-\gamma\right)  ^{\frac{3}{2}}\left\vert 10110\right\rangle -\left(
1-\gamma\right)  \left\vert 01010\right\rangle -\left(  1-\gamma\right)
\left\vert 00101\right\rangle
\end{array}
\right]  \text{,} \label{e2}%
\end{equation}
respectively. The normalization factors $N_{1}\left(  \gamma\right)  $ and
$N_{2}\left(  \gamma\right)  $ are given by,%
\begin{equation}
N_{1}\left(  \gamma\right)  \overset{\text{def}}{=}\frac{1}{\sqrt{1+\left(
1-\gamma\right)  ^{4}+4\left(  1-\gamma\right)  ^{3}+2\left(  1-\gamma\right)
^{2}}}\text{ and, }N_{2}\left(  \gamma\right)  \overset{\text{def}}{=}\frac
{1}{\sqrt{\left(  1-\gamma\right)  ^{5}+2\left(  1-\gamma\right)
^{3}+4\left(  1-\gamma\right)  ^{2}+\left(  1-\gamma\right)  }}\text{.}%
\end{equation}
The next ten orthonormal vectors can be constructed by taking into
consideration the action of the five weight-$1$ error operators $A_{10000}$,
$A_{01000}$, $A_{00100}$, $A_{00010}$, $A_{00001}$ on the codewords. The
action of $A_{10000}$ on the codewords $\left\vert 0_{L}\right\rangle $ and
$\left\vert 1_{L}\right\rangle $ leads to,%
\begin{equation}
\left\vert e_{3}\right\rangle \overset{\text{def}}{=}\frac{-\left\vert
00011\right\rangle +\left\vert 01100\right\rangle +\left\vert
00101\right\rangle +\left\vert 01010\right\rangle }{\sqrt{4}}\text{,}%
\end{equation}
and,%
\begin{equation}
\left\vert e_{4}\right\rangle \overset{\text{def}}{=}\frac{-\left(
1-\gamma\right)  ^{2}\left\vert 01111\right\rangle +\left\vert
00000\right\rangle +\left(  1-\gamma\right)  \left\vert 01001\right\rangle
+\left(  1-\gamma\right)  \left\vert 00110\right\rangle }{\sqrt{1+\left(
1-\gamma\right)  ^{4}+2\left(  1-\gamma\right)  ^{2}}}\text{,}%
\end{equation}
respectively. The action of $A_{01000}$ on the codewords $\left\vert
0_{L}\right\rangle $ and $\left\vert 1_{L}\right\rangle $ yields,%
\begin{equation}
\left\vert e_{5}\right\rangle \overset{\text{def}}{=}\frac{\left(
1-\gamma\right)  ^{\frac{3}{2}}\left\vert 00111\right\rangle +\left(
1-\gamma\right)  \left\vert 10100\right\rangle +\sqrt{1-\gamma}\left\vert
00001\right\rangle +\left(  1-\gamma\right)  \left\vert 10010\right\rangle
}{\sqrt{\left(  1-\gamma\right)  ^{3}+2\left(  1-\gamma\right)  ^{2}+\left(
1-\gamma\right)  }}\text{,}%
\end{equation}
and,%
\begin{equation}
\left\vert e_{6}\right\rangle \overset{\text{def}}{=}\frac{-\left(
1-\gamma\right)  ^{2}\left\vert 10111\right\rangle +\sqrt{1-\gamma}\left\vert
00100\right\rangle +\left(  1-\gamma\right)  \left\vert 10001\right\rangle
-\sqrt{1-\gamma}\left\vert 00010\right\rangle }{\sqrt{\left(  1-\gamma\right)
^{4}+\left(  1-\gamma\right)  ^{2}+2\left(  1-\gamma\right)  }}\text{,}%
\end{equation}
respectively. The action of $A_{00100}$ on the codewords $\left\vert
0_{L}\right\rangle $ and $\left\vert 1_{L}\right\rangle $ leads to,%
\begin{equation}
\left\vert e_{7}\right\rangle \overset{\text{def}}{=}\frac{\left(
1-\gamma\right)  ^{\frac{3}{2}}\left\vert 01011\right\rangle +\left(
1-\gamma\right)  \left\vert 11000\right\rangle +\left(  1-\gamma\right)
\left\vert 10001\right\rangle +\sqrt{1-\gamma}\left\vert 00010\right\rangle
}{\sqrt{\left(  1-\gamma\right)  ^{3}+2\left(  1-\gamma\right)  ^{2}+\left(
1-\gamma\right)  }}\text{,}%
\end{equation}
and,%
\begin{equation}
\left\vert e_{8}\right\rangle \overset{\text{def}}{=}\frac{-\left(
1-\gamma\right)  ^{2}\left\vert 11011\right\rangle +\sqrt{1-\gamma}\left\vert
01000\right\rangle +\left(  1-\gamma\right)  \left\vert 10010\right\rangle
-\sqrt{1-\gamma}\left\vert 00001\right\rangle }{\sqrt{\left(  1-\gamma\right)
^{4}+\left(  1-\gamma\right)  ^{2}+2\left(  1-\gamma\right)  }}\text{,}%
\end{equation}
respectively. The action of $A_{00010}$ on the codewords $\left\vert
0_{L}\right\rangle $ and $\left\vert 1_{L}\right\rangle $ yields,%
\begin{equation}
\left\vert e_{9}\right\rangle \overset{\text{def}}{=}\frac{\left(
1-\gamma\right)  ^{\frac{3}{2}}\left\vert 01101\right\rangle -\left(
1-\gamma\right)  \left\vert 10001\right\rangle +\sqrt{1-\gamma}\left\vert
00100\right\rangle +\left(  1-\gamma\right)  \left\vert 11000\right\rangle
}{\sqrt{\left(  1-\gamma\right)  ^{3}+2\left(  1-\gamma\right)  ^{2}+\left(
1-\gamma\right)  }}\text{,}%
\end{equation}
and,%
\begin{equation}
\left\vert e_{10}\right\rangle \overset{\text{def}}{=}\frac{-\left(
1-\gamma\right)  ^{2}\left\vert 11101\right\rangle -\sqrt{1-\gamma}\left\vert
00001\right\rangle +\left(  1-\gamma\right)  \left\vert 10100\right\rangle
-\sqrt{1-\gamma}\left\vert 01000\right\rangle }{\sqrt{\left(  1-\gamma\right)
^{4}+\left(  1-\gamma\right)  ^{2}+2\left(  1-\gamma\right)  }}\text{,}%
\end{equation}
respectively. Finally, the action of $A_{00001}$ on the codewords $\left\vert
0_{L}\right\rangle $ and $\left\vert 1_{L}\right\rangle $ gives,%
\begin{equation}
\left\vert e_{11}\right\rangle \overset{\text{def}}{=}\frac{\left(
1-\gamma\right)  ^{\frac{3}{2}}\left\vert 01110\right\rangle -\left(
1-\gamma\right)  \left\vert 10010\right\rangle +\sqrt{1-\gamma}\left\vert
01000\right\rangle +\left(  1-\gamma\right)  \left\vert 10100\right\rangle
}{\sqrt{\left(  1-\gamma\right)  ^{3}+2\left(  1-\gamma\right)  ^{2}+\left(
1-\gamma\right)  }}\text{,}%
\end{equation}
and,%
\begin{equation}
\left\vert e_{12}\right\rangle \overset{\text{def}}{=}\frac{-\left(
1-\gamma\right)  ^{2}\left\vert 11110\right\rangle -\sqrt{1-\gamma}\left\vert
00010\right\rangle +\left(  1-\gamma\right)  \left\vert 11000\right\rangle
-\sqrt{1-\gamma}\left\vert 00100\right\rangle }{\sqrt{\left(  1-\gamma\right)
^{4}+\left(  1-\gamma\right)  ^{2}+2\left(  1-\gamma\right)  }}\text{,}%
\end{equation}
respectively. The next ten orthonormal vectors can be constructed by
considering the action of the five weight-$1$ error operators $A_{30000}$,
$A_{03000}$, $A_{00300}$, $A_{00030}$, $A_{00003}$ on the codewords. The
action of $A_{30000}$ on the codewords $\left\vert 0_{L}\right\rangle $ and
$\left\vert 1_{L}\right\rangle $ leads to,%
\begin{equation}
\left\vert e_{13}\right\rangle \overset{\text{def}}{=}\frac{-\left\vert
10000\right\rangle +\left(  1-\gamma\right)  ^{2}\left\vert 11111\right\rangle
+\left(  1-\gamma\right)  \left\vert 10110\right\rangle +\left(
1-\gamma\right)  \left\vert 11001\right\rangle }{\sqrt{1+\left(
1-\gamma\right)  ^{4}+2\left(  1-\gamma\right)  ^{2}}}\text{,}%
\end{equation}
and,%
\begin{equation}
\left\vert e_{14}\right\rangle \overset{\text{def}}{=}\frac{\left\vert
11100\right\rangle -\left\vert 10011\right\rangle -\left\vert
11010\right\rangle -\left\vert 10101\right\rangle }{\sqrt{4}}\text{,}%
\end{equation}
respectively. The action of $A_{03000}$ on the codewords $\left\vert
0_{L}\right\rangle $ and $\left\vert 1_{L}\right\rangle $ yields,%
\begin{equation}
\left\vert e_{15}\right\rangle \overset{\text{def}}{=}\frac{-\left\vert
01000\right\rangle -\left(  1-\gamma\right)  ^{\frac{3}{2}}\left\vert
11011\right\rangle +\left(  1-\gamma\right)  \left\vert 01110\right\rangle
+\left(  1-\gamma\right)  ^{\frac{3}{2}}\left\vert 11101\right\rangle }%
{\sqrt{1+2\left(  1-\gamma\right)  ^{3}+\left(  1-\gamma\right)  ^{2}}%
}\text{,}%
\end{equation}
and,%
\begin{equation}
\left\vert e_{16}\right\rangle \overset{\text{def}}{=}\frac{\sqrt{1-\gamma
}\left\vert 11000\right\rangle -\left(  1-\gamma\right)  \left\vert
01011\right\rangle +\left(  1-\gamma\right)  ^{\frac{3}{2}}\left\vert
11110\right\rangle -\left(  1-\gamma\right)  \left\vert 01101\right\rangle
}{\sqrt{\left(  1-\gamma\right)  ^{3}+2\left(  1-\gamma\right)  ^{2}+\left(
1-\gamma\right)  }}\text{,}%
\end{equation}
respectively. The action of $A_{00300}$ on the codewords $\left\vert
0_{L}\right\rangle $ and $\left\vert 1_{L}\right\rangle $ leads to,%
\begin{equation}
\left\vert e_{17}\right\rangle \overset{\text{def}}{=}\frac{-\left\vert
00100\right\rangle -\left(  1-\gamma\right)  ^{\frac{3}{2}}\left\vert
10111\right\rangle +\left(  1-\gamma\right)  \left\vert 01101\right\rangle
+\left(  1-\gamma\right)  ^{\frac{3}{2}}\left\vert 11110\right\rangle }%
{\sqrt{1+2\left(  1-\gamma\right)  ^{3}+\left(  1-\gamma\right)  ^{2}}%
}\text{,}%
\end{equation}
and,%
\begin{equation}
\left\vert e_{18}\right\rangle \overset{\text{def}}{=}\frac{\sqrt{1-\gamma
}\left\vert 10100\right\rangle -\left(  1-\gamma\right)  \left\vert
00111\right\rangle +\left(  1-\gamma\right)  ^{\frac{3}{2}}\left\vert
11101\right\rangle -\left(  1-\gamma\right)  \left\vert 01110\right\rangle
}{\sqrt{\left(  1-\gamma\right)  ^{3}+2\left(  1-\gamma\right)  ^{2}+\left(
1-\gamma\right)  }}\text{,}%
\end{equation}
respectively. The action of $A_{00030}$ on the codewords $\left\vert
0_{L}\right\rangle $ and $\left\vert 1_{L}\right\rangle $ yields,%
\begin{equation}
\left\vert e_{19}\right\rangle \overset{\text{def}}{=}\frac{-\left\vert
00010\right\rangle +\left(  1-\gamma\right)  ^{\frac{3}{2}}\left\vert
11110\right\rangle +\left(  1-\gamma\right)  \left\vert 01011\right\rangle
+\left(  1-\gamma\right)  ^{\frac{3}{2}}\left\vert 10111\right\rangle }%
{\sqrt{1+2\left(  1-\gamma\right)  ^{3}+\left(  1-\gamma\right)  ^{2}}%
}\text{,}%
\end{equation}
and,%
\begin{equation}
\left\vert e_{20}\right\rangle \overset{\text{def}}{=}\frac{\sqrt{1-\gamma
}\left\vert 10010\right\rangle +\left(  1-\gamma\right)  \left\vert
01110\right\rangle +\left(  1-\gamma\right)  ^{\frac{3}{2}}\left\vert
11011\right\rangle -\left(  1-\gamma\right)  \left\vert 00111\right\rangle
}{\sqrt{\left(  1-\gamma\right)  ^{3}+2\left(  1-\gamma\right)  ^{2}+\left(
1-\gamma\right)  }}\text{,}%
\end{equation}
respectively. Finally, The action of $A_{00003}$ on the codewords $\left\vert
0_{L}\right\rangle $ and $\left\vert 1_{L}\right\rangle $ leads to,%
\begin{equation}
\left\vert e_{21}\right\rangle \overset{\text{def}}{=}\frac{-\left\vert
00001\right\rangle +\left(  1-\gamma\right)  ^{\frac{3}{2}}\left\vert
11101\right\rangle +\left(  1-\gamma\right)  \left\vert 00111\right\rangle
+\left(  1-\gamma\right)  ^{\frac{3}{2}}\left\vert 11011\right\rangle }%
{\sqrt{1+2\left(  1-\gamma\right)  ^{3}+\left(  1-\gamma\right)  ^{2}}%
}\text{,}%
\end{equation}
and,%
\begin{equation}
\left\vert e_{22}\right\rangle \overset{\text{def}}{=}\frac{\sqrt{1-\gamma
}\left\vert 10001\right\rangle +\left(  1-\gamma\right)  \left\vert
01101\right\rangle +\left(  1-\gamma\right)  ^{\frac{3}{2}}\left\vert
10111\right\rangle -\left(  1-\gamma\right)  \left\vert 01011\right\rangle
}{\sqrt{\left(  1-\gamma\right)  ^{3}+2\left(  1-\gamma\right)  ^{2}+\left(
1-\gamma\right)  }}\text{,}%
\end{equation}
respectively. We have $22$ orthonormal vectors and we need $10$ more. From
Eqs. (\ref{e1}) and (\ref{e2}), it turns out that the following six linearly
independent orthonormal vectors are orthogonal to both $\left\vert
e_{1}\right\rangle $ and $\left\vert e_{2}\right\rangle $,%
\begin{align}
&  \left\vert e_{23}\right\rangle \overset{\text{def}}{=}\frac{\left\vert
10110\right\rangle -\left\vert 11001\right\rangle }{\sqrt{2}}\text{,
}\left\vert e_{24}\right\rangle \overset{\text{def}}{=}\frac{\left\vert
11100\right\rangle +\left\vert 10011\right\rangle }{\sqrt{2}}\text{,
}\left\vert e_{25}\right\rangle \overset{\text{def}}{=}\frac{\left\vert
11010\right\rangle -\left\vert 10101\right\rangle }{\sqrt{2}}\text{,}%
\nonumber\\
& \nonumber\\
&  \left\vert e_{26}\right\rangle \overset{\text{def}}{=}\frac{\left\vert
00011\right\rangle +\left\vert 01100\right\rangle }{\sqrt{2}}\text{,
}\left\vert e_{27}\right\rangle \overset{\text{def}}{=}\frac{\left\vert
00101\right\rangle -\left\vert 01010\right\rangle }{\sqrt{2}}\text{,
}\left\vert e_{28}\right\rangle \overset{\text{def}}{=}\frac{\left\vert
01001\right\rangle -\left\vert 00110\right\rangle }{\sqrt{2}}\text{.}%
\end{align}
Indeed, it can be explicitly checked that $\left\langle e_{k}\left.
e_{k^{\prime}}\right.  \right\rangle =\delta_{kk^{\prime}}$ for any $k$ and
$k^{\prime}$ in $\left\{  1\text{,..., }28\right\}  $. The last four vectors
are uncovered as follows: first, we add four linearly independent vectors in
such a manner to have a basis for $\mathcal{H}_{2}^{5}$ and then we apply the
Gram-Schmidt orthonormalization procedure to obtain an orthogonal basis. It
finally turns out that the remaining four vectors needed are given by,%
\begin{align}
&  \left\vert e_{29}\right\rangle \overset{\text{def}}{=}\frac{\left\vert
11000\right\rangle -\left(  1-\gamma\right)  \left\vert e_{7}\right\rangle
-\left(  1-\gamma\right)  \left\vert e_{9}\right\rangle -\left(
1-\gamma\right)  \left\vert e_{12}\right\rangle -\sqrt{1-\gamma}\left\vert
e_{16}\right\rangle }{\sqrt{1-3\left(  1-\gamma\right)  ^{2}-\left(
1-\gamma\right)  }}\text{,}\nonumber\\
& \nonumber\\
&  \left\vert e_{30}\right\rangle \overset{\text{def}}{=}\frac{\left\vert
10100\right\rangle -\left(  1-\gamma\right)  \left\vert e_{5}\right\rangle
-\left(  1-\gamma\right)  \left\vert e_{10}\right\rangle -\left(
1-\gamma\right)  \left\vert e_{11}\right\rangle -\sqrt{1-\gamma}\left\vert
e_{18}\right\rangle }{\sqrt{1-3\left(  1-\gamma\right)  ^{2}-\left(
1-\gamma\right)  }}\text{,}\nonumber\\
& \nonumber\\
&  \left\vert e_{31}\right\rangle \overset{\text{def}}{=}\frac{\left\vert
10010\right\rangle -\left(  1-\gamma\right)  \left\vert e_{5}\right\rangle
-\left(  1-\gamma\right)  \left\vert e_{8}\right\rangle +\left(
1-\gamma\right)  \left\vert e_{11}\right\rangle -\sqrt{1-\gamma}\left\vert
e_{20}\right\rangle }{\sqrt{1-3\left(  1-\gamma\right)  ^{2}-\left(
1-\gamma\right)  }}\text{,}\nonumber\\
& \nonumber\\
&  \left\vert e_{32}\right\rangle \overset{\text{def}}{=}\frac{\left\vert
10001\right\rangle -\left(  1-\gamma\right)  \left\vert e_{6}\right\rangle
-\left(  1-\gamma\right)  \left\vert e_{7}\right\rangle +\left(
1-\gamma\right)  \left\vert e_{9}\right\rangle -\sqrt{1-\gamma}\left\vert
e_{22}\right\rangle }{\sqrt{1-3\left(  1-\gamma\right)  ^{2}-\left(
1-\gamma\right)  }}\text{.}%
\end{align}
In conclusion, $\left\{  \left\vert e_{k}\right\rangle \right\}  $ with
$k\in\left\{  1\text{,..., }32\right\}  $ is a suitable orthonormal basis for
$\mathcal{H}_{2}^{5}$.

\section{Estimating the entanglement fidelity}

The entanglement fidelity $\mathcal{F}\left(  \gamma\text{, }\varepsilon
\right)  $ of a $\left(  \left(  n,K,d\right)  \right)  $ quantum code
$\mathcal{C}$ for enlarged GAD errors $A_{a}^{\prime}$ with $a\in\left\{
0\text{,..., }2^{2n}-1\right\}  $ and recovery operation $\mathcal{R}%
\overset{\text{def}}{=}\left\{  R_{1}\text{,..., }R_{r}\text{,..., }%
R_{s}\text{, }R_{s+1}\overset{\text{def}}{=}\hat{O}\right\}  $ reads,%
\begin{equation}
\mathcal{F}\left(  \gamma\text{, }\varepsilon\right)  \overset{\text{def}}%
{=}\frac{1}{K^{2}}\left[
%TCIMACRO{\dsum \limits_{a=0}^{2^{2n}-1}}%
%BeginExpansion
{\displaystyle\sum\limits_{a=0}^{2^{2n}-1}}
%EndExpansion%
%TCIMACRO{\dsum \limits_{r=1}^{s}}%
%BeginExpansion
{\displaystyle\sum\limits_{r=1}^{s}}
%EndExpansion
\left\vert \text{Tr}\left(  R_{r}A_{a}^{\prime}\right)  _{\mathcal{C}%
}\right\vert ^{2}+%
%TCIMACRO{\dsum \limits_{a=0}^{2^{2n}-1}}%
%BeginExpansion
{\displaystyle\sum\limits_{a=0}^{2^{2n}-1}}
%EndExpansion
\left\vert \text{Tr}\left(  \hat{O}A_{a}^{\prime}\right)  _{\mathcal{C}%
}\right\vert ^{2}\right]  \text{,} \label{fa1}%
\end{equation}
where the recovery operator $\hat{O}$ is defined as,%
\begin{equation}
\hat{O}\overset{\text{def}}{=}%
%TCIMACRO{\dsum \limits_{b=1}^{2^{n}-2s}}%
%BeginExpansion
{\displaystyle\sum\limits_{b=1}^{2^{n}-2s}}
%EndExpansion
\left\vert o_{b}\right\rangle \left\langle o_{b}\right\vert \text{.}
\label{fa3}%
\end{equation}
The $2s$ orthonormal vectors employed to construct the recovery operators
$R_{r}$ in $\mathcal{R}\backslash\left\{  \hat{O}\right\}  $ together with the
$2^{n}-2s$ orthonormal vectors used to define the recovery operator $\hat{O}$
form a orthonormal basis of the complex Hilbert space $\mathcal{H}_{2}^{2n}$.
We observe that the entanglement fidelity $\mathcal{F}\left(  \gamma\text{,
}\varepsilon\right)  $ in Eq. (\ref{fa1}) can be regarded as the sum of two
contributions,%
\begin{equation}
\mathcal{F}\left(  \gamma\text{, }\varepsilon\right)  \overset{\text{def}}%
{=}\mathcal{F}_{\mathcal{R}\backslash\left\{  \hat{O}\right\}  }\left(
\gamma\text{, }\varepsilon\right)  +\mathcal{F}_{\hat{O}}\left(  \gamma\text{,
}\varepsilon\right)  \text{,}%
\end{equation}
where,%
\begin{equation}
\mathcal{F}_{\mathcal{R}\backslash\left\{  \hat{O}\right\}  }\left(
\gamma\text{, }\varepsilon\right)  \overset{\text{def}}{=}\frac{1}{K^{2}}%
%TCIMACRO{\dsum \limits_{a=0}^{2^{2n}-1}}%
%BeginExpansion
{\displaystyle\sum\limits_{a=0}^{2^{2n}-1}}
%EndExpansion%
%TCIMACRO{\dsum \limits_{r=1}^{s}}%
%BeginExpansion
{\displaystyle\sum\limits_{r=1}^{s}}
%EndExpansion
\left\vert \text{Tr}\left(  R_{r}A_{a}^{\prime}\right)  _{\mathcal{C}%
}\right\vert ^{2}\text{ and, }\mathcal{F}_{\hat{O}}\left(  \gamma\text{,
}\varepsilon\right)  \overset{\text{def}}{=}\frac{1}{K^{2}}%
%TCIMACRO{\dsum \limits_{a=0}^{2^{2n}-1}}%
%BeginExpansion
{\displaystyle\sum\limits_{a=0}^{2^{2n}-1}}
%EndExpansion
\left\vert \text{Tr}\left(  \hat{O}A_{a}^{\prime}\right)  _{\mathcal{C}%
}\right\vert ^{2}\text{.} \label{fa2}%
\end{equation}
In what follows, we shall describe our reasoning for the analytic estimates of
entanglement fidelities.

\subsection{Part A}

Let us first consider $\mathcal{F}_{\mathcal{R}\backslash\left\{  \hat
{O}\right\}  }\left(  \gamma\text{, }\varepsilon\right)  $ in Eq. (\ref{fa2}).
Recall that for $r\in\left\{  1\text{,..., }s\right\}  $, the recovery
operators $R_{r}$ read%
\begin{equation}
R_{r}\overset{\text{def}}{=}%
%TCIMACRO{\dsum \limits_{j=0}^{K-1}}%
%BeginExpansion
{\displaystyle\sum\limits_{j=0}^{K-1}}
%EndExpansion
\frac{\left\vert j_{L}\right\rangle \left\langle j_{L}\right\vert
A_{r}^{\prime\dagger}}{\sqrt{\left\langle j_{L}\left\vert A_{r}^{\prime
\dagger}A_{r}^{\prime}\right\vert j_{L}\right\rangle }}\text{.} \label{faa}%
\end{equation}
Using Eq. (\ref{faa}), $\mathcal{F}_{\mathcal{R}\backslash\left\{  \hat
{O}\right\}  }\left(  \gamma\text{, }\varepsilon\right)  $ becomes%
\begin{equation}
\mathcal{F}_{\mathcal{R}\backslash\left\{  \hat{O}\right\}  }\left(
\gamma\text{, }\varepsilon\right)  \overset{\text{def}}{=}\frac{1}{K^{2}}%
%TCIMACRO{\dsum \limits_{a=0}^{2^{2n}-1}}%
%BeginExpansion
{\displaystyle\sum\limits_{a=0}^{2^{2n}-1}}
%EndExpansion%
%TCIMACRO{\dsum \limits_{r=1}^{s}}%
%BeginExpansion
{\displaystyle\sum\limits_{r=1}^{s}}
%EndExpansion
\left\vert
%TCIMACRO{\dsum \limits_{i=0}^{K-1}}%
%BeginExpansion
{\displaystyle\sum\limits_{i=0}^{K-1}}
%EndExpansion
\frac{\left\langle i_{L}\left\vert A_{r}^{\prime\dagger}A_{a}^{\prime
}\right\vert i_{L}\right\rangle }{\sqrt{\left\langle i_{L}\left\vert
A_{r}^{\prime\dagger}A_{r}^{\prime}\right\vert i_{L}\right\rangle }%
}\right\vert ^{2}\text{.}%
\end{equation}
Let us denote with $\mathcal{K}$ the index set of all the enlarged GAD error
operators. This set has cardinality $2^{2n}$ and can be decomposed in two
parts,%
\begin{equation}
\mathcal{K}\overset{\text{def}}{=}\mathcal{K}_{\mathcal{R}\backslash\left\{
\hat{O}\right\}  }\oplus\mathcal{K}^{\prime}\text{.}%
\end{equation}
The set $\mathcal{K}_{\mathcal{R}\backslash\left\{  \hat{O}\right\}  }$ is the
index set of all the enlarged GAD errors $\left\{  A_{a}^{\prime}\right\}  $
recovered by means of the recovery operators $\left\{  R_{r}\right\}  $ in
$\mathcal{R}\backslash\left\{  \hat{O}\right\}  $ and it has cardinality $s$.
The cardinality of the index set $\mathcal{K}^{\prime}$ is $2^{2n}-s$ and
denotes the number of all the remaining potentially contributing enlarged
error operators. Since we only aim at finding estimates of $\mathcal{F}\left(
\gamma\text{, }\varepsilon\right)  $ with $\mathcal{F}\left(  \gamma\text{,
}\varepsilon\right)  \gtrsim1-\mathcal{O}\left(  2\right)  $ and assuming as
working hypothesis $0\leq\varepsilon\ll\gamma\ll1$, we only need to consider a
subset of $\mathcal{K}^{\prime}$ for the estimation of $\mathcal{F}%
_{\mathcal{R}\backslash\left\{  \hat{O}\right\}  }\left(  \gamma\text{,
}\varepsilon\right)  $. In the worst scenario, we have to consider the subset
$\mathcal{K}_{\text{weight-}2}\subset$ $\mathcal{K}^{\prime}$ of cardinality
$3^{2}\binom{n}{2}$ of all the weight-$2$ enlarged GAD\ errors, and%
\begin{equation}
\mathcal{F}_{\mathcal{R}\backslash\left\{  \hat{O}\right\}  }\left(
\gamma\text{, }\varepsilon\right)  \approx\frac{1}{K^{2}}%
%TCIMACRO{\dsum \limits_{a\in\mathcal{K}^{\prime\prime}}}%
%BeginExpansion
{\displaystyle\sum\limits_{a\in\mathcal{K}^{\prime\prime}}}
%EndExpansion%
%TCIMACRO{\dsum \limits_{r=1}^{s}}%
%BeginExpansion
{\displaystyle\sum\limits_{r=1}^{s}}
%EndExpansion
\left\vert
%TCIMACRO{\dsum \limits_{i=0}^{K-1}}%
%BeginExpansion
{\displaystyle\sum\limits_{i=0}^{K-1}}
%EndExpansion
\frac{\left\langle i_{L}\left\vert A_{r}^{\prime\dagger}A_{a}^{\prime
}\right\vert i_{L}\right\rangle }{\sqrt{\left\langle i_{L}\left\vert
A_{r}^{\prime\dagger}A_{r}^{\prime}\right\vert i_{L}\right\rangle }%
}\right\vert ^{2}\text{,}%
\end{equation}
with $\mathcal{K}^{\prime\prime}\overset{\text{def}}{=}\mathcal{K}%
_{\mathcal{R}\backslash\left\{  \hat{O}\right\}  }\cup\mathcal{K}%
_{\text{weight-}2}$. In addition, among these $3^{2}\binom{n}{2}$ errors, some
of them are more likely to occur than others. Denote with $\left[  ik\right]
$ the set of cardinality $\binom{n}{2}$ with weight-$2$ enlarged error
operators acting on $n$-qubit quantum states defined by means of single-qubit
errors of type $A_{i}$ and $A_{k}$,%
\begin{equation}
\left[  ik\right]  \overset{\text{def}}{=}\left\{  A_{ik0\text{...}0}\text{,
}A_{i0k0\text{...}0}\text{,..., }A_{00\text{...}0ik}\text{ }\right\}  \text{.}%
\end{equation}
It turns out that the probabilities of occurrence Pr$\left(  \left[
ik\right]  \right)  \overset{\text{def}}{=}$Pr$_{ik}\left(  \gamma\text{,
}\varepsilon\right)  $ of errors in the set $\left[  ik\right]  $ scale in
terms of the perturbation parameters $\gamma$ and $\varepsilon$ as follows,%
\begin{equation}
\left(
\begin{array}
[c]{ccc}%
\text{Pr}_{11} & \text{Pr}_{12} & \text{Pr}_{13}\\
\text{Pr}_{21} & \text{Pr}_{22} & \text{Pr}_{23}\\
\text{Pr}_{31} & \text{Pr}_{32} & \text{Pr}_{33}%
\end{array}
\right)  \approx\left(
\begin{array}
[c]{ccc}%
\gamma^{2} & \varepsilon\gamma & \varepsilon\gamma^{2}\\
\varepsilon\gamma & \varepsilon^{2} & \varepsilon^{2}\gamma\\
\varepsilon\gamma^{2} & \varepsilon^{2}\gamma & \varepsilon^{2}\gamma^{2}%
\end{array}
\right)  \text{.}%
\end{equation}
Therefore, in addition to limit our attention to these $3^{2}\binom{n}{2}$
weight-$2$ errors in $\mathcal{K}_{\text{weight-}2}$, we also rank the
relevance of each of these nine possible subsets $\left[  ik\right]  $ of
cardinality $\binom{n}{2}$ according to the $\gamma$- and $\varepsilon
$-dependences of the probabilities of occurrence of their errors.

\subsection{Part B}

Let us now take into consideration $\mathcal{F}_{\hat{O}}\left(  \gamma\text{,
}\varepsilon\right)  $ in Eq. (\ref{fa2}). We notice that,%
\begin{equation}
\mathcal{F}_{\hat{O}}\left(  \gamma\text{, }\varepsilon\right)  \overset
{\text{def}}{=}\frac{1}{K^{2}}%
%TCIMACRO{\dsum \limits_{a=0}^{2^{2n}-1}}%
%BeginExpansion
{\displaystyle\sum\limits_{a=0}^{2^{2n}-1}}
%EndExpansion
\left\vert \text{Tr}\left(  \hat{O}A_{a}^{\prime}\right)  _{\mathcal{C}%
}\right\vert ^{2}=\frac{1}{K^{2}}%
%TCIMACRO{\dsum \limits_{a=0}^{2^{2n}-1}}%
%BeginExpansion
{\displaystyle\sum\limits_{a=0}^{2^{2n}-1}}
%EndExpansion
\left\vert
%TCIMACRO{\dsum \limits_{i=0}^{K-1}}%
%BeginExpansion
{\displaystyle\sum\limits_{i=0}^{K-1}}
%EndExpansion
\left\langle i_{L}\left\vert \hat{O}A_{a}^{\prime}\right\vert i_{L}%
\right\rangle \right\vert ^{2}\leq\frac{2^{2n}}{K}\left\vert \left\langle
\bar{\imath}_{L}\left\vert \hat{O}A_{\bar{a}}^{\prime}\right\vert \bar{\imath
}_{L}\right\rangle \right\vert ^{2}\text{,} \label{fa6}%
\end{equation}
with,%
\begin{equation}
\left\langle \bar{\imath}_{L}\left\vert \hat{O}A_{\bar{a}}^{\prime}\right\vert
\bar{\imath}_{L}\right\rangle \overset{\text{def}}{=}\max_{i,a}\left\{
\left\langle i_{L}\left\vert \hat{O}A_{a}^{\prime}\right\vert i_{L}%
\right\rangle \right\}  \text{.}%
\end{equation}
If the index $\bar{a}$ labels an enlarged GAD\ error recovered by a recovery
operation $R_{r}$ in $\mathcal{R}\backslash\left\{  \hat{O}\right\}  $, then
$\mathcal{F}_{\hat{O}}\left(  \gamma\text{, }\varepsilon\right)  $ is
identically zero by construction. Therefore, let us assume that $\bar{a}$ is
not such an index. Using Eq. (\ref{fa3}), $\left\langle \bar{\imath}%
_{L}\left\vert \hat{O}A_{\bar{a}}^{\prime}\right\vert \bar{\imath}%
_{L}\right\rangle $ becomes,%
\begin{equation}
\left\langle \bar{\imath}_{L}\left\vert \hat{O}A_{\bar{a}}^{\prime}\right\vert
\bar{\imath}_{L}\right\rangle =%
%TCIMACRO{\dsum \limits_{b=1}^{2^{n}-2s}}%
%BeginExpansion
{\displaystyle\sum\limits_{b=1}^{2^{n}-2s}}
%EndExpansion
\left\langle \bar{\imath}_{L}\left\vert o_{b}\right.  \right\rangle
\left\langle o_{b}\left\vert A_{\bar{a}}^{\prime}\right\vert \bar{\imath}%
_{L}\right\rangle \text{.} \label{fa5}%
\end{equation}
We observe that $\left\langle \bar{\imath}_{L}\left\vert o_{b}\right.
\right\rangle $ would be identically zero for any $b\in\left\{  1\text{,...,
}2^{n}-2s\right\}  $ if the enlarged identity error operator belonged to the
error model. Unfortunately, this is not our case. Nevertheless, it turns out
that%
\begin{equation}
\left\langle \bar{\imath}_{L}\left\vert o_{b}\right.  \right\rangle
\approx\gamma\left\langle \bar{\imath}_{L}\left\vert T^{\prime}\right\vert
o_{b}\right\rangle \text{,} \label{fa4}%
\end{equation}
where the operator $T^{\prime}$ acting on $n$-qubit quantum states is defined
as,%
\begin{equation}
T^{\prime}\overset{\text{def}}{=}T^{1}\otimes I^{2}\otimes\text{...}\otimes
I^{n-1}\otimes I^{n}+I^{1}\otimes T^{2}\otimes\text{...}\otimes I^{n-1}\otimes
I^{n}+I^{1}\otimes I^{2}\otimes\text{...}\otimes I^{n-1}\otimes T^{n}\text{,}%
\end{equation}
with,%
\begin{equation}
T^{k}\overset{\text{def}}{=}\frac{1}{4}\left(  I^{k}-\sigma_{z}^{k}\right)
\text{.}%
\end{equation}
Substituting Eq. (\ref{fa4}) into Eq. (\ref{fa5}), we get%
\begin{equation}
\left\langle \bar{\imath}_{L}\left\vert \hat{O}A_{\bar{a}}^{\prime}\right\vert
\bar{\imath}_{L}\right\rangle \approx\gamma%
%TCIMACRO{\dsum \limits_{b=1}^{2^{n}-2s}}%
%BeginExpansion
{\displaystyle\sum\limits_{b=1}^{2^{n}-2s}}
%EndExpansion
\left\langle \bar{\imath}_{L}\left\vert T^{\prime}\right\vert o_{b}%
\right\rangle \left\langle o_{b}\left\vert A_{\bar{a}}^{\prime}\right\vert
\bar{\imath}_{L}\right\rangle \text{.} \label{fa7}%
\end{equation}
We remark that both $\left\vert \bar{\imath}_{L}\right\rangle $ and
$T^{\prime}$ are $\gamma$-independent quantities while the states $\left\vert
o_{b}\right\rangle $ are the sum-decomposition of $n$-qubit quantum states
where $\gamma$-dependent expansion coefficients may appear. However, as we
have noticed from our explicit construction in the previous appendix, these
coefficients do not exhibit nontrivial $\gamma$-dependence in the limit of
$\gamma$ approaching zero. Therefore, terms like $\left\langle \bar{\imath
}_{L}\left\vert T^{\prime}\right\vert o_{b}\right\rangle $ do not possess
relevant scaling laws in the damping probability parameter $\gamma$
approaching zero. On the contrary, terms like $\left\langle o_{b}\left\vert
A_{\bar{a}}^{\prime}\right\vert \bar{\imath}_{L}\right\rangle $ do exhibit
important $\gamma$-scaling laws,%
\begin{equation}
\left\langle o_{b}\left\vert A_{\bar{a}}^{\prime}\right\vert \bar{\imath}%
_{L}\right\rangle \approx\gamma^{\frac{\text{wt}\left(  A_{\bar{a}}^{\prime
}\right)  }{2}}\left\langle o_{b}\left\vert \tilde{A}_{\bar{a}}^{\prime
}\right\vert \bar{\imath}_{L}\right\rangle \text{,} \label{fa8}%
\end{equation}
where wt$\left(  A_{\bar{a}}^{\prime}\right)  $ denotes the weight of the
operator $A_{\bar{a}}^{\prime}$ and $\left\langle o_{b}\left\vert \tilde
{A}_{\bar{a}}^{\prime}\right\vert \bar{\imath}_{L}\right\rangle $ is redefined
in such a manner to have no relevant $\gamma$-dependence. Substituting Eqs.
(\ref{fa7}) and (\ref{fa8}) into Eq. (\ref{fa6}), we finally obtain%
\begin{equation}
\mathcal{F}_{\hat{O}}\left(  \gamma\text{, }\varepsilon\right)  \lesssim
\frac{2^{2n}}{K}\gamma^{2+\text{wt}\left(  A_{\bar{a}}^{\prime}\right)
}\left\vert
%TCIMACRO{\dsum \limits_{b=1}^{2^{n}-2s}}%
%BeginExpansion
{\displaystyle\sum\limits_{b=1}^{2^{n}-2s}}
%EndExpansion
\left\langle \bar{\imath}_{L}\left\vert T^{\prime}\right\vert o_{b}%
\right\rangle \left\langle o_{b}\left\vert \tilde{A}_{\bar{a}}^{\prime
}\right\vert \bar{\imath}_{L}\right\rangle \right\vert ^{2}\text{.}
\label{fa9}%
\end{equation}
From Eq. (\ref{fa9}), we conclude that while $\mathcal{F}_{\hat{O}}\left(
\gamma\text{, }\varepsilon\right)  $ could be, in principle, nonvanishing and
contribute to the computation of the entanglement fidelity $\mathcal{F}\left(
\gamma\text{, }\varepsilon\right)  $, its contribution is negligible given to
order of approximations chosen for our analytic estimates.

In what follows, we report a more explicit example. For the sake of reasoning,
consider the CSS seven-qubit code and AD errors. Does any weight-$2$ enlarged
error operator contribute to the computation of the entanglement fidelity? In
general, errors $A_{l}$ for which no corresponding recovery operator
$R_{A_{l}\text{ }}$is constructed can contribute to the computation of the
entanglement fidelity via the expression given by,%
\begin{equation}
\mathcal{F}_{\hat{O}}^{\left[  \left[  7,1,3\right]  \right]  }\left(
\gamma\right)  \overset{\text{def}}{=}\frac{1}{4}\sum_{l,k}\left[
\left\langle 0_{L}\left\vert v_{k}\right.  \right\rangle \left\langle
v_{k}\left\vert A_{l}\right\vert 0_{L}\right\rangle +\left\langle
1_{L}\left\vert v_{k}\right.  \right\rangle \left\langle v_{k}\left\vert
A_{l}\right\vert 1_{L}\right\rangle \right]  ^{2}\text{,} \label{fo713}%
\end{equation}
where the operator $\hat{O}$ reads,%
\begin{equation}
\hat{O}\overset{\text{def}}{=}\sum_{k}\left\vert v_{k}\right\rangle
\left\langle v_{k}\right\vert \text{.}%
\end{equation}
We notice that state vectors $\left\{  \left\vert v_{k}\right\rangle \right\}
$ in Eq. (\ref{fo713}) lead to non-vanishing contributions provided that: i)
$\left\vert v_{k}\right\rangle $ has some non-zero component along $\left\vert
0_{L}\right\rangle $ and/or $\left\vert 1_{L}\right\rangle $; ii) $\left\vert
v_{k}\right\rangle $ has some non-zero component along $A_{l}\left\vert
0_{L}\right\rangle $ and/or $A_{l}\left\vert 1_{L}\right\rangle $; iii)
$\left\{  \left\vert v_{k}\right\rangle \text{, }A_{l}^{\text{correctable}%
}\left\vert i_{L}\right\rangle \right\}  $ forms an orthonormal basis of
$\mathcal{H}_{2}^{7}$.

Before proceeding along this line of reasoning, we would like to emphasize
that we could have proceeded along a alternative route. Observe that
$D\overset{\text{def}}{=}\dim_{%
%TCIMACRO{\U{2102} }%
%BeginExpansion
\mathbb{C}
%EndExpansion
}\mathcal{H}_{2}^{7}=2^{7}=128$. Furthermore, the cardinality $c^{\prime}$ of
the set $S^{\prime}$ of linearly independent and orthogonal vectors in the
sum-decomposition of the correctable weight-$0$ and weight-$1$ enlarged errors
is given by,%

\begin{equation}
c^{\prime}\overset{\text{def}}{=}2\times\binom{7}{0}\times8+2\times\binom
{7}{1}\times4=72\text{.}%
\end{equation}
Therefore, there also exists a new set $S^{\prime\prime}$ that consists of
$c^{\prime\prime}\overset{\text{def}}{=}D-c^{\prime}=56$ linearly independent
and orthogonal vectors such that vectors in $S^{\prime}$ and $S^{\prime\prime
}$ form an orthonormal basis of $\mathcal{H}_{2}^{7}$. If any of these $56$
vectors is in the vector sum-decomposition of any of the $\binom{7}{2}=21$
weight-$2$ errors, then some weight-$2$ error could contribute to the
entanglement fidelity provided that the selected vector in $S^{\prime\prime}$
has some non-zero component along $\left\vert 0_{L}\right\rangle $ and/or
$\left\vert 1_{L}\right\rangle $ (see Eq. (\ref{fo713})). We checked that all
the $2\times\binom{7}{2}\times2=84$ state vectors in the sum-decomposition of
$A_{l}\left\vert i_{L}\right\rangle $ with $A_{l}$ any weight-$2$ error
operator are orthogonal to $\left\vert i_{L}\right\rangle $ with $i\in\left\{
0\text{, }1\right\}  $. Therefore, we cannot take this short-cut and are
forced to proceed in a more general manner. As a consequence, constructing
these vectors $\left\vert v_{k}\right\rangle $ is going to be more involved,
since we wish to provide a constructive explanation avoiding numerics that
could obscure the construction itself.

Let us then return to the more general approach. For the sake of clarity,
fucus on the possible partial recovery of the weight-$2$ error $A_{1100000}$.
This error is not correctable because it is incompatible with the weight-$1$
error $A_{0010000}$. Having observed this and recalling the three conditions
for good state vectors $\left\{  \left\vert v_{k}\right\rangle \right\}  $, it
turns out that a suitable vector $\left\vert v_{\bar{k}}\right\rangle $ in the
definition of $\hat{O}$ \ so that $\left\vert v_{\bar{k}}\right\rangle
\left\langle v_{\bar{k}}\right\vert $ can partially recover $A_{1100000}$
reads,%
\begin{equation}
\left\vert v_{\bar{k}}\right\rangle \overset{\text{def}}{=}\frac{\left\vert
0000110\right\rangle -\left\vert 1100000\right\rangle +\left\vert
0110011\right\rangle -\left(  1-\gamma\right)  ^{2}\left\vert
0000000\right\rangle }{\sqrt{3+\left(  1-\gamma\right)  ^{4}}}\text{.}%
\end{equation}
This contribution of $A_{1100000}$ to the computation of the entanglement
fidelity becomes,

\begin{figure}[ptb]
\centering
\includegraphics[width=0.5\textwidth]{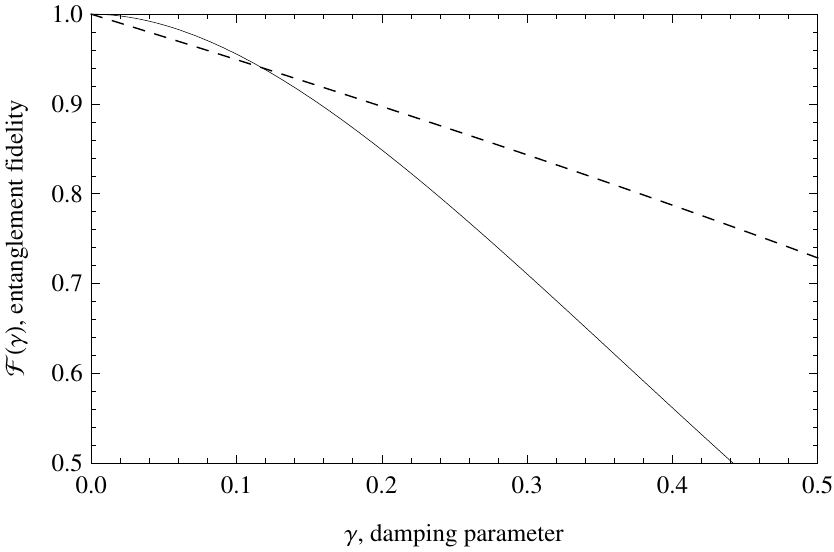}\caption{ \textit{Performance of the
CSS seven-qubit code}. The non-truncated expression of the entanglement
fidelity $\mathcal{F}\left(  \gamma\right)  $ vs. the amplitude damping
parameter $\gamma$ in the presence (thin solid line) and absence (dashed line)
of quantum error correction.}%
\label{figure7}%
\end{figure}%

\begin{equation}
\mathcal{F}_{\left\vert v_{\bar{k}}\right\rangle \left\langle v_{\bar{k}%
}\right\vert }^{\left[  \left[  7,1,3\right]  \right]  }\left(  \gamma\right)
\overset{\text{def}}{=}\frac{1}{4}\left(  \left\langle 0_{L}\left\vert
v_{\bar{k}}\right.  \right\rangle \left\langle v_{\bar{k}}\left\vert
A_{1100000}\right\vert 0_{L}\right\rangle \right)  ^{2}=\frac{1}{32}\left(
\frac{1-\left(  1-\gamma\right)  ^{2}}{\sqrt{3+\left(  1-\gamma\right)  ^{4}}%
}\times\frac{\gamma\left(  1-\gamma\right)  }{\sqrt{3+\left(  1-\gamma\right)
^{4}}}\right)  ^{2}\approx\gamma^{4}\text{.} \label{vediaoh}%
\end{equation}
From Eq. (\ref{vediaoh}), we conclude that although $A_{1100000}$ contributes
to the computation of the entanglement fidelity, its contribution is
negligible given our chosen order of approximation. Analogously, for each
weight-$2$ enlarged error operator, a similar line of reasoning can be carried out.

\section{On the CSS seven-qubit code}

For $\varepsilon=0$, the non-truncated expression for the entanglement
fidelity reads,%
\begin{equation}
\mathcal{F}_{\text{non-truncated}}^{\left[  \left[  7,1,3\right]  \right]
}\left(  \gamma\right)  \overset{\text{def}}{=}\frac{1}{4}\left(  \sqrt
{\frac{1+7\left(  1-\gamma\right)  ^{4}}{8}}+\sqrt{\frac{\left(
1-\gamma\right)  ^{7}+7\left(  1-\gamma\right)  ^{3}}{8}}\right)  ^{2}%
+\frac{7}{4}\left(  \sqrt{\frac{4\gamma\left(  1-\gamma\right)  ^{3}}{8}%
}+\sqrt{\frac{\gamma\left(  1-\gamma\right)  ^{6}+3\gamma\left(
1-\gamma\right)  ^{2}}{8}}\right)  ^{2}\text{.}%
\end{equation}
The Taylor-expansion of $\mathcal{F}^{\left[  \left[  7,1,3\right]  \right]
}\left(  \gamma\right)  $ up to the $10$th-order is given by,%
\begin{equation}
\mathcal{F}_{\text{non-truncated}}^{\left[  \left[  7,1,3\right]  \right]
}\left(  \gamma\right)  \approx1-\frac{21}{4}\gamma^{2}+\frac{35}{4}\gamma
^{3}-\frac{63}{8}\gamma^{4}+\frac{609}{128}\gamma^{5}-\frac{315}{256}%
\gamma^{6}-\frac{51}{256}\allowbreak\gamma^{7}-\frac{63}{256}\gamma^{8}%
+\frac{1701}{8192}\gamma^{9}+O\left(  \gamma^{10}\right)  \text{,}%
\end{equation}
while the $1$-qubit baseline performance is given by,%
\begin{equation}
\mathcal{F}_{\text{baseline}}^{1\text{-qubit}}\left(  \gamma\right)
\overset{\text{def}}{=}2^{-2}\left(  1+\sqrt{1-\gamma}\right)  ^{2}\text{.}%
\end{equation}
We checked the good overlap between our results (non truncated fidelity
expressions with and without error correction) and the ones plotted in
\cite{andy3}. See also Fig. $7$. \begin{figure}[ptb]
\centering
\includegraphics[width=0.5\textwidth]{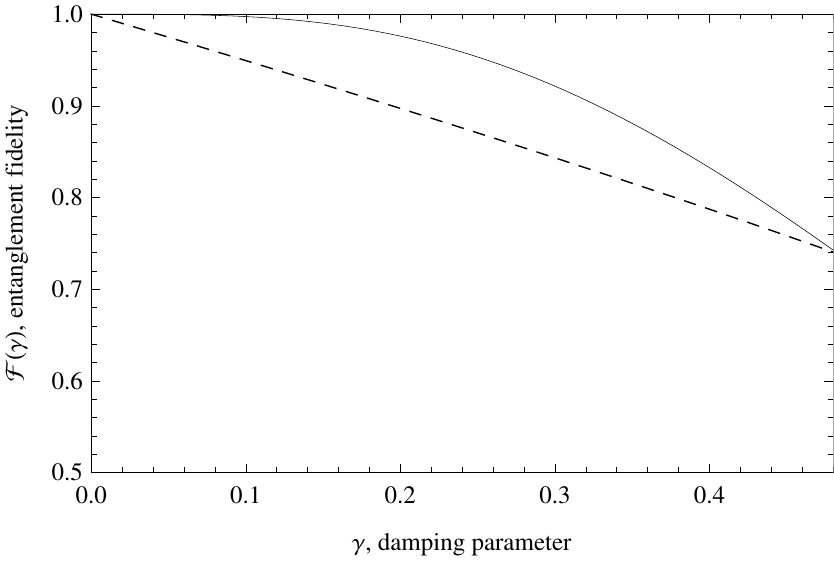}\caption{\textit{Performance of the
Shor nine-qubit code}. The non-truncated expression of the entanglement
fidelity $\mathcal{F}\left(  \gamma\right)  $ vs. the amplitude damping
parameter $\gamma$ in the presence (thin solid line) and absence (dashed line)
of quantum error correction.}%
\label{figure8}%
\end{figure}

\section{On the Shor nine-qubit code}

For $\varepsilon=0$, the non-truncated expression for the entanglement
fidelity reads,%
\begin{equation}
\mathcal{F}_{\text{non-truncated}}^{\left[  \left[  9,1,3\right]  \right]
}\left(  \gamma\right)  \overset{\text{def}}{=}1-\frac{3}{2}\gamma^{3}%
-\frac{135}{8}\allowbreak\gamma^{4}+\frac{513}{8}\gamma^{5}-\frac{201}%
{2}\gamma^{6}+\frac{675}{8}\gamma^{7}-\frac{297}{8}\gamma^{8}+\frac{53}%
{8}\gamma^{9}\text{,}%
\end{equation}
while the $1$-qubit baseline performance is given by,%
\begin{equation}
\mathcal{F}_{\text{baseline}}^{1\text{-qubit}}\left(  \gamma\right)
\overset{\text{def}}{=}2^{-2}\left(  1+\sqrt{1-\gamma}\right)  ^{2}\text{.}%
\end{equation}
We checked the good overlap between our results (non truncated fidelity
expressions with and without error correction) and the ones plotted in
\cite{andy1}. See also Fig. $8$.

\end{document}